\documentclass[useAMS,usenatbib]{mn2e}
\usepackage{graphicx}
\usepackage{lscape}
\usepackage{listings}
\usepackage{url}
\usepackage{array}
\usepackage{float}
\usepackage{color}
\usepackage{listings}

\newfloat{Query}{H}{myc}

\newcommand{\Msun}{\mbox{M$_{\odot}$}}
\newcommand{\Lsun}{\mbox{$L_{\odot}$}} 
\newcommand{\Rsun}{\mbox{$R_{\odot}$}} 
\newcommand{\ltsimeq}{\raisebox{-0.6ex}{$\,\stackrel
        {\raisebox{-.2ex}{$\textstyle <$}}{\sim}\,$}}
\newcommand{\gtsimeq}{\raisebox{-0.6ex}{$\,\stackrel
        {\raisebox{-.2ex}{$\textstyle >$}}{\sim}\,$}}
 
\title[76 T dwarfs]{Seventy six T~dwarfs from the UKIDSS LAS: benchmarks, kinematics and an updated space density}
\author[Ben Burningham et al]{Ben Burningham$^{1,2}$\thanks{E-mail:
    B.Burningham@herts.ac.uk}, C. V. Cardoso$^{3,1}$, L. Smith$^{1}$, S. K. Leggett$^{4}$, R. L. Smart$^{3}$,
\newauthor
A. W. Mann$^{5}$, S. Dhital$^{6}$, P.W. Lucas$^{1}$, C. G. Tinney$^{7,8}$, D. J. Pinfield$^{1}$, Z. Zhang$^{1}$, 
\newauthor
C. Morley$^{9}$, D.Saumon$^{10}$, K. Aller$^{5}$, S. P. Littlefair$^{11}$, D. Homeier$^{12,13}$, N. Lodieu$^{14,15}$, 
\newauthor
N. Deacon$^{16}$, M. S.Marley$^{17}$,  L. van Spaandonk$^{1}$, D. Baker$^{1}$, F. Allard$^{13}$, A. H. Andrei$^{2,3,18,19}$, 
\newauthor
J. Canty$^{1}$, J. Clarke$^{1}$, A. C. Day-Jones$^{20,1}$, T. Dupuy$^{21}$, J. J. Fortney$^{9}$,
\newauthor
  J. Gomes$^{1}$, M. Ishii$^{22}$, H. R. A. Jones$^{1}$,  M. Liu$^{5}$, A. Magazz\'u$^{23}$, F. Marocco$^{1}$, 
\newauthor
D. N. Murray$^{1}$, B. Rojas-Ayala$^{24}$, M. Tamura$^{25}$ 
\\
$^{1}$ Centre for Astrophysics Research, Science and Technology Research Institute, University of Hertfordshire, Hatfield AL10 9AB, UK\\
$^{2}$ Observat\'orio Nacional,  Rua General Jos\'e Cristino, 77 - S\~ao Crist\'ov\~ao, Rio de Janeiro - RJ, 20921-400, Brazil\\
$^{3}$ Istituto Nazionale di Astrofisica, Osservatorio Astrofisico di Torino, Strada Osservatorio 20, 10025 Pino Torinese, Italy \\
$^{4}$ Gemini Observatory, 670 N. A'ohoku Place, Hilo, HI 96720, USA \\
$^{5}$ Institute for Astronomy, University of Hawai`i, 2680 Woodlawn Drive, Honolulu, HI 96822, USA \\
$^{6}$ Department of Astronomy, Boston University,725 Commonwealth Ave, Boston MA 02215, USA \\
$^{7}$  Australian Centre for Astrobiology, University of New South Wales, 2052, Australia \\
$^{8}$  School of Physics, University of New South Wales, 2052, Australia\\
$^{9}$  Department of Astronomy and Astrophysics, University of California, Santa Cruz, CA 95064, USA\\
$^{10}$ Los Alamos National Laboratory, P.O. Box 1663, MS F663, Los Alamos, NM 87545, USA \\
$^{11}$ Department of Physics and Astronomy, University of Sheffield, Sheffield S3 7RH, UK\\
$^{12}$ Institut fur Astrophysik, Georg-August-Universitat, Friedrich-Hund-Platz 1, 37077 Gottingen, Germany \\
$^{13}$ C.R.A.L. (UMR 5574 CNRS), Ecole Normale Superieure, 69364 Lyon Cedex 07, France \\
$^{14}$ Instituto de Astrof\'isica de Canarias (IAC), Calle V\'ia L\'actea s/n, E-38200 La Laguna, Tenerife, Spain\\
$^{15}$ Instituto de Astrof\'isica de Canarias, 38200 La Laguna, Spain\\
$^{16}$ Max Planck Institute for Astronomy, Koenigstuhl 17, D-69117 Heidelberg, Germany\\
$^{17}$ NASA Ames Research Center, Mail Stop 245-3, Moffett Field, CA 94035, USA \\
$^{18}$  Shanghai Astronomical Observatory/CAS, 80 Nandan Road, Shanghia 200030, China\\
$^{19}$ Observat\'orio do Valongo/UFRJ, Ladeira Pedro Antonio 43, Rio de Janeiro - RJ, 20080-090, Brazil\\
$^{20}$ Universidad de Chile,Camino el Observatorio \# 1515, Santiago, Chile, Casilla 36-D\\
$^{21}$ Harvard-Smithsonian Center for Astrophysics, 60 Garden Street, Cambridge, MA 02138, USA  \\
$^{22}$Subaru Telescope, 650 North A'ohoku Place, Hilo, Hi 96720, USA \\
$^{23}$ Fundaci\'on Galileo Galilei - INAF, Apartado 565, E-38700 Santa Cruz de La Palma, Spain\\
$^{24}$ Department of Astrophysics, American Museum of Natural History, Central Park West at 79th Street, New York, NY 10024, USA \\
$^{25}$ National Astronomical Observatory, Mitaka, Tokyo 181-8588\\
}

\begin{document}
%
%  These Macros are taken from the AAS TeX macro package version 4.0.
%  Include this file in your LaTeX source only if you are not using
%  the AAS TeX macro package and need to resolve the macro definitions
%  in the BibTeX entries returned by the ADS abstract service.
%
%  If you plan not to use this file to resolve the journal macros
%  rather than the whole AAS TeX macro package, you should save the
%  file as ``aas_macros.sty'' and then include it in your paper by
%  using a construct such as:
%	\documentstyle[11pt,aas_macros]{article}
%
%  For more information on the AASTeX macro package, please see the URL
%	http://www.aas.org/publications/aastex.html
%  For more information about ADS abstract server, please see the URL
%	http://adswww.harvard.edu/ads_abstracts.html
%

% Abbreviations for journals.  The object here is to provide authors
% with convenient shorthands for the most "popular" (often-cited)
% journals; the author can use these markup tags without being concerned
% about the exact form of the journal abbreviation, or its formatting.
% It is up to the keeper of the macros to make sure the macros expand
% to the proper text.  If macro package writers agree to all use the
% same TeX command name, authors only have to remember one thing, and
% the style file will take care of editorial preferences.  This also
% applies when a single journal decides to revamp its abbreviating
% scheme, as happened with the ApJ (Abt 1991).

\def\aj{\rm{AJ}}                   % Astronomical Journal
\def\araa{\rm{ARA\&A}}             % Annual Review of Astron and Astrophys
\def\apj{\rm{ApJ}}                 % Astrophysical Journal
\def\apjl{\rm{ApJ}}                % Astrophysical Journal, Letters
\def\apjs{\rm{ApJS}}               % Astrophysical Journal, Supplement
\def\ao{\rm{Appl.~Opt.}}           % Applied Optics
\def\apss{\rm{Ap\&SS}}             % Astrophysics and Space Science
\def\aap{\rm{A\&A}}                % Astronomy and Astrophysics
\def\aapr{\rm{A\&A~Rev.}}          % Astronomy and Astrophysics Reviews
\def\aaps{\rm{A\&AS}}              % Astronomy and Astrophysics, Supplement
\def\azh{\rm{AZh}}                 % Astronomicheskii Zhurnal
\def\baas{\rm{BAAS}}               % Bulletin of the AAS
\def\jrasc{\rm{JRASC}}             % Journal of the RAS of Canada
\def\memras{\rm{MmRAS}}            % Memoirs of the RAS
\def\mnras{\rm{MNRAS}}             % Monthly Notices of the RAS
\def\pra{\rm{Phys.~Rev.~A}}        % Physical Review A: General Physics
\def\prb{\rm{Phys.~Rev.~B}}        % Physical Review B: Solid State
\def\prc{\rm{Phys.~Rev.~C}}        % Physical Review C
\def\prd{\rm{Phys.~Rev.~D}}        % Physical Review D
\def\pre{\rm{Phys.~Rev.~E}}        % Physical Review E
\def\prl{\rm{Phys.~Rev.~Lett.}}    % Physical Review Letters
\def\pasp{\rm{PASP}}               % Publications of the ASP
\def\pasj{\rm{PASJ}}               % Publications of the ASJ
\def\qjras{\rm{QJRAS}}             % Quarterly Journal of the RAS
\def\skytel{\rm{S\&T}}             % Sky and Telescope
\def\solphys{\rm{Sol.~Phys.}}      % Solar Physics
\def\sovast{\rm{Soviet~Ast.}}      % Soviet Astronomy
\def\ssr{\rm{Space~Sci.~Rev.}}     % Space Science Reviews
\def\zap{\rm{ZAp}}                 % Zeitschrift fuer Astrophysik
\def\nat{\rm{Nature}}              % Nature
\def\iaucirc{\rm{IAU~Circ.}}       % IAU Cirulars
\def\aplett{\rm{Astrophys.~Lett.}} % Astrophysics Letters
\def\apspr{\rm{Astrophys.~Space~Phys.~Res.}}
                % Astrophysics Space Physics Research
\def\bain{\rm{Bull.~Astron.~Inst.~Netherlands}} 
                % Bulletin Astronomical Institute of the Netherlands
\def\fcp{\rm{Fund.~Cosmic~Phys.}}  % Fundamental Cosmic Physics
\def\gca{\rm{Geochim.~Cosmochim.~Acta}}   % Geochimica Cosmochimica Acta
\def\grl{\rm{Geophys.~Res.~Lett.}} % Geophysics Research Letters
\def\jcp{\rm{J.~Chem.~Phys.}}      % Journal of Chemical Physics
\def\jgr{\rm{J.~Geophys.~Res.}}    % Journal of Geophysics Research
\def\jqsrt{\rm{J.~Quant.~Spec.~Radiat.~Transf.}}
                % Journal of Quantitiative Spectroscopy and Radiative Transfer
\def\memsai{\rm{Mem.~Soc.~Astron.~Italiana}}
                % Mem. Societa Astronomica Italiana
\def\nphysa{\rm{Nucl.~Phys.~A}}   % Nuclear Physics A
\def\physrep{\rm{Phys.~Rep.}}   % Physics Reports
\def\physscr{\rm{Phys.~Scr}}   % Physica Scripta
\def\planss{\rm{Planet.~Space~Sci.}}   % Planetary Space Science
\def\procspie{\rm{Proc.~SPIE}}   % Proceedings of the SPIE

\let\astap=\aap
\let\apjlett=\apjl
\let\apjsupp=\apjs
\let\applopt=\ao

\maketitle

\begin{abstract}
We report the discovery of 76 new T~dwarfs from the UKIDSS Large Area Survey (LAS). Near-infrared broad and narrow-band photometry and spectroscopy are presented for the new objects, along with WISE  and warm-Spitzer photometry. Proper motions for 128 UKIDSS T~dwarfs are presented from a new two epoch LAS proper motion catalogue. We use these motions to identify two new benchmark systems: LHS~6176AB, a T8+M4 pair and HD118865AB, a T5.5+F8 pair. Using age constraints from the primaries and evolutionary models to constrain the radii we have estimated their physical properties from their bolometric luminosity.  We compare the colours and properties of known benchmark T~dwarfs to the latest model atmospheres and draw two principal conclusions. Firstly, it appears that the $H-{\rm [4.5]}$ and $J - W2$ colours are more sensitive to metallicity than has previously been recognised, such that differences in metallicity may dominate over differences in $T_{\rm eff}$ when considering relative properties of cool objects using these colours.  Secondly, the previously noted apparent dominance of young objects in the late-T dwarf sample  is no longer apparent when using the new model grids and the expanded sample of late-T~dwarfs and benchmarks. This is supported by the apparently similar distribution of late-T~dwarfs and earlier-type T~dwarfs on reduced proper motion diagrams that we present. Finally, we present updated space densities for the late-T~dwarfs, and compare our values to simulation predictions and those from WISE.
\end{abstract}


\section{Introduction}
\label{sec:intro}

The current generation of wide field surveys is bringing about a step change in our understanding of the coolest and lowest mass components of the Solar neighbourhood. The total number of cool T~dwarfs, substellar objects with $1400$~K$\gtsimeq T_{\rm eff} \gtsimeq 500$~K, has been taken into the hundreds by infrared surveys such as the UKIRT Infrared Deep Sky Survey \citep[UKIDSS; ][]{ukidss}, the Canada-France Brown Dwarf Survey \citep[CFBDS; e.g. ][]{delorme10} and most recently the Wide field Infrared Survey Explorer \citep[WISE; ][]{wise}. 
The last of these, which is an all-sky mid-infrared survey,  has extended the substellar census to well below $T_{\rm eff} = 500$~K, and the adoption of a new spectral class ``Y" has been suggested to classify these new extremely cool objects \citep{cushing2011,kirkpatrick2012}. 
The VISTA Hemisphere Survey (McMahon et al 2012, in prep) and VIKING survey are now also adding to the census \citep{pinfield2012,lod12b}. 

Our exploitation of the UKIDSS Large Area Survey (LAS) has focused on using the photometric characteristics of mid-to late-T~dwarfs at red optical and near-infrared wavelengths \citep[see e.g. ][ for a review of the L and T spectral classes]{kirkpatrick05}  to select a statistically complete sample of T~dwarfs across the T6--T8+ range \citep{lod07,pinfield08,ben10b}. This allowed us to identify an apparent dearth of late-T~dwarfs in the Solar neighbourhood compared to Monte Carlo simulations based on functional forms of the initial mass function (IMF) that have been fitted to observations of the substellar component of young clusters \citep{pinfield08,ben10b}.  

Interpreting this result is hampered by the inherently indirect nature of the observations, and the problems associated with determining the properties of a mixed-age population of brown dwarfs, which by their nature have no single mass-radius relationship. \citet{sandy10} found that colours of the late-T~dwarfs that were identified in UKIDSS suggest they are a predominantly young and low-mass population from comparisons to the model atmospheres of \citet{sm08}. This surprising result could have significant bearing on the interpretation of observed space densities of late-T~dwarfs.

Benchmark brown dwarfs provide the opportunity to break the degeneracies in age, mass and metallicity that hamper the characterisation of cool substellar objects \citep{pinfield06}. As part of our search of the UKIDSS LAS, and more recently WISE and VISTA we have identified several wide binary systems that allow fiducial constraints to be placed on the properties of the T~dwarf secondary component \citep{ben09,ben2011b,adj2011,pinfield2012}.  Comparisons of these objects with both the BT Settl \citep{btsettlCS16} model colours, and those of \citet{sm08} also appear to support the result of \citet{sandy10}.

In this paper we present the results of the extension of our search of the UKIDSS LAS up to and including the sky available in Data Release 9 (DR9), with near complete follow-up of Data Release 8 (DR8). We have used the significantly enhanced sample of late~T~dwarfs, along with 2 epoch UKIDSS LAS proper motions from the catalogue of Smith et al (in prep) to perform a systematic search for wide binary benchmark objects, and compare the observed properties of the benchmark sample and the wider UKIDSS sample with the latest model prediction from \citet{saumon2012} and \citet{morley2012}.  We also use the proper motions for our sample to briefly investigate the kinematics of UKIDSS T~dwarfs.  Finally, we provide an updated estimate of the space densities of T6--T8+ dwarfs, and discuss possible routes to reconciling the observations of the field and young clusters.

\section{Candidate selection}
\label{sec:sel}
Our initial candidate selection followed a similar method to that
described in \citet{pinfield08} and \citet{ben10b}, which we summarise
here.  The selection process consists of two channels: 1) those sources
that are detected in the three UKIDSS LAS $YJH$ bands (the $YJH$ channel);
and 2) those sources that are only detected in the $YJ$ bands (the
$YJ$-only channel).  We did not employ WISE data for guiding our selection since the all-sky catalogue became available part way through our follow-up campaign, and consistent selection is crucial for establishing a well characterised statistical sample. Additionally, for fainter T6 and earlier dwarfs,  the WISE faint limits are effectively shallower than the UKIDSS LAS.

\subsection{The $YJH$ selection channel}
\label{sec:yjhsel}

Our $YJH$ selection channel requires sources to lie within the UKIDSS
LAS sky that overlaps with the SDSS DR8 footprint and have the following
photometric characteristics:

\begin{itemize}
\item $J-H < 0.1$
\item $J-K < 0.1$ or $K$ band non-detection
\item $z' - J > 2.5$ or no SDSS detection within 2\arcsec
\end{itemize}

\begin{figure}
\includegraphics[height=250pt, angle=90]{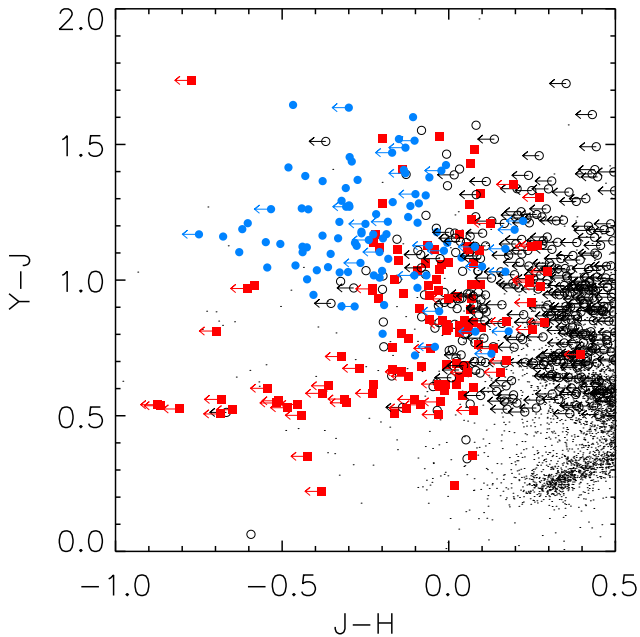}
\caption{$YJH$ colour-colour plot showing UKIDSS LAS photometry of candidate T~dwarfs from DR5--DR8.  Confirmed T~dwarfs are shown with blue filled circles, with arrows indicating limits on $J-H$ colours for candidates from the $YJ$-only channel. Limits are for illustrative purposes, and are based on the canonical 5$\sigma$ depth  for the LAS of $H = 18.8$. Rejected candidates are shown with filled red squares, whilst yet-to-be followed up targets are shown with black open circles. For reference, field stars from a randomly selected 1 square degree region of LAS sky are shown as black dots. 
}
\label{fig:colsel}
\end{figure}

\begin{figure}
\includegraphics[height=250pt, angle=90]{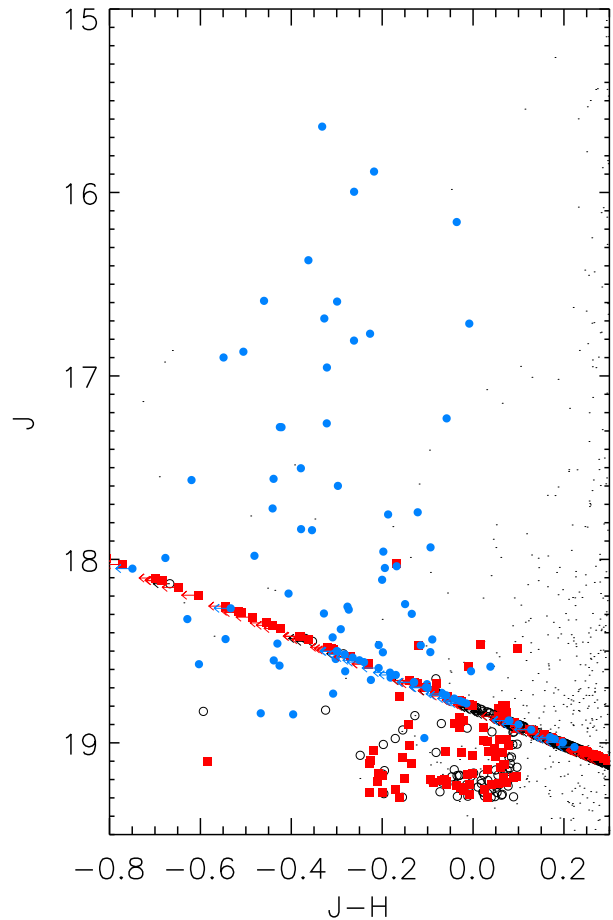}
\caption{$JH$ colour-magnitude plot showing UKIDSS LAS photometry of candidate T~dwarfs from DR5--DR8.  Confirmed T~dwarfs are shown with blue filled circles, with arrows indicating limits on $J-H$ colours for candidates from the $YJ$-only channel. Limits are for illustrative purposes, and are based on the canonical 5$\sigma$ depth  for the LAS of $H = 18.8$, this results in these candidates forming a straight diagonal sequence across the plot. Rejected candidates are shown with filled red squares, whilst yet-to-be followed up targets are shown with open circles. For reference, field stars from a randomly selected 1 square degree region of LAS sky are shown as black dots. 
}
\label{fig:cmdsel}
\end{figure}

We also imposed a number of data quality constraints to minimise
contamination from artefacts and poor signal-to-noise data for which we
refer the reader to the Appendix, which includes the SQL queries we
used to accesss the LAS via the WFCAM Science Archive \citep[WSA; ][]{wsa}. 
The epoch difference between the UKIDSS data and SDSS data is variable and ranges up to 6 years.  Our inclusion of UKIDSS sources with no SDSS counterpart within 2\arcsec~introduced sources with $z'-J < 2.5$ and proper motions above $\sim$300~mas/yr to our candidate list. This source of contamination was relatively small and in most cases such objects were identified as fast moving earlier type objects prior to detailed follow-up.
Figures~\ref{fig:colsel} and~\ref{fig:cmdsel} show  the UKIDSS LAS $YJH$ photometry of our selected candidates from DR5--DR8. The greatest degree of contamination in our $YJH$ selection channel is from photometrically scattered M dwarfs, with the greatest frequency of contaminants found at faint $J$, blue $Y-J$ and red $J-H$.

\subsection{The $YJ$-only selection channel}
\label{sec:yjonly}

Our $YJ$-only selection channel ensures that we do not exclude bona
fide late-T~dwarfs that are fainter than the LAS $H$ band detection
limit due to the inherently blue $J-H$ colours of such objects. 
To minimise contamination from scattered and blue M dwarfs we impose a
$Y-J$ constraint for this selection, such that our photometric
criteria are:

\begin{itemize}
\item $Y-J > 0.5$ or $J < 18.5$
\item $H$ and $K$ band non-detection
\item $z' - J > 2.5$ or no SDSS detection within 2\arcsec
\end{itemize}

We applied the same data quality restrictions to the $YJ$-only channel
as were applied to $YJH$ selection channel. In addition, to minimise contamination from Solar System Objects (SSOs), which can
appear as non-detections in the $H$ and $K$ bands due to different
epochs of observation, we also imposed a criterion that $Y$ and $J$
band coordinates must agree to within 0.75\arcsec~for observations
taken within a day of each other.  For observations taken more than a day apart we remove this requirement to avoid excluding bona fide candidates with high proper motion. We again refer the reader to the
Appendix for details of the SQL queries that we used to access the WSA.
The additional SSO contamination present in the $YJ$-only selection channel can be seen in Figure~\ref{fig:cmdsel} as relatively bright and blue $Y-J$ contaminants with very blue limits on their $J-H$ colours.

\section{Follow-up photometry}
\label{sec:phot}

To remove contaminants such as photometrically scattered M dwarfs and SSOs we used a combination of near-infrared and optical photometry. We followed two distinct strategies in removing contaminants. Prior to December 2010 we continued the strategy previously described in \citet{ben10b}. Briefly, this involved obtaining higher SNR $H$ band photometry to remove early-type objects that had been scattered into our $J-H < 0.1$ selection (or to fill in the missing data for $H$ band drop-outs in the $YJ$-only channel), and repeat $J$ band observations to remove SSOs. Whilst adequate for drawing a roughly complete sample of T4+ dwarfs, this method was unable to effectively prioritise objects with spectral types of T6 and later for follow-up. Since the T6--T9 region is most useful for constraining the form of the field mass function \citep[e.g. ][]{burgasser04}, we revised our strategy to allow the rejection of most of the T4 dwarfs that dominated our \citet{ben10b} sample. 

Our revised strategy involved using relatively short exposure (15 - 30 minute) $z'$ band imaging to confirm red $z' - J > 2.5$ colours for targets with no SDSS detection and for which the limits were not sufficient to rule out a bluer $z' - J$ colour. For targets from the $YJ$-only channel this step was preceded by short $J$ band observations to remove SSO contaminants. Targets with confirmed $z' - J > 2.5$ were then targeted with CH$_4$ imaging to identify late-type T~dwarfs.  Targets with $\rm{CH_{4}}s - \rm{CH_{4}}l < -0.5$  were prioritised for spectroscopy based on the methane colours of T~dwarfs reported by \citet{tinney2005}.

In the following subsections we outline the photometric observations and data reduction that were carried out for this follow-up programme. Details of the observations carried out for each target are given in the Appendix.

\subsection{Broad band photometry}
\label{sec:broadband}

Our broad band near-infrared photometry was obtained using the UKIRT Fast Track Imager \citep[UFTI; ][]{roche03}  and WFCAM \citep{wfcam}, both mounted on UKIRT across a number of observing runs spanning 2009 to the end of 2010. UFTI data  were dark subtracted, flatfield corrected, sky subtracted and mosaiced using the ORAC-DR pipeline\footnote{\url{http://www.oracdr.org/}}.
WFCAM data were processed using the WFCAM science pipeline by the Cambridge Astronomical Surveys Unit (CASU) \citep{irwin04}, and
archived at the WFCAM Science Archive \citep[WSA; ][]{wsa}. Observations consisted of a three point jitter pattern in the $Y$ and
$J$ bands, and five point jitter patterns in the $H$ and $K$ bands repeated twice. All data were acquired with 2x2 microstepping.  The WFCAM and UFTI filters are on the Mauna Kea Observatories (MKO) photometric system \citep{mko}.

The majority of our $z'$ photometry was taken using the Device Optimized for the LOw RESolution \citep[DOLORES; ][]{molinari1997} at the Telescopio Nazionale Galileo (TNG).
The observations were taken under program AOT22~TAC 96 spanning from 2010 to 2012. DOLORES is equipped with a  $2048 \times 2048$~pixels CCD with a field of view of $8.6 \times 8.6$~arcmin with a 0.252 arcsec/pixel scale. The observations were taken with the $z'$  Sloan filter. 
A small number of targets were observed in the $z'$ band using the using the ESO Faint
Object Spectrograph and Camera (EFOSC2) mounted on the New Technology Telescope (NTT; program 082.C-0399) and using the Auxiliary-port Camera (ACAM) on the William Herschel Telescope (WHT).
For each epoch a set of standard calibration flatfields and darks observations were taken. The images were dark subtracted, flatfielded and in the case of multiple exposures combined using standard IRAF routines. The data were taken in different observing conditions, from photometric conditions to cirrus.  No attempts to perform defringing to the images were made. The E2V4240 CCD detector in use in DOLORES has a low fringing level, the science object was also normally located in the top right section of the CCD where fringing is even smaller.  
Photometry was performed with IRAF using a fixed circular aperture with radius 2\arcsec. 
The photometric zero point was calibrated using the non-saturated SDSS stars present in the field of view.

The SDSS $z'$ band filter is slightly peculiar in that it has no red cut-off. Instead the red cut-off is defined by the detector sensitivity. So, although the DOLORES and ACAM data were taken through an SDSS $z'$ band filter, this does not trivially lead to the photometry being on the SDSS system, since there may be differences in the detectors' long wavelength responses.
To check the consistency of the DOLORES and ACAM photometric systems with SDSS we have compared synthetic photometry for a set of reference stellar spectra convolved with the combined filter and detector response curves for each of the  systems. The difference was found to be much smaller than the typical scatter in the zero-point from the reference stars, which dominate our quoted errors, and so we did not correct the SDSS reference stars magnitudes before calibration. 
For objects as red as T~dwarfs, however, the difference can be more significant. Synthesised photometry using template-T dwarf spectra found a mean offset of close to zero for DOLORES ($z'(DOLORES) - z'(SDSS) = -0.02 \pm 0.02$). 
For ACAM we found a small offset of $z'(ACAM) - z'(SDSS) = +0.09 \pm 0.03$.

For the EFOSC2 observations a Gunn $z$-band filter (ESO
Z\#623) was used and we used the transform given in \citet{ben09} to calculate
$z_{EFOSC2}$ for the SDSS secondary calibrators.
To place the resulting $z_{EFOSC2}(AB)$ photometry on the sloan
$z'(AB)$ system, we used the transform determined in \citet{ben10b}: 
$z(EFOSC2) - z'(SDSS)= -0.19 \pm 0.02$.

The best available  broad band photometry for all targets presented here is given in Table~\ref{tab:photom}. 

\subsection{$CH_4$ photometry}
\label{sec:ch4}

Differential methane photometry was obtained using the Near Infrared Camera Spectrometer \citep[NICS; ][]{baffa2001} mounted on the TNG under program AOT22 TAC 96 spanning from 2010 to 2012. NICS contains a set of Mauna Kea Observatories near-infrared filters as specified by \citet{mko}. More information about the narrow-band and intermediate-band sets are given by A. Tokunaga\footnote{See \url{http://www.ifa.hawaii.edu/~tokunaga/NB_special_ordersorting.html}}. The methane filters used in this work are denoted as $\rm{CH_{4}}s$ and $\rm{CH_{4}}l$. The comparison of these two filters provides information about the strength of the methane absorption bands in late-T~dwarfs. $\rm{CH_{4}}l$ samples the methane absorption bands present between 1.6 and 1.8 \micron, while the $\rm{CH_{4}}s$ samples a pseudo-continuum outside the methane band.

The final image mosaics were produced using the Speedy Near-infrared data Automatic Pipeline (SNAP) provided by TNG (version 1.3). SNAP is an automated wrapper of existing pieces of software (IRDR, IRAF, Sextractor and Drizzle) to perform a full reduction with a single command.
SNAP performs flat-fielding, computes the offsets between the dithered images, creates a mosaic image with double-pass sky subtraction and correction for field distortion.  

The data were taken in different observing conditions, from photometric conditions to cirrus. 
Photometry was performed with IMCORE, part of CASUTOOLS (version 1.0.21), using a fixed circular aperture of 2\arcsec.  CASUTOOLS\footnote{\url{http://apm49.ast.cam.ac.uk/surveys-projects/software-release}} is a suite of programmes developed and used by the Cambridge Astronomical Surveys Unit (CASU) for survey data reduction tasks associated with the UKIDSS and VISTA surveys, amongst others.

Differential photometric calibration of the methane colour $\rm{CH_{4}}s - \rm{CH_{4}}l$ was performed using the UKIDSS field stars present in the field, and the method defined by \cite{tinney2005} in their Section 2.5.
\cite{tinney2005} only provides the parameterisation for the 2MASS system. Using the information available on their Table 3 we performed the parameterisation for the UKIDSS system avoiding the region $0.48 < (J-H)_{\rm{MKO}} < 0.512$ where the sequence is degenerate. The sequence was fitted with two separate quadratics to the regions $-0.050 <  (J-H)_{\rm{MKO}}  < 0.480$,

\begin{equation}
\rm{CH_{4}}s - \rm{CH_{4}}l = +0.00046 - 0.01259 (J-H) +0.31817 (J-H)^2
\end{equation}

and $0.512 <  (J-H)_{\rm{MKO}}  < 1.000$, 

\begin{equation}
\rm{CH_{4}}s - \rm{CH_{4}}l = -0.17317 + 0.92744 (J-H) -0.58969 (J-H)^2
\end{equation}

An estimate of the spectral type was obtained using the conversion defined by equation (2) from \cite{tinney2005}. The resulting methane colours for all targets with spectra presented here are given in Table~\ref{tab:photom}, along with spectroscopically determined spectral types (see Section~\ref{sec:spec}) and photometric spectral types. We also present CH$_{4}$ photometry for several DR8 targets which have $\rm{CH_{4}}s - \rm{CH_{4}}l < -0.5$, but which for various reasons were not followed-up with spectroscopy. These targets are included as they form part of our UKIDSS DR8 space density estimate outlined in Section~\ref{sec:density}. A full summary of all the CH$_{4}$ photometry obtained, including earlier type objects and the extension of our analysis to include photometrically confirmed T~dwarfs in DR9, along with a more detailed description of the CH$_4$ calibration can be found in Cardoso et al. (2013; in prep).

For a small number of targets we obtained differential methane photometry using the Long-slit Infrared Imaging Spectrograph \citep[LIRIS;][]{Manchado98} mounted on the WHT. These data were were flatfield corrected, sky subtracted and mosaiced using LIRIS-DR\footnote{\url{http://www.iac.es/galeria/jap/lirisdr/LIRIS_DATA_REDUCTION.html}}.
 In these cases the methane colours is constructed as $H-[\rm CH_{4}]l$, and calibrated assuming that the average $H-[\rm CH_{4}]l$ of bright secondary calibrators in the field was zero \citep[see also ][]{kendall07,pinfield08}. Since no calibration for spectral type is yet determined for the $H-{\rm CH_{4}}l$ colour we do not present photometric estimates for the spectral types from these data.

\begin{landscape}
\begin{table}
{\scriptsize
\begin{tabular}{ c  c c  >{$}c<{$}  >{$}c<{$}  >{$}c<{$}  >{$}c<{$}  >{$}c<{$}  >{$}c<{$}    c  c  c  c }
\hline
Name & $\alpha$  & $\delta$ & z' & Y_{MKO} & J_{MKO} & H_{MKO} & K_{MKO} &  {\rm CH_4}s - {\rm CH_4}l & ${\rm CH_4}$ Type & ${\rm CH_4}$ Type  & ${\rm CH_4}$ Type & SpType \\
& (J2000) & (J2000) & & & & & & & & (min) & (max) & \\
\hline 
  ULAS~J000734.90+011247.1 & 00:07:34.90 & +01:12:47.10 & - & 19.22 \pm 0.07 & 18.05 \pm 0.04 & - & - & -0.91 \pm 0.13 & T6.1 & T5.8 & T6.4 & T7\\
  ULAS~J012735.66+153905.9 & 01:27:35.66 & +15:39:05.90 & - & 19.47 \pm 0.13 & 18.22 \pm 0.07 & 18.62 \pm 0.14 & - & -0.88 \pm 0.17 & T6.0 & T5.5 & T6.4 & T6.5\\
  ULAS~J012855.07+063357.0 & 01:28:55.07 & +06:33:57.00 & 22.73 \pm 0.40^A & 19.66 \pm 0.14 & 18.93 \pm 0.12 & - & - & -0.81 \pm 0.14 & T5.8 & T5.4 & T6.2 & T6\\
  ULAS~J013017.79+080453.9 & 01:30:17.79 & +08:04:53.90 & - & 19.06 \pm 0.03^W & 17.93 \pm 0.02^W & 18.21 \pm 0.02^W & 18.35 \pm 0.04^W & -0.72 \pm 0.07 & T5.5 & T5.3 & T5.7 & T6\\
  CFBDS~J013302+023128$^{1}$   & 01:33:02.48 & +02:31:28.90 & - & 19.36 \pm 0.11 & 18.34 \pm 0.08 & 18.51 \pm 0.15 & - & <-1.19 & N/A & N/A & N/A & T8\\
  ULAS~J013950.51+150307.6 & 01:39:50.51 & +15:03:07.60 & - & 19.72 \pm 0.17 & 18.44 \pm 0.1 & 18.53 \pm 0.18 & - & -0.83 \pm 0.11 & T5.9 & T5.6 & T6.2 & T7\\
  ULAS~J020013.18+090835.2 & 02:00:13.18 & +09:08:35.20 & - & 18.98 \pm 0.07 & 17.81 \pm 0.04 & 18.18 \pm 0.11 & 18.18 \pm 0.2 & -0.83 \pm 0.11 & T5.9 & T5.5 & T6.1 & T6\\
  ULAS~J022603.18+070231.4 & 02:26:03.18 & +07:02:31.40 & - & 19.62 \pm 0.05^W & 18.52 \pm 0.04^W & 18.82 \pm 0.03^W & 18.79 \pm 0.06^W & - & N/A & N/A & N/A & T7\\
  ULAS~J024557.88+065359.4 & 02:45:57.88 & +06:53:59.40 & - & 19.43 \pm 0.1 & 18.36 \pm 0.04^W & 18.95 \pm 0.17 & - & - & N/A & N/A & N/A & T7\\
  ULAS~J025545.28+061655.8 & 02:55:45.28 & +06:16:55.80 & - & 19.15 \pm 0.07 & 18.04 \pm 0.03^W & 18.4 \pm 0.02^W & - & - & N/A & N/A & N/A & T6\\
  ULAS~J032920.22+043024.5 & 03:29:20.22 & +04:30:24.50 & 20.75 \pm 0.17 & 18.55 \pm 0.02^W & 17.55 \pm 0.02^W & 17.89 \pm 0.02^W & 18.4 \pm 0.04^W & - & N/A & N/A & N/A & T5\\
  ULAS~J074502.79+233240.3 & 07:45:02.79 & +23:32:40.30 & - & 20.0 \pm 0.15 & 18.88 \pm 0.07 & - & - & -1.62 \pm 0.17 & T7.6 & T7.4 & T7.9 & T9$^{a}$\\
  ULAS~J074616.98+235532.2 & 07:46:16.98 & +23:55:32.20 & >20.99^D & 20.18 \pm 0.19 & 19.0 \pm 0.08 & - & - & -0.87  \pm 0.17^L & N/A & N/A & N/A & T7\\
  ULAS~J074720.07+245516.3 & 07:47:20.07 & +24:55:16.30 & - & 19.35 \pm 0.05^W & 18.17 \pm 0.05^W & 18.5 \pm 0.04^W & 18.53 \pm 0.07^W & - & N/A & N/A & N/A & T6.5\\
  ULAS~J075829.83+222526.7 & 07:58:29.83 & +22:25:26.70 & - & 18.68 \pm 0.04 & 17.62 \pm 0.02^W & 17.91 \pm 0.02^W & 17.87 \pm 0.12 & - & N/A & N/A & N/A & T6.5\\
  ULAS~J075937.75+185555.0 & 07:59:37.75 & +18:55:55.00 & 23.32 \pm 0.09^D & 20.21 \pm 0.18 & 18.7 \pm 0.07 & - & - & -0.95 \pm 0.12 & T6.2 & T5.9 & T6.5 & T6\\
  ULAS~J080048.27+190823.8 & 08:00:48.27 & +19:08:23.80 & 22.05 \pm 0.16^D & 19.76 \pm 0.12 & 18.55 \pm 0.06 & - & - & -0.73 \pm 0.28^L & N/A & N/A & N/A & N/A\\
  ULAS~J080918.41+212615.2 & 08:09:18.41 & +21:26:15.20 & - & 19.65 \pm 0.09 & 18.58 \pm 0.03^W & 18.99 \pm 0.03^W & 18.65 \pm 0.22 & - & N/A & N/A & N/A & T8\\
  ULAS~J081110.86+252931.8 & 08:11:10.86 & +25:29:31.80 & - & 18.76 \pm 0.03 & 17.57 \pm 0.02 & 18.19 \pm 0.12 & 18.02 \pm 0.19 & -1.03 \pm 0.13 & T6.4 & T6.1 & T6.7 & T7\\
  ULAS~J081407.51+245200.9 & 08:14:07.51 & +24:52:00.90 & >21.05^D & 19.6 \pm 0.1 & 18.54 \pm 0.05 & - & - & -0.30  \pm 0.10^L & N/A & N/A & N/A & T5p\\
  ULAS~J081507.26+271119.2 & 08:15:07.26 & +27:11:19.20 & - & 19.48 \pm 0.1 & 18.31 \pm 0.03^W & 18.6 \pm 0.03^W & - & - & N/A & N/A & N/A & T7p\\
  ULAS~J081918.58+210310.4 & 08:19:18.58 & +21:03:10.40 & 21.93 \pm 0.08^E & 18.25 \pm 0.03 & 16.95 \pm 0.01 & 17.28 \pm 0.03 & 17.18 \pm 0.06 & - & N/A & N/A & N/A & T6\\
  ULAS~J082155.49+250939.6 & 08:21:55.49 & +25:09:39.60 & - & 18.61 \pm 0.04 & 17.23 \pm 0.01^W & 17.24 \pm 0.01^W & 17.23 \pm 0.09 & - & N/A & N/A & N/A & T4.5\\
  ULAS~J084743.93+035040.2 & 08:47:43.93 & +03:50:40.20 & 21.90 \pm 0.10^A & 19.61 \pm 0.05^W & 18.53 \pm 0.04^W & 18.71 \pm 0.03^W & 18.99 \pm 0.08^W & -0.65 \pm 0.14 & T5.3 & T4.7 & T5.7 & N/A\\
  ULAS~J092608.82+040239.7 & 09:26:08.82 & +04:02:39.70 & - & 19.7 \pm 0.09 & 18.59 \pm 0.06 & - & - & -0.69 \pm 0.14 & T5.4 & T4.9 & T5.8 & T6\\
  ULAS~J092744.20+341308.7 & 09:27:44.20 & +34:13:08.70 &  >21.8^D & 19.66 \pm 0.14 & 18.77 \pm 0.11 & - & - & -1.27 \pm 0.28 & T7.0 & T6.4 & T7.5 & T5.5\\
  WISEP~J092906.77+040957.9$^{2}$& 09:29:06.75 & +04:09:57.70 & - & 17.89 \pm 0.01^W & 16.87 \pm 0.01^W & 17.24 \pm 0.01^W & 17.61 \pm 0.02^W & -0.92 \pm 0.07 & T6.1 & T6.0 & T6.3 & T7\\
  ULAS~J093245.48+310206.4 & 09:32:45.48 & +31:02:06.40 & - & 20.0 \pm 0.09 & 18.73 \pm 0.05 & 19.04 \pm 0.23 & - & - & N/A & N/A & N/A & T2\\
  ULAS~J095047.28+011734.3 & 09:50:47.28 & +01:17:34.30 & - & 18.9 \pm 0.03^W & 18.02 \pm 0.03^W & 18.4 \pm 0.03^W & 18.85 \pm 0.07^W & - & N/A & N/A & N/A & T8\\
  ULAS~J095429.90+062309.6 & 09:54:29.90 & +06:23:09.60 & - & 17.73 \pm 0.01^W & 16.6 \pm 0.01^W & 16.87 \pm 0.01^W & 17.05 \pm 0.01^W & -0.58 \pm 0.09 & T5.0 & T4.6 & T5.3 & T5\\
  ULAS~J102144.87+054446.1 & 10:21:44.87 & +05:44:46.10 & - & 18.82 \pm 0.03^W & 17.66 \pm 0.02^W & 17.96 \pm 0.02^W & 17.97 \pm 0.03^W & -0.99 \pm 0.26 & T6.3 & T5.6 & T6.9 & T6\\
  ULAS~J102305.44+044739.2 & 10:23:05.44 & +04:47:39.20 & 23.50 \pm 0.17^D & 19.49 \pm 0.05^W & 18.39 \pm 0.04^W & 18.73 \pm 0.04^W & 18.58 \pm 0.07^W & -0.52 \pm 0.12^L & N/A & N/A & N/A & T6.5\\
  ULAS~J102940.52+093514.6 & 10:29:40.52 & +09:35:14.60 & - & 18.24 \pm 0.02^W & 17.28 \pm 0.01^W & 17.63 \pm 0.01^W & 17.64 \pm 0.02^W & -1.56 \pm 0.17 & T7.5 & T7.3 & T7.8 & T8\\
  ULAS~J104224.20+121206.8 & 10:42:24.20 & +12:12:06.80 & - & 19.58 \pm 0.09 & 18.52 \pm 0.06 & 18.9 \pm 0.12 & - & -0.93 \pm 0.15 & T6.2 & T5.8 & T6.5 & T7.5\\
  ULAS~J104355.37+104803.4 & 10:43:55.37 & +10:48:03.40 & - & 19.21 \pm 0.03^W & 18.23 \pm 0.02^W & 18.58 \pm 0.02^W & 18.66 \pm 0.05^W & -1.36 \pm 0.22 & T7.2 & T6.7 & T7.5 & T8\\
  ULAS~J105134.32-015449.8 & 10:51:34.32 & -01:54:49.80 & - & 18.85 \pm 0.03^W & 17.75 \pm 0.02^W & 18.07 \pm 0.02^W & 18.27 \pm 0.04^W & -0.56 \pm 0.14 & T4.9 & T4.4 & T5.4 & T6\\
  ULAS~J105334.64+015719.7 & 10:53:34.64 & +01:57:19.70 & - & 19.77 \pm 0.1 & 18.5 \pm 0.06 & - & - & -1.10 \pm 0.16 & T6.6 & T6.2 & T6.9 & T6.5\\
  ULAS~J111127.77+051855.5 & 11:11:27.77 & +05:18:55.50 & - & 19.87 \pm 0.1 & 18.74 \pm 0.07 & - & - & - & N/A & N/A & N/A & T4.5\\
  ULAS~J113717.17+112657.2 & 11:37:17.17 & +11:26:57.20 & >22.24^D & 20.14 \pm 0.21 & 18.5 \pm 0.09 & - & - & -0.51 \pm 0.16 & T4.7 & T4.0 & T5.3 & N/A\\
  ULAS~J115229.68+035927.3 & 11:52:29.68 & +03:59:27.30 & - & 18.54 \pm 0.03 & 17.28 \pm 0.02 & 17.7 \pm 0.05 & 17.77 \pm 0.12 & -0.69 \pm 0.06 & T5.4 & T5.2 & T5.6 & T6\\
  ULAS~J115239.94+113407.6 & 11:52:39.94 & +11:34:07.60 & - & 19.3 \pm 0.06 & 18.26 \pm 0.04 & 18.66 \pm 0.1 & 18.32 \pm 0.17 & <-1.38 & N/A & N/A & N/A & T8.5$^a$\\
  ULAS~J115508.39+044502.3 & 11:55:08.39 & +04:45:02.30 & - & 19.38 \pm 0.07 & 18.33 \pm 0.05 & - & -  & - & N/A & N/A & N/A & T7\\
  ULAS~J120444.67-015034.9 & 12:04:44.67 & -01:50:34.90 & - & 17.99 \pm 0.03 & 16.74 \pm 0.02^U & 17.1 \pm 0.02^U & 17.29 \pm 0.09 & - & N/A & N/A & N/A & T4.5\\
  ULAS~J120621.03+101802.9 & 12:06:21.03 & +10:18:02.90 & - & 20.57 \pm 0.23 & 19.11 \pm 0.15^W & 19.53 \pm 0.09^W & - & - & N/A & N/A & N/A & T5\\
  ULAS~J121226.80+101007.4 & 12:12:26.80 & +10:10:07.40 & - & 20.48 \pm 0.25 & 18.69 \pm 0.09^W & 19.06 \pm 0.08^W & - & - & N/A & N/A & N/A & T5\\
  ULAS~J122343.35-013100.7 & 12:23:43.35 & -01:31:00.70 & - & 19.71 \pm 0.13 & 18.7 \pm 0.09 & - & - & -0.66 \pm 0.16 & T5.3 & T4.7 & T5.8 & T6\\
 ULAS~J125446.35+122215.7 & 12:54:46.35 & +12:22:15.70 & - & 19.51 \pm 0.11 & 18.29 \pm 0.06 & 18.62 \pm 0.17 & 18.26 \pm 0.2 & -0.56 \pm 0.19 & T4.9 & T4.1 & T5.6 & N/A\\
  ULAS~J125835.97+030736.1 & 12:58:35.97 & +03:07:36.10 & - & 19.7 \pm 0.14 & 18.38 \pm 0.05^W & 18.59 \pm 0.05^W & - & - & N/A & N/A & N/A & T5\\
  ULAS~J125939.44+293322.4 & 12:59:39.44 & +29:33:22.40 & - & 19.65 \pm 0.09 & 18.39 \pm 0.06 & 18.55 \pm 0.14 & - & -0.59 \pm 0.13 & T5.0 & T4.5 & T5.4 & T5\\
  \hline
  \multicolumn{11}{l}{$^{1}$ \citet{albert2011}; UKIDSS designation: ULAS~J013302.48+023128.9}\\
  \multicolumn{11}{l}{$^{2}$ \citet{kirkpatrick2011}; UKIDSS designation: ULAS~J092906.75+04:0957.7}\\
  \multicolumn{11}{l}{$^{a}$ on the spectral typing system of \citet{ben08}.}\\  
  \end{tabular}
}
\caption{Best available near-infrared photometry for our sample. No superscript on a broad band photometric value indicates  UKIDSS survey photometry for $YJHK$, SDSS DR8 for $z'$ band. Unless indicated otherwise, all CH$_{4}$ photometry is from TNG/NICS. Superscripts refer to the following instruments: A =  ACAM (WHT); D = DOLORES (TNG);  E = EFOSC2 (NTT); L = LIRIS(WHT);  U =  UFTI (UKIRT); W = WFCAM (UKIRT). $z'$ band photometry has been converted to the SDSS system as described in the text.
\label{tab:photom}
}
\end{table}
\end{landscape}

\begin{landscape}
\begin{table}
\addtocounter{table}{-1}
{\scriptsize
\begin{tabular}{ c c c  >{$}c<{$}  >{$}c<{$}  >{$}c<{$}  >{$}c<{$}  >{$}c<{$}  >{$}c<{$}    c  c  c  c }
\hline
Name & $\alpha$  & $\delta$ & z' & Y_{MKO} & J_{MKO} & H_{MKO} & K_{MKO} &  {\rm CH_4}s - {\rm CH_4}l & ${\rm CH_4}$ Type & ${\rm CH_4}$ Type  & ${\rm CH_4}$ Type & SpType \\
& (J2000) & (J2000) & & & & & & & & (min) & (max) & \\
\hline
 ULAS~J130227.54+143428.0 & 13:02:27.54 & +14:34:28.00 & >19.12^D & 19.75 \pm 0.13 & 18.6 \pm 0.04^W & 18.8 \pm 0.04^W & - & -0.23 \pm 0.10 & T3.1 & T2.4 & T3.7 & T4.5\\
  ULAS~J133502.11+150653.5 & 13:35:02.11 & +15:06:53.50 & - & 19.03 \pm 0.03^U & 17.97 \pm 0.02^U & 18.3 \pm 0.03^U & 18.23 \pm 0.14 & -0.60 \pm 0.09 & T5.1 & T4.8 & T5.4 & T6\\
  ULAS~J133828.69-014245.4 & 13:38:28.69 & -01:42:45.40 & - & 19.57 \pm 0.08^W & 18.69 \pm 0.1^W & 19.14 \pm 0.09^W & 19.21 \pm 0.12^W & -  & N/A & N/A & N/A & T7.5\\
  ULAS~J133933.64-005621.1 & 13:39:33.64 & -00:56:21.10 & - & 19.21 \pm 0.05^W & 18.24 \pm 0.05^W & 18.48 \pm 0.04^W & 18.39 \pm 0.05^W & - & N/A & N/A & N/A & T7\\
  ULAS~J133943.79+010436.4 & 13:39:43.79 & +01:04:36.40 & - & 19.15 \pm 0.05^W & 18.08 \pm 0.04 & 18.39 \pm 0.13 & 18.39 \pm 0.05^W & - & N/A & N/A & N/A & T5\\
  ULAS~J141756.22+133045.8 & 14:17:56.22 & +13:30:45.80 & 20.42 \pm 0.16 & 17.94 \pm 0.03 & 16.77 \pm 0.01 & 17.0 \pm 0.03 & 17.0 \pm 0.04 & -0.51 \pm 0.08 & T4.7 & T4.4 & T5.0 & T5\\
  ULAS~J142145.63+013619.0 & 14:21:45.63 & +01:36:19.00 & - & 19.31 \pm 0.12 & 18.52 \pm 0.04^W & 18.54 \pm 0.03^W & - & - & N/A & N/A & N/A & T4.5\\
  ULAS~J142536.35+045132.3 & 14:25:36.35 & +04:51:32.30 & >21.87^D & 20.02 \pm 0.14 & 18.7 \pm 0.09 & - & - & -0.93 \pm 0.12 & T6.2 & T5.9 & T6.4 & T6.5\\
  ULAS~J144902.02+114711.4 & 14:49:02.02 & +11:47:11.40 & - & 18.35 \pm 0.04 & 17.36 \pm 0.02 & 17.73 \pm 0.07 & 18.1 \pm 0.15 & -0.48 \pm 0.13 & T4.6 & T4.0 & T5.1 & T5.5\\
  ULAS~J151637.89+011050.1 & 15:16:37.89 & +01:10:50.10 & - & 19.48 \pm 0.12 & 18.41 \pm 0.05^W & 18.67 \pm 0.06^W & 18.49 \pm 0.2 & -0.96 \pm 0.20 & T6.3 & T5.7 & T6.7 & T6.5\\
  WISE~J151721.13+052929.3$^{3}$ & 15:17:21.12 & +05:29:29.03 & - & 19.57 \pm 0.07 & 18.54 \pm 0.05 & 18.85 \pm 0.15 & - & - & N/A & N/A & N/A & T8\\
  ULAS~J153406.06+055643.9 & 15:34:06.06 & +05:56:43.90 & 22.52 \pm 0.15^D & 20.24 \pm 0.19 & 19.02 \pm 0.1 & - & - & -0.56 \pm 0.13 & T4.9 & T4.4 & T5.3 & T5\\
  ULAS~J153653.80+015540.6 & 15:36:53.80 & +01:55:40.60 & - & 19.15 \pm 0.08 & 17.93 \pm 0.05 & 18.03 \pm 0.1 & 18.01 \pm 0.16 & - & N/A & N/A & N/A & T5\\
  ULAS~J154914.45+262145.6 & 15:49:14.45 & +26:21:45.60 & - & 19.15 \pm 0.07 & 18.05 \pm 0.03^W & 18.29 \pm 0.03^W & - & -0.60 \pm 0.12 & T5.1 & T4.6 & T5.5 & T5\\
  ULAS~J160143.75+264623.4 & 16:01:43.75 & +26:46:23.40 & 21.35 \pm 0.05^D & 19.48 \pm 0.08^W & 18.43 \pm 0.05^W & 18.82 \pm 0.07^W & 18.75 \pm 0.08^W & - & N/A & N/A & N/A & T6.5\\
  ULAS~J161436.96+244230.1 & 16:14:36.96 & +24:42:30.10 & 22.36 \pm 0.35^D & 19.42 \pm 0.08 & 18.52 \pm 0.04 & - & - & -1.01 \pm 0.15 & T6.4 & T6.0 & T6.7 & T7\\
  ULAS~J161710.39+235031.4 & 16:17:10.39 & +23:50:31.40 & - & 18.99 \pm 0.05 & 17.72 \pm 0.02 & 18.16 \pm 0.08 & - & -0.70 \pm 0.09 & T5.4 & T5.1 & T5.7 & T6\\
  ULAS~J161934.78+235829.3 & 16:19:34.78 & +23:58:29.30 & - & 19.72 \pm 0.11 & 18.62 \pm 0.06^W & 18.91 \pm 0.06^W & - & -0.75 \pm 0.11 & T5.6 & T5.3 & T5.9 & T6\\
  ULAS~J161938.12+300756.4 & 16:19:38.12 & +30:07:56.40 & - & 19.84 \pm 0.11 & 18.61 \pm 0.07^W & 18.79 \pm 0.06^W & - & -0.43 \pm 0.09 & T4.3 & T3.9 & T4.7 & T5\\
  ULAS~J162655.04+252446.8 & 16:26:55.04 & +25:24:46.80 & - & 19.82 \pm 0.11 & 18.4 \pm 0.04^W & 18.62 \pm 0.04^W & - & - & N/A & N/A & N/A & T5\\
  ULAS~J163931.52+323212.7 & 16:39:31.52 & +32:32:12.70 & 20.30 \pm 0.11  & 18.14 \pm 0.02 & 16.71 \pm 0.01 & 16.72 \pm 0.03 & 16.8 \pm 0.06 & - & N/A & N/A & N/A & T3\\
  ULAS~J211616.26-010124.3 & 21:16:16.26 & -01:01:24.30 & >22.10^D & 19.53 \pm 0.12 & 18.27 \pm 0.07 & - & - & -1.10 \pm 0.31 & T6.6 & T5.8 & T7.2 & T6\\
  ULAS~J223728.91+064220.1 & 22:37:28.91 & +06:42:20.10 & - & 19.79 \pm 0.08^W & 18.78 \pm 0.05^W & 19.23 \pm 0.04^W & 19.94 \pm 0.18^W & - & N/A & N/A & N/A & T6.5p\\
  ULAS~J230049.08+070338.0 & 23:00:49.08 & +07:03:38.00 & 21.6 \pm 0.12^D & 18.97 \pm 0.04^W & 17.67 \pm 0.02^W & 17.77 \pm 0.03^W & 17.74 \pm 0.05^W & -0.43 \pm 0.08 & T4.3 & T3.9 & T4.7 & T4.5\\
  ULAS~J231536.93+034422.7 & 23:15:36.93 & +03:44:22.70 & - & 19.89 \pm 0.12 & 18.79 \pm 0.08 & - & - & < -0.77 & N/A & N/A & N/A & T7\\
  ULAS~J231856.24+043328.5 & 23:18:56.24 & +04:33:28.50 & 25.6 \pm 0.32^D & 20.18 \pm 0.13 & 18.78 \pm 0.07 & - & - & < -0.95 & N/A & N/A & N/A & T7.5\\
  ULAS~J232600.40+020139.2 & 23:26:00.40 & +02:01:39.20 & - & 19.4 \pm 0.08 & 17.98 \pm 0.04 & 18.46 \pm 0.12 & 18.41 \pm 0.2 & -1.64 \pm 0.16 & T7.7 & T7.5 & T7.9 & T8\\
  ULAS~J232624.07+050931.6 & 23:26:24.07 & +05:09:31.60 & 21.85 \pm 0.21^D & 19.75 \pm 0.15 & 18.61 \pm 0.1 & 18.61 \pm 0.14 & - & -0.51 \pm 0.12 & T4.7 & T4.2 & T5.1 & N/A\\
  ULAS~J233104.12+042652.6 & 23:31:04.12 & +04:26:52.60 & 22.12 \pm 0.26^D & 20.16 \pm 0.14 & 18.67 \pm 0.08 & - & - & - & N/A & N/A & N/A & T4\\
  ULAS~J234228.97+085620.1 & 23:42:28.97 & +08:56:20.10 & 20.15 \pm 0.12 & 17.37 \pm 0.01^W & 16.39 \pm 0.01^W & 16.77 \pm 0.01^W & 17.1 \pm 0.02^W & - & N/A & N/A & N/A & T6\\
  ULAS~J235204.62+124444.9 & 23:52:04.62 & +12:44:44.90 & - & 19.64 \pm 0.11 & 18.27 \pm 0.05 & 18.55 \pm 0.16 & 18.41 \pm 0.21 & -0.95 \pm 0.12 & T6.2 & T5.9 & T6.5 & T6.5\\
  ULAS~J235715.98 +0:3:40. & 23:57:15.98  & +01:32:40.30 & 21.42 \pm 0.22^D & 19.78 \pm 0.06^W & 18.5 \pm 0.04^W & 18.68 \pm 0.03^W & 18.6 \pm 0.05^W & - & N/A & N/A & N/A & T5.5p\\
\hline
  \multicolumn{11}{l}{$^{3}$ \citet{mace2013}; UKIDSS designation: ULAS~J151721.12+052929.0 }\\
  \multicolumn{11}{l}{$^{a}$ on the spectral typing system of \citet{ben08}.}\\  
\end{tabular}
}
\caption{(Continued) Best available near-infrared photometry for our sample. No superscript on a broad band photometric value indicates  UKIDSS survey photometry for $YJHK$, SDSS DR8 for $z'$ band. Unless indicated otherwise, all CH$_{4}$ photometry is from TNG/NICS. Superscripts refer to the following instruments: A =  ACAM (WHT); D = DOLORES (TNG);  E = EFOSC2 (NTT); L = LIRIS(WHT);  U =  UFTI (UKIRT); W = WFCAM (UKIRT). $z'$ band photometry has been converted to the SDSS system as described in the text.
\label{tab:photom}
}

\end{table}
\end{landscape}

\section{Spectroscopic confirmation}
\label{sec:spec}

Spectroscopic confirmation of most T dwarf candidates that survived the
photometric follow-up program was achieved using the Near InfraRed
Imager and Spectrometer \citep[NIRI;][]{hodapp03} and the Gemini Near Infrared spectrograph \citep[GNIRS;][]{gnirs} on the Gemini North
Telescope\footnote{under programs GN-2009A-Q-16, GN-2009B-Q-62, GN-2009B-Q-99, GN-2010A-Q-44, GN-2010B-Q-41, GN-2011A-Q-73, GN-2011B-Q-5, GN-2011B-Q-43 and GN-2012A-Q-84}, and  the InfraRed Camera and
Spectrograph \citep[IRCS;][]{IRCS2000} on the Subaru telescope, both
on Mauna Kea, Hawaii.
In addition a smaller number of spectra were obtained using the Folded
port InfraRed 
Echellette (FIRE) spectrograph \citep{fire08,fire10} mounted on the Baade 6.5m
Magellan telescope at Las Campanas Observatory.  
We also obtained spectroscopy for a single target using XSHOOTER \citep{xshooter} on UT2 of the VLT (Program ID: 086.C-0450).

All observations were made up of a set of sub-exposures in an ABBA
jitter pattern to facilitate effective background subtraction, with a
slit width of 1 arcsec for NIRI, GNIRS and IRCS, whilst 0.6 arcsec was
used for the FIRE observations. 
The length of the A-B jitter was 10 arcsecs.
For targets brighter than $J=18.5$ total integrations were typically 4x300s for NIRI, GNIRS and IRCS observations, whilst fainter targets were typically integrated for 8x300s. 
FIRE integrations were: 2x120s for $J<18.0$, 4x150s for $18.0 < J < 18.5$, 6x150s for $18.50 < J < 18.6$ and 8x150s for our faintest targets.
The program numbers and dates of individual observations are summarised in the Appendix.

The NIRI and GNIRS observations were reduced using standard IRAF
Gemini packages {\citep{cooke2005}. 
The Subaru IRCS spectra were extracted using standard IRAF
packages. The AB pairs were subtracted using generic IRAF tools,
and median stacked.  
The NIRI, GNIRS and IRCS spectra were calibrated in a similar manner.
Comparison argon arc frames were
used to obtain dispersion solutions, which were then applied to the
pixel coordinates in the dispersion direction on the images.
The resulting wavelength-calibrated subtracted pairs had a low-level
of residual sky emission removed by fitting and subtracting this
emission with a set of polynomial functions fit to each pixel row
perpendicular to the dispersion direction, and considering pixel data
on either side of the target spectrum only. 
The spectra were then extracted using a linear aperture, and cosmic
rays and bad pixels removed using a sigma-clipping algorithm.
Telluric correction was achieved by dividing each extracted target
spectrum by that of an early A or F type standard star observed just before or
after the target and at a similar airmass.
Prior to division, hydrogen lines were removed from the standard star
spectrum by interpolating the stellar
continuum.
Relative flux calibration was then achieved by multiplying through by a
blackbody spectrum of the appropriate $T_{\rm eff}$. This $T_{\rm eff}$ was taken from \citet{masana2006} where available, or else was estimated from the spectral type of the telluric standard. Since the near-infared region is well into the Rayleigh-Jeans tail of an A or F star's spectrum, very little systematic error is likely to be introduced from a crude estimate of this $T_{\rm eff}$.

The FIRE spectra were extracted using the low-dispersion version of the FIREHOSE
pipeline, which is based on the MASE pipeline \citep{mase,fire10}. The
pipeline uses a flat field constructed from two quartz lamp images
taken with the lamp at high (2.5 V) and low (1.5 V) voltage
settings. The data were divided by this pixel flat before being
wavelength calibrated. The pipeline
performs sky subtraction following the method outlined in
\citet{bochanski2011}, adapted for the low-dispersion configuration of
the spectrograph.  The spectra were optimally extracted before being
combined using a weighted mean, using an adaptation of the
xcombspec routine from SpexTool 
\citep{cushing2004}. The T dwarf spectra were then corrected for
telluric absorption and flux calibrated using a FIRE specific version
of the xtellcor routine \citep{vacca2003}. Finally, residual outlying
points due to cosmic rays and bad pixels were removed using a simple
3-sigma clipping algorithm. 

The X-Shooter data were reduced using the ESO pipeline (version 1.3.7). The pipeline removes non-linear pixels, subtracts the bias (in the VIS arm) or dark frames (in the NIR arm) and divides the raw frames by flat fields. Images are pair-wise subtracted to remove sky background. The pipeline then extracts and merges the different orders in each arm, rectifying them using a multi-pinhole arc lamp (taken during the day-time calibration) and correcting for the flexure of the instrument using single-pinhole arc lamps (taken at night, one for each object observed). Telluric stars are reduced in the same way, except that sky subtraction is done by fitting the background (as tellurics are not observed in nodding mode). The spectra were telluric corrected and flux calibrated using IDL routines, following a standard procedure: first the telluric spectrum is cleared of HI absorption lines (by interpolating over them) and scaled to match the measured magnitudes; then is divided by a blackbody curve for the appropriate temperature (estimated from the telluric standard's spectral type), to obtain the instrument$+$atmosphere response curve; finally the target spectra are multiplied by the response curve obtained to flux calibrate it. The arms (VIS and NIR) were then merged by matching the flux level in the overlapping regions between them.

Complete details of the spectroscopic observations obtained for each
 of the T~dwarfs presented here are given in the Appendix.
The resulting spectra are shown in Figure~\ref{fig:specs}. This includes one T dwarf (ULAS~J0929+0409) confirmed by \citet{kirkpatrick2011}, three (ULAS~J0954+0623, ULAS J1204-0150, ULAS~J1152+0359) confirmed by \citet{scholz2012}  and one (ULAS~J1517+0529) confirmed by \citet{mace2013} since our spectroscopic follow-up, but prior to this publication. 

\begin{figure*}
\includegraphics[width=700pt, angle=90]{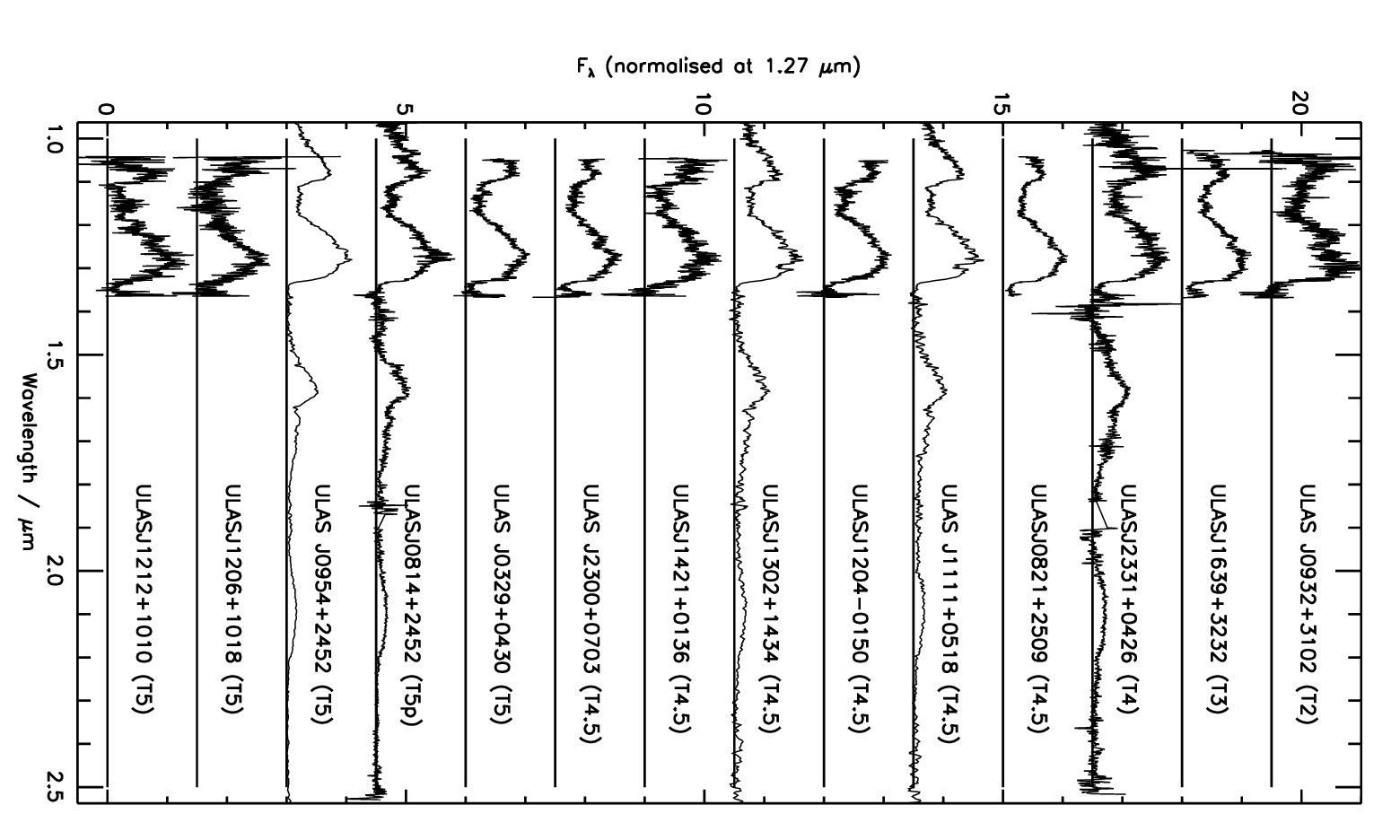}
\caption{Spectra of the 76 T~dwarfs presented here. Each spectrum is
  normalised at $1.27 \pm 0.005  \mu m$ and offset for clarity. The GNIRS, NIRI and IRCS spectra have been rebinned by a factor of three as a compromise to maximise signal-to-noise and not sacrifice resolution.}
\label{fig:specs}
\end{figure*}

\addtocounter{figure}{-1}
\begin{figure*}
\includegraphics[width=700pt, angle=90]{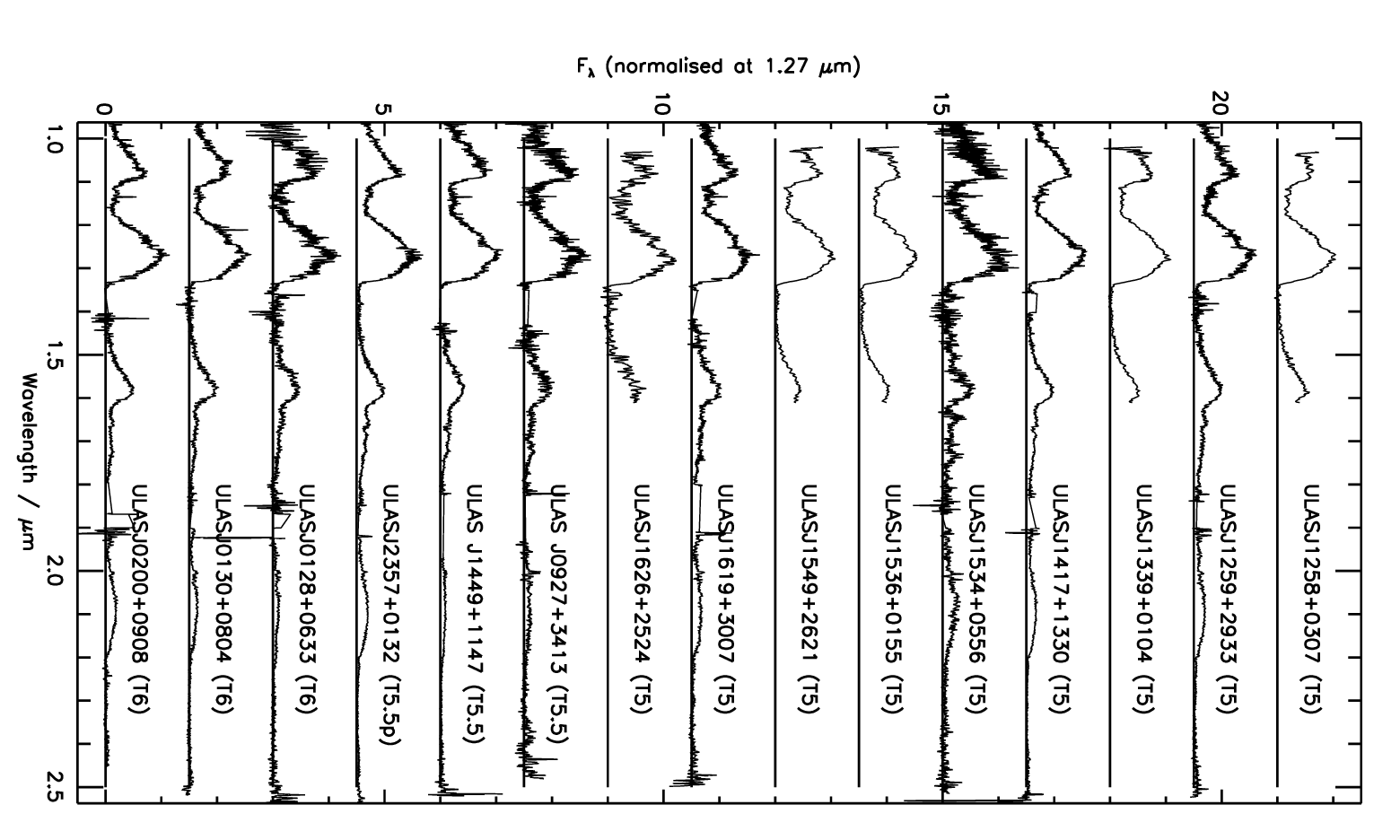}
\caption{Continued}
\end{figure*}

\addtocounter{figure}{-1}
\begin{figure*}
\includegraphics[width=700pt, angle=90]{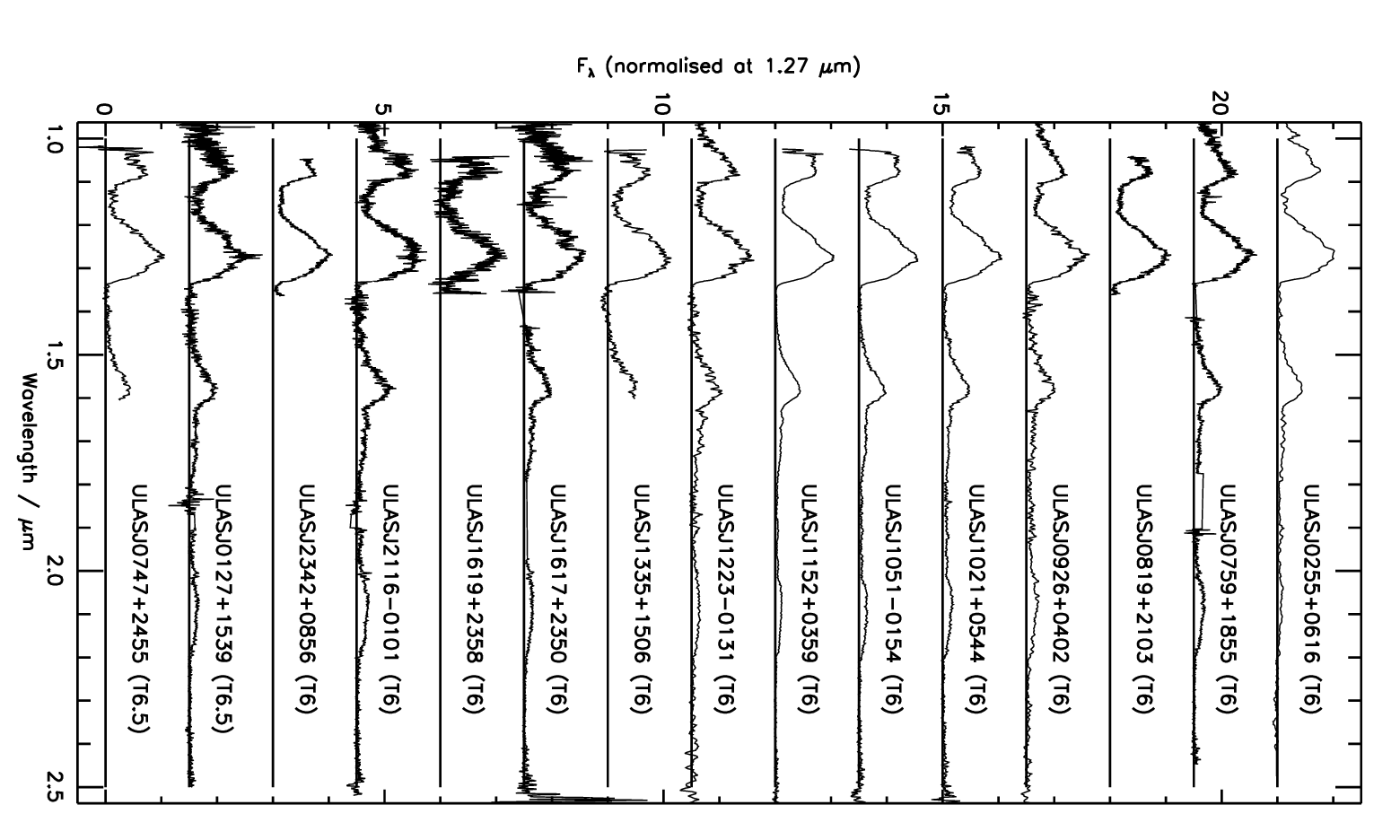}
\caption{Continued.}
\end{figure*}

\addtocounter{figure}{-1}
\begin{figure*}
\includegraphics[width=700pt, angle=90]{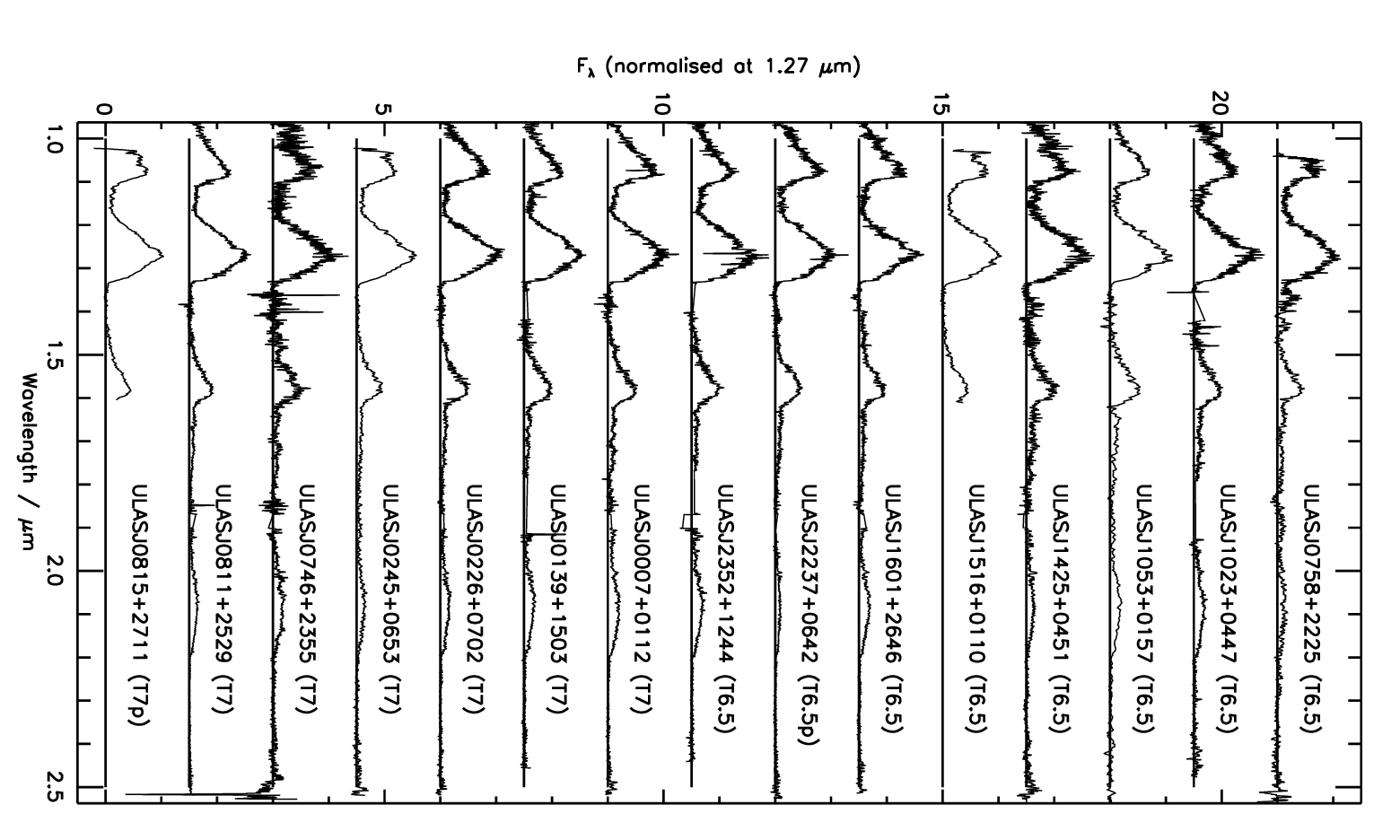}
\caption{Continued}
\end{figure*}

\addtocounter{figure}{-1}
\begin{figure*}
\includegraphics[width=700pt, angle=90]{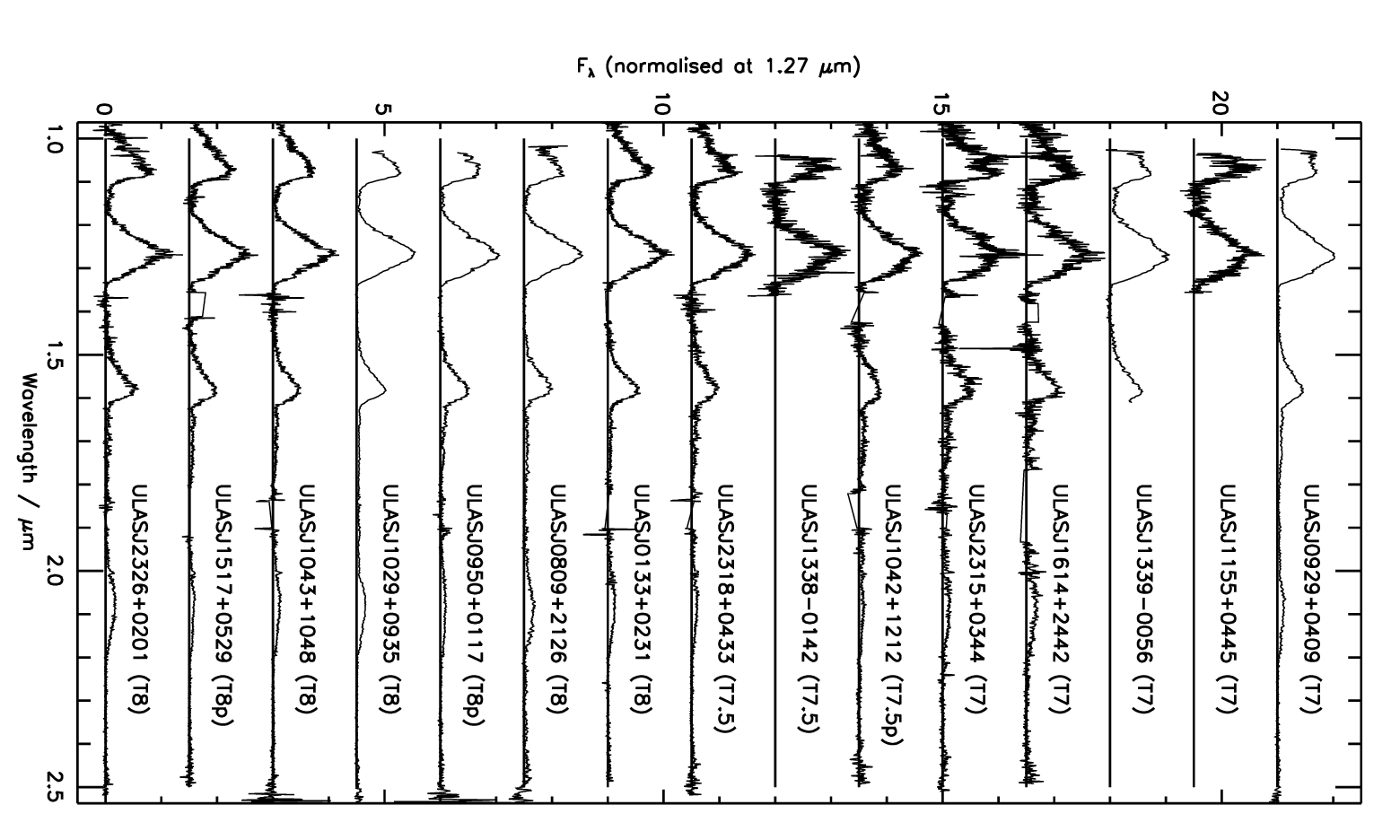}
\caption{Continued.}
\end{figure*}

\addtocounter{figure}{-1}
\begin{figure*}
\includegraphics[width=250pt, angle=90]{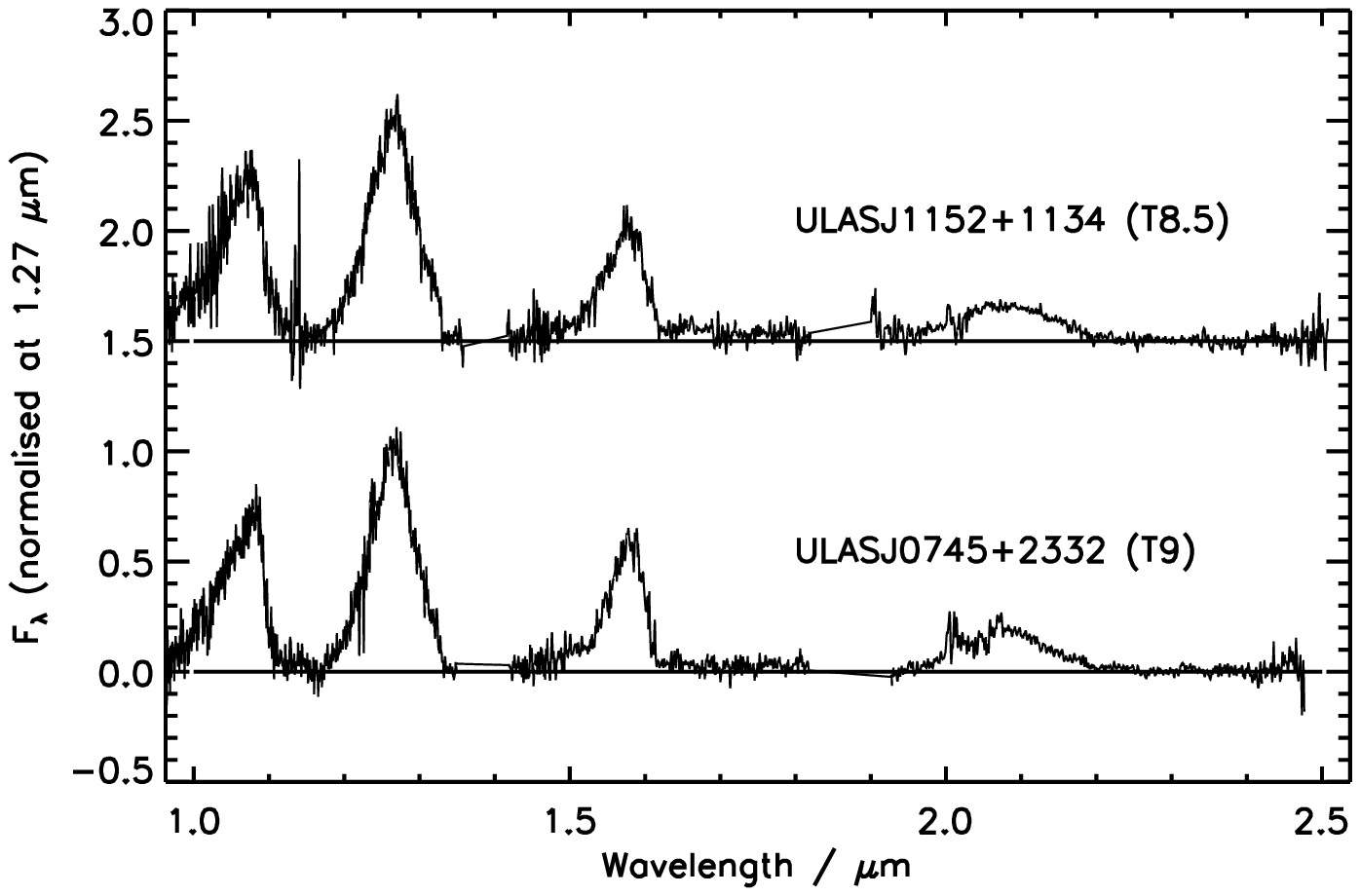}
\caption{Continued.}
\end{figure*}

\subsection{Spectral types}
\label{subsec:sptype}

We have assigned spectral types following the scheme of \citet{burgasser06} for types as late as T8, and the extension of \citet{ben08} for types beyond T8.  We have adopted this scheme in this work for two reasons. Firstly, it provides continuity with our previous work \citep{ben08,ben09,ben10b}, allowing a meaningful update to our previous space density estimate. Secondly, as was discussed in detail in \citet{ben08}, this scheme provides excellent continuity with the evolution of spectral index values from earlier types. This does not diminish the scheme's fundamentally empirical nature; it is anchored to template objects. However, it does seek to minimise the subjectivity as to the degree of spectral difference required for the distinction of two subtypes. 
We have only identified two new T8+ dwarfs in this paper so this issue is of minor importance, and we have indicated the spectral types of these two objects on the \citet{cushing2011} extension scheme in the notes column for completeness, and to avoid any future confusion. Objects that show substantial discrepancy either between the spectral types indicated by their spectral indices, or in comparison to their best fitting spectral template have been classified as peculiar, and are denoted with the suffix `p'. The most common feature leading to this designation is the suppression or enhancement of the $K$ band peak relative to the template. Table~\ref{tab:types} summarises our spectral type measurements and adopted classifications.

In Figure~\ref{fig:yjhcol} we have plotted our full UKIDSS LAS sample of 146 spectroscopically confirmed T~dwarfs with $YJH$ photometry (from a total of 171) on a $YJH$ colour-colour diagram. Spectral types are distinguished by coloured symbols in whole subtype bins.  Although there is significant scatter, a general trend from top right to lower left is apparent for the T4 -- T8 dwarfs, with the earliest type objects dominating the top-right of the plot, whilst the later type objects dominate the lower-left.  This is consistent with the reduced $H$ band flux due to deepening CH$_{4}$ absorption, and a general trend to bluer $Y-J$ colours with decreasing $T_{\rm eff}$ that have previously been noted for the T spectral sequence \citep{ben10b,sandy10,liu2012}.

\begin{figure}
\includegraphics[width=220pt, angle=90]{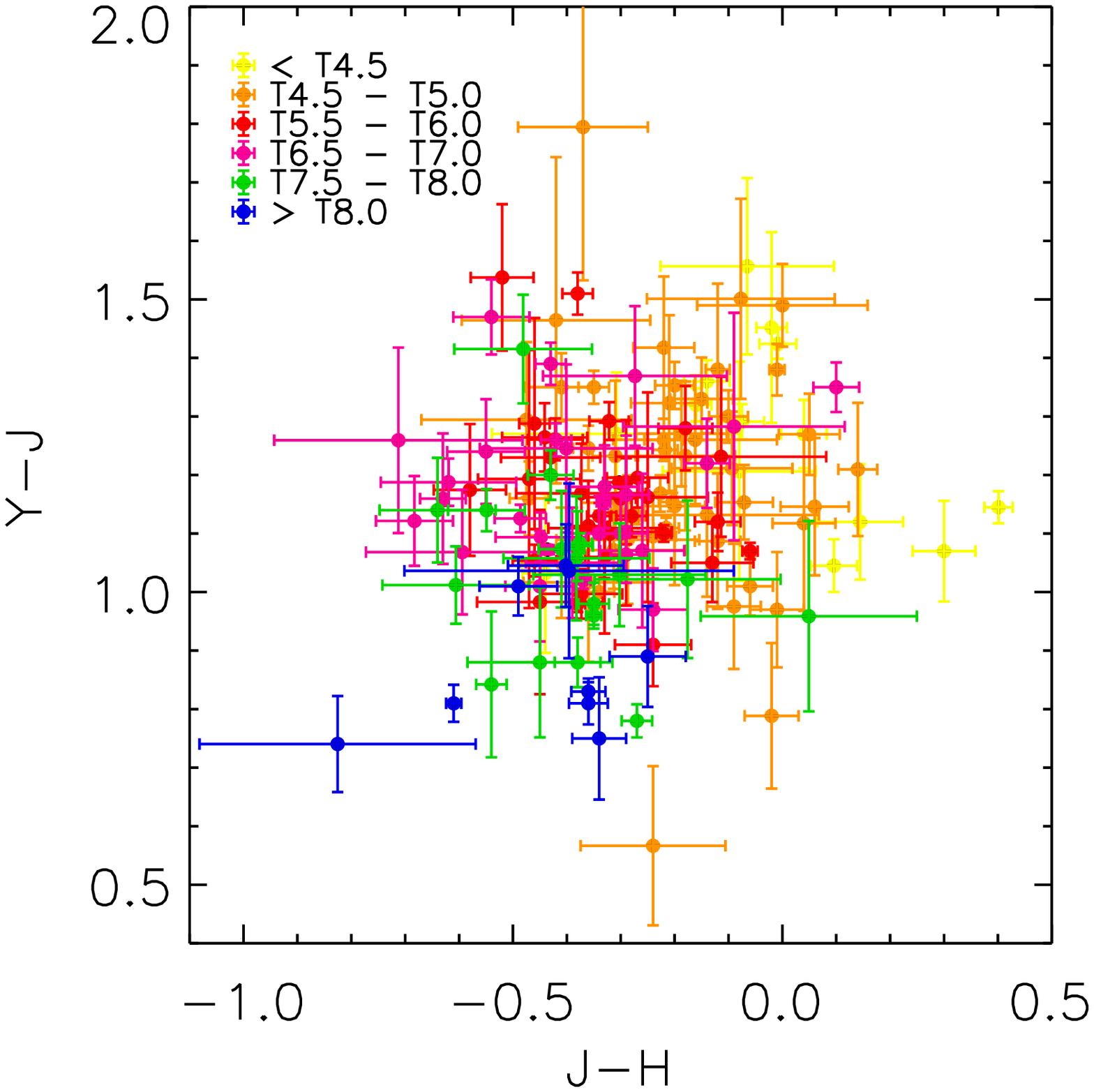}
\caption{A $Y-J$ vs $J-H$ colour-colour diagram for 146 spectroscopically confirmed T~dwarfs in the UKIDSS LAS with $YJH$ photometry. Spectral types are indicated by coloured symbols.}
\label{fig:yjhcol}
\end{figure}

\begin{landscape}
\begin{table}
{\scriptsize
\begin{tabular}{ c  c c c c c c c c c c c c c }
\hline
Name & Adopted & Templ. & H$_2$O-$J$ & CH$_4$-$J$ & $W_J$ & H$_2$O-$H$ & CH$_4$-$H$ & CH$_4$-$K$ & Note \\
\hline
ULAS~J0007+0112	 &	T7 &      T7	&    $0.103 \pm  0.008$ (T7)   & $0.286 \pm  0.007$ (T6/7)   & $0.398 \pm 0.015$ (T6/7)  & $0.320 \pm  0.018$  (T5/6) & $0.233 \pm  0.007$  (T7) & $0.118 \pm  0.010$  ($>$T6) & \\
ULAS~J0127+1539	 &	T6.5 &    T6.5  &    $0.179 \pm  0.011$ (T5/6) & $0.307 \pm  0.007$ (T6)   & $0.451 \pm  0.021$ ($<$T7) & $0.264 \pm  0.016$  (T6/7) & $0.280 \pm  0.011$  (T6) & $0.127 \pm  0.012$  (T6/7) & \\
ULAS~J0128+0633	 &   	T6   &	  T6	&    $0.142 \pm  0.009$ (T6)   & $0.320 \pm  0.008$ (T6)   & $0.459 \pm  0.009$ ($<$T7) & $0.298 \pm  0.021$  (T6) & $0.392 \pm  0.014$  (T5) & $0.172 \pm  0.023$  (T5/6) & \\
ULAS~J0130+0804	 &   	T6   & 	  T6	&    $0.170 \pm  0.011$ (T5/6) & $0.338 \pm  0.005$ (T6)   & $0.497 \pm  0.014$ ($<$T7) & $0.275 \pm  0.010$  (T6) & $0.291 \pm  0.009$  (T6) & $0.100 \pm  0.012$  ($>$T6) & T8.5 A11 \\
CFBDS~J0133+0231	 &   	T8   &	  T8	&    $0.051 \pm  0.020$ (T7/8) & $0.191 \pm  0.018$ ($>$T7)  & $0.301 \pm  0.009$ (T8) & $0.163 \pm  0.011$  (T8) & $0.113 \pm  0.006$  ($>$T7) & $0.030 \pm  0.027$  ($>$T6) & \\
ULAS~J0139+1503	 &	T7   & 	  T7	&    $0.122 \pm  0.015$ (T6/7) & $0.268 \pm  0.014$ (T6/7) & $0.386 \pm  0.008$ (T7) & $0.257 \pm  0.030$  (T6/7) & $0.209 \pm  0.010$  (T7) & $0.084 \pm  0.019$  ($>$T6) & \\
ULAS~J0200+0908	 &   	T6   &	  T6	&    $0.146 \pm  0.011$ (T6)   & $0.285 \pm  0.008$ (T6/7) & $0.453 \pm  0.005$ ($<$T7) & $0.294 \pm  0.008$  (T6) & $0.237 \pm  0.008$  (T7) & $0.079 \pm  0.013$  ($>$T6) & \\
ULAS~J0226+0702	 &   	T7   &	  T7	&    $0.082 \pm  0.005$ (T7)   & $0.230 \pm  0.004$ (T7)   & $0.374 \pm  0.005$ (T7) & $0.273 \pm  0.010$  (T6) & $0.219 \pm  0.008$  (T7) & $0.108 \pm  0.010$  ($>$T6) & \\
ULAS~J0245+0653	 &   	T7   &    T7	&    $0.101 \pm  0.004$ (T7)   & $0.260 \pm  0.003$ (T7)   & $0.394 \pm  0.003$ (T7) & $0.271 \pm  0.009$  (T6) & $0.229 \pm  0.008$  (T7) & $0.056 \pm  0.025$  ($>$T6) & \\
ULAS~J0255+0616	 &	T6   &	  T6	&    $0.154 \pm  0.002$ (T6)   & $0.282 \pm  0.002$ (T6/7) & $0.481 \pm  0.003$ ($<$T7) & $0.337 \pm  0.003$  (T5) & $0.251 \pm  0.002$  (T6/7) & $0.027 \pm  0.012$  ($>$T6) & \\
ULAS~J0329+0430	 &	T5   &	  T5	&    $0.238 \pm  0.007$ (T5)   & $0.452 \pm  0.007$ (T4/5) & $0.540 \pm  0.006$ ($<$T7) &        -           &        -           &        -           & \\
ULAS~J0745+2332	 &	T9   &	  T9	&    $0.024 \pm  0.010$ ($>$T7)  & $0.130 \pm  0.039$ ($>$T7)  & $0.231 \pm  0.012$ (T9) & $0.172 \pm  0.022$  (T8) & $0.069 \pm  0.012$  ($>$T7) & $0.102 \pm  0.023$  ($>$T6) & T8.5 \\
ULAS~J0746+2355	 &	T7   &	  T7	&    $0.111 \pm  0.012$ (T7)   & $0.226 \pm  0.009$ (T7)   & $0.379 \pm  0.011$ (T7) & $0.261 \pm  0.023$  (T6/7) & $0.183 \pm  0.016$  (T7) & $0.126 \pm  0.017$  (T6/7) & \\
ULAS~J0747+2455	 &   	T6.5 &	  T6.5  &    $0.126 \pm  0.008$ (T6/7) & $0.284 \pm  0.005$ (T6/7) & $0.443 \pm  0.005$ ($<$T7) & $0.295 \pm  0.012$  (T6) & $0.058 \pm  0.025$  ($>$T7) &        -           & \\
ULAS~J0758+2225	 &	T6.5 &    T6.5  &    $0.130 \pm  0.005$ (T6/7) & $0.340 \pm  0.006$ (T6)   & $0.410 \pm  0.004$ ($<$T7) & $0.236 \pm  0.017$  (T7) & $0.273 \pm  0.016$  (T6) & $0.260 \pm  0.035$  (T4/5) & \\
ULAS~J0759+1855	 &	T6   &	  T6	&    $0.191 \pm  0.006$ (T5)   & $0.324 \pm  0.004$ (T6)   & $0.467 \pm  0.005$ ($<$T7) & $0.318 \pm  0.012$  (T5/6) & $0.274 \pm  0.008$  (T6) & $0.090 \pm  0.008$  ($>$T6) & \\
ULAS~J0809+2126	 &	T8   &	  T8	&    $0.026 \pm  0.005$ ($>$T7)  & $0.193 \pm  0.003$ ($>$T7)  & $0.304 \pm  0.003$ (T8) & $0.202 \pm  0.007$  (T7/8) & $0.097 \pm  0.009$  ($>$T7) & $0.175 \pm  0.022$  (T5/6) & \\
ULAS~J0811+2529	 &	T7   &	  T7	&    $0.113 \pm  0.003$ (T7)   & $0.259 \pm  0.002$ (T7)   & $0.399 \pm  0.002$ (T6/7) & $0.263 \pm  0.004$  (T6/7) & $0.229 \pm  0.003$  (T7) & $0.091 \pm  0.004$  ($>$T6) & \\
ULAS~J0814+2452	 &	T5p  & 	  T5	&    $0.237 \pm  0.007$ (T5)   & $0.408 \pm  0.007$ (T5)   & $0.533 \pm  0.008$ ($<$T7) & $0.189 \pm  0.015$  (T7/8) & $0.464 \pm  0.011$  (T5) & $0.152 \pm  0.011$  (T6) & \\
ULAS~J0815+2711	 &	T7p  & 	  T7	&    $0.112 \pm  0.004$ (T7)   & $0.243 \pm  0.002$ (T7)   & $0.412 \pm  0.002$ ($<$T7) & $0.327 \pm  0.005$  (T5) & $0.050 \pm  0.017$  ($>$T7) &        -           & \\
ULAS~J0819+2103	 &	T6   &	  T6	&    $0.183 \pm  0.004$ (T5/6) & $0.350 \pm  0.004$ (T6)   & $0.437 \pm  0.003$ ($<$T7) &        -           &        -           &        -           & \\
ULAS~J0821+2509	 &	T4.5 & 	  T4.5  &    $0.326 \pm  0.004$ (T4)   & $0.508 \pm  0.004$ (T4)   & $0.574 \pm  0.004$ ($<$T7) &        -           &        -           &        -           & \\
ULAS~J0926+0402	 &   	T6   &	  T6	&    $0.186 \pm  0.008$ (T5/6) & $0.317 \pm  0.007$ (T6)   & $0.451 \pm  0.009$ ($<$T7) & $0.392 \pm  0.045$  (T4/5) & $0.309 \pm  0.019$  (T6) & $0.266 \pm  0.025$  (T4) & \\
ULAS~J0927+3413	 &	T5.5 & 	  T5.5  &    $0.200 \pm  0.016$ (T5)   & $0.347 \pm  0.011$ (T6)   & $0.509 \pm  0.015$ ($<$T7) & $0.313 \pm  0.032$  (T5/6) & $0.334 \pm  0.022$  (T6) & $0.166 \pm  0.044$  (T6/7) & \\
WISEP~J0929+0409	 &	T7   &	  T7	&    $0.087 \pm  0.003$ (T7)   & $0.276 \pm  0.002$ (T7)   & $0.374 \pm  0.002$ (T7) & $0.243 \pm  0.004$  (T7) & $0.204 \pm  0.004$  (T7) & $0.092 \pm  0.014$  ($>$T6) & T6.5 K11 \\
ULAS~J0932+3102	 &	T2   &	  T2	&    $0.431 \pm  0.023$ (T2/3) & $0.643 \pm  0.027$ (T2)   & $0.726 \pm  0.023$ ($<$T7) &        -           &        -           &        -           & \\
ULAS~J0950+0117	 &	T8p   &	  T8p	&    $0.043 \pm  0.004$ ($>$T7)  & $0.218 \pm  0.003$ (T7)   & $0.337 \pm  0.003$ (T8) & $0.156 \pm  0.005$  (T8) & $0.187 \pm  0.003$  (T7) & $0.168 \pm  0.014$  (T5/6) & \\
ULAS~J0954+0623	 &	T5   &	  T5	&    $0.200 \pm  0.003$ (T5)   & $0.370 \pm  0.003$ (T5)   & $0.517 \pm  0.003$ ($<$T7) & $0.314 \pm  0.005$  (T6) & $0.407 \pm  0.005$  (T5) & $0.209 \pm  0.009$  (T5) & T5.5 S12 \\
ULAS~J1021+0544	 &	T6   & 	  T6	&    $0.167 \pm  0.004$ (T6)   & $0.331 \pm  0.002$ (T6)   & $0.448 \pm  0.003$ ($<$T7) & $0.346 \pm  0.006$  (T5) & $0.292 \pm  0.005$  (T6) & $0.113 \pm  0.012$  ($>$T6) & \\
ULAS~J1023+0447	 &   	T6.5 &    T6.5  &    $0.100 \pm  0.012$ (T7)   & $0.279 \pm  0.010$ (T6/7) & $0.375 \pm  0.010$ (T7) & $0.282 \pm  0.021$  (T6) & $0.266 \pm  0.014$  (T6) & $0.103 \pm  0.015$  ($>$T6) & \\
ULAS~J1029+0935	 &	T8   &	  T8	&    $0.049 \pm  0.002$ ($>$T7)  & $0.182 \pm  0.001$ ($>$T7)  & $0.303 \pm  0.001$ (T8) & $0.191 \pm  0.003$  (T8) & $0.117 \pm  0.002$  ($>$T7) & $0.071 \pm  0.010$  ($>$T6) & \\
ULAS~J1042+1212	 &	T7.5p & 	  T7.5p  &    $0.064 \pm  0.009$ (T7/8) & $0.286 \pm  0.007$ (T6/7) & $0.333 \pm  0.007$ (T8) & $0.144 \pm  0.018$  (T8/9) & $0.263 \pm  0.013$  (T6) & $0.279 \pm  0.028$  (T4) & \\
ULAS~J1043+1048	 &   	T8   &    T8	&    $0.047 \pm  0.004$ ($>$T7)  & $0.221 \pm  0.003$ (T7)   & $0.320 \pm  0.004$ (T8) & $0.203 \pm  0.009$  (T7/8) & $0.173 \pm  0.006$  (T7) & $0.066 \pm  0.009$  ($>$T6) & \\
ULAS~J1051-0154	 &	T6   &	  T6	&    $0.146 \pm  0.003$ (T6)   & $0.354 \pm  0.002$ (T6)   & $0.436 \pm  0.002$ ($<$T7) & $0.290 \pm  0.005$  (T6) & $0.301 \pm  0.004$  (T6) & $0.095 \pm  0.013$  ($>$T6) & \\
ULAS~J1053+0157	 &	T6.5 &    T6.5  &    $0.092 \pm  0.013$ (T7)   & $0.245 \pm  0.029$ (T7)   & $0.390 \pm  0.046$ (T7/8) & $0.295 \pm  0.028$  (T5/6) & $0.267 \pm  0.012$  (T6) & $0.130 \pm  0.035$  (T6/7) & \\
ULAS~J1111+0518	 &	T4.5 &    T4.5  &    $0.292 \pm  0.037$ (T4/5) & $0.408 \pm  0.051$ (T5/6) & $0.569 \pm  0.071$ ($<$T7) & $0.343 \pm  0.033$  (T5/6) & $0.553 \pm  0.018$  (T4) & $0.259 \pm  0.068$  (T4/5) & \\
ULAS~J1152+0359	 &	T6   &	  T6	&    $0.142 \pm  0.002$ (T6)   & $0.345 \pm  0.001$ (T6)   & $0.448 \pm  0.001$ ($<$T7) & $0.295 \pm  0.003$  (T6) & $0.287 \pm  0.002$  (T6) & $0.158 \pm  0.009$  (T6) & T6 S12 \\
ULAS~J1152+1134	 &	T8.5 & 	  T8.5  &    $0.038 \pm  0.006$ ($>$T7)  & $0.147 \pm  0.004$ ($>$T7)  & $0.260 \pm  0.005$ (T9) & $0.168 \pm  0.011$  (T8) & $0.134 \pm  0.008$  ($>$T7) & $0.052 \pm  0.011$  ($>$T6) & T8.5 \\
ULAS~J1155+0445	 &	T7   & 	  T7	&    $0.126 \pm  0.016$ (T6/7) & $0.319 \pm  0.017$ (T6)   & $0.379 \pm  0.013$ (T7) &        -           &        -           &        -           & \\
ULAS~J1204-0150	 &	T4.5 & 	  T4.5  &    $0.348 \pm  0.008$ (T4)   & $0.426 \pm  0.008$ (T5)   & $0.611 \pm  0.006$ ($<$T7) &        -           &        -           &        -           & T5.5 S12 \\
ULAS~J1206+1018  &	T5   &	  T5	&    $0.227 \pm  0.019$ (T5)   & $0.430 \pm  0.020$ (T5)   & $0.426 \pm  0.015$ ($<$T7) &        -           &        -           &        -           & \\
ULAS~J1212+1010	 &	T5   &	  T5	&    $0.333 \pm  0.020$ (T4/5) & $0.455 \pm  0.022$ (T4/5) & $0.460 \pm  0.016$ ($<$T7) &        -           &        -           &        -           & \\
ULAS~J1223-0131	 &   	T6   &	  T6	&    $0.147 \pm  0.019$ (T6/7) & $0.315 \pm  0.037$ (T6/7) & $0.466 \pm  0.055$ ($<$T7) & $0.312 \pm  0.027$  (T5/6) & $0.434 \pm  0.018$  (T5) & $0.293 \pm  0.058$  (T4/5) & \\
ULAS~J1258+0307	 &	T5   &	  T5	&    $0.187 \pm  0.006$ (T5)   & $0.383 \pm  0.004$ (T5)   & $0.462 \pm  0.004$ ($<$T7) & $0.332 \pm  0.008$  (T5) &        -           &        -           & \\
ULAS~J1259+2933  &	T5   &	  T5	&    $0.227 \pm  0.005$ (T5)   & $0.408 \pm  0.004$ (T5)   & $0.530 \pm  0.004$ ($<$T7) & $0.382 \pm  0.009$  (T4) & $0.415 \pm  0.006$  (T5) & $0.193 \pm  0.007$  (T5) & \\
ULAS~J1302+1434	 &	T4.5 & 	  T4.5  &    $0.276 \pm  0.049$ (T4/5) & $0.432 \pm  0.076$ (T5/6) & $0.561 \pm  0.098$ ($<$T7) & $0.378 \pm  0.028$  (T4/5) & $0.516 \pm  0.018$  (T4) & $0.317 \pm  0.054$  (T3/4) & \\
ULAS~J1335+1506	 &	T6   &	  T6	&    $0.148 \pm  0.010$ (T6)   & $0.371 \pm  0.007$ (T5)   & $0.450 \pm  0.007$ ($<$T7) & $0.289 \pm  0.015$  (T6) &        -           &        -           & \\
ULAS~J1338-0142	 &	T7.5 & 	  T7.5  &    $0.010 \pm  0.045$ ($>$T7)  & $0.237 \pm  0.045$ (T7/8) & $0.237 \pm  0.033$ (T9) &        -           &        -           &        -           & \\
ULAS~J1339-0056	 &	T7   &	  T7	&    $0.073 \pm  0.004$ (T7/8) & $0.275 \pm  0.003$ (T7)   & $0.354 \pm  0.003$ (T7) & $0.226 \pm  0.006$  (T7) &        -           &        -           & \\
ULAS~J1339+0104	 &	T5   &	  T5	&    $0.215 \pm  0.004$ (T5)   & $0.402 \pm  0.003$ (T5)   & $0.500 \pm  0.004$ ($<$T7) & $0.288 \pm  0.006$  (T6) &        -           &        -           & \\
\hline\end{tabular}
}
\caption{Spectral typing ratios for the confirmed T~dwarfs as set out
  by \citet{burgasser06,ben09}, along with the types from by-eye comparison to template spectral standards
  and the final adopted types. The notes column indicates spectral types determined by authors where: A11 = \citet{albert2011}; K11 = \citet{kirkpatrick2011}; S12 = \citet{scholz2012}. In the case of T8+ objects, the notes column indicates the spectral type using the \citet{cushing2011} system.
\label{tab:types}
}

\end{table}
\end{landscape}

\begin{landscape}
\addtocounter{table}{-1}
\begin{table}
{\scriptsize
\begin{tabular}{ c  c c c c c c c c c c c c c }
  \hline
Name & Adopted & Templ. & H$_2$O-$J$ & CH$_4$-$J$ & $W_J$ & H$_2$O-$H$ & CH$_4$-$H$ & CH$_4$-$K$ & Note \\
\hline
ULAS~J1417+1330	 &	T5   &	  T5	&    $0.204 \pm  0.004$ (T5)   & $0.381 \pm  0.003$ (T5)   & $0.491 \pm  0.002$ ($<$T7) & $0.306 \pm  0.003$  (T6) & $0.378 \pm  0.002$  (T5) & $0.174 \pm  0.002$  (T6) & \\
ULAS~J1421+0136	 &	T4.5 & 	  T4.5  &    $0.374 \pm  0.015$ (T3/4) & $0.610 \pm  0.018$ (T2)   & $0.582 \pm  0.014$ ($<$T7) &        -           &        -           &        -           & \\
ULAS~J1425+0451	 &	T6.5 & 	  T6.5  &    $0.122 \pm  0.009$ (T6/7) & $0.284 \pm  0.008$ (T6/7) & $0.424 \pm  0.010$ ($<$T7) & $0.263 \pm  0.016$  (T6/7) & $0.282 \pm  0.012$  (T6) & $0.097 \pm  0.021$  ($>$T6) & \\
ULAS~J1449+1147	 &	T5.5 & 	  T5.5  &    $0.235 \pm  0.004$ (T5)   & $0.378 \pm  0.003$ (T5)   & $0.522 \pm  0.003$ ($<$T7) & $0.301 \pm  0.009$  (T6) & $0.389 \pm  0.005$  (T5) & $0.254 \pm  0.010$  (T4) & \\
ULAS~J1516+0110	 &	T6.5 &	  T6.5  &    $0.126 \pm  0.007$ (T6/7) & $0.323 \pm  0.004$ (T6)   & $0.437 \pm  0.005$ ($<$T7) & $0.301 \pm  0.009$  (T6) &        -           &        -           & \\
WISE~J1517+0529	 &	T8p   &	  T8p	&    $0.045 \pm  0.004$ ($>$T7)  & $0.220 \pm  0.003$ (T7)   & $0.331 \pm  0.004$ (T8) & $0.224 \pm  0.010$  (T7) & $0.168 \pm  0.006$  (T7) & $0.201 \pm  0.018$  (T5) & T8 M13 \\
ULAS~J1534+0556	 &	T5   &	  T5	&    $0.256 \pm  0.016$ (T5)   & $0.395 \pm  0.015$ (T5)   & $0.515 \pm  0.016$ ($<$T7) & $0.351 \pm  0.025$  (T4/5) & $0.442 \pm  0.022$  (T5) & $0.237 \pm  0.021$  (T4/5) & \\
ULAS~J1536+0155	 &	T5   &	  T5	&    $0.298 \pm  0.004$ (T5)   & $0.475 \pm  0.003$ (T4)   & $0.570 \pm  0.003$ ($<$T7) & $0.348 \pm  0.004$  (T5) &        -           &        -           & \\
ULAS~J1549+2621	 &	T5   &	  T5	&    $0.233 \pm  0.005$ (T5)   & $0.381 \pm  0.004$ (T5)   & $0.543 \pm  0.004$ ($<$T7) & $0.318 \pm  0.007$  (T5/6) &        -           &        -           & \\
ULAS~J1601+2646	 &	T6.5 & 	  T6.5  &    $0.084 \pm  0.005$ (T7)   & $0.224 \pm  0.004$ (T7)   & $0.310 \pm  0.004$ (T8) & $0.278 \pm  0.012$  (T6) & $0.205 \pm  0.009$  (T7) & $0.080 \pm  0.008$  ($>$T6) & \\
ULAS~J1614+2442	 &   	T7   &	  T7	&    $0.096 \pm  0.012$ (T7)   & $0.289 \pm  0.010$ (T6/7) & $0.375 \pm  0.011$ (T7) & $0.062 \pm  0.030$  (T9) & $0.160 \pm  0.014$  (T7/8) & $0.088 \pm  0.023$  ($>$T6) & \\
ULAS~J1617+2350	 &	T6   &	  T6	&    $0.181 \pm  0.006$ (T5/6) & $0.383 \pm  0.003$ (T5)   & $0.522 \pm  0.004$ ($<$T7) & $0.313 \pm  0.010$  (T5/6) & $0.348 \pm  0.006$  (T6) & $0.170 \pm  0.007$  (T6) & \\
ULAS~J1619+2358	 &	T6   &	  T7	&    $0.074 \pm  0.017$ (T7/8) & $0.273 \pm  0.018$ (T6/7) & $0.362 \pm  0.012$ (T7) &        -           &        -           &        -           & \\
ULAS~J1619+3007	 &	T5   &	  T5	&    $0.321 \pm  0.007$ (T4/5) & $0.441 \pm  0.005$ (T5)   & $0.605 \pm  0.006$ ($<$T7) & $0.389 \pm  0.015$  (T4) & $0.402 \pm  0.010$  (T5) & $0.199 \pm  0.010$  (T5) & \\
ULAS~J1626+2524	 &	T5   &	  T5	&    $0.243 \pm  0.020$ (T5)   & $0.468 \pm  0.014$ (T4)   & $0.472 \pm  0.014$ ($<$T7) & $0.294 \pm  0.023$  (T6) &        -           &        -           & \\
ULAS~J1639+3232	 &	T3   &	  T3	&    $0.387 \pm  0.005$ (T3)   & $0.584 \pm  0.007$ (T2/3) & $0.631 \pm  0.005$ ($<$T7) &        -           &        -           &        -           & \\
ULAS~J2116-0101	 &	T6   &	  T6	&    $0.175 \pm  0.010$ (T5/6) & $0.377 \pm  0.011$ (T5)   & $0.511 \pm  0.014$ ($<$T7) & $0.291 \pm  0.014$  (T6) & $0.339 \pm  0.010$  (T6) & $0.138 \pm  0.013$  (T6/7) & \\
ULAS~J2237+0642	 &	T6.5p &   T6.5p &    $0.128 \pm  0.006$ (T6/7) & $0.295 \pm  0.005$ (T6)   & $0.430 \pm  0.007$ ($<$T7) & $0.190 \pm  0.014$  (T7/8) & $0.241 \pm  0.010$  (T6/7) & $0.139 \pm  0.016$  (T6/7) & \\
ULAS~J2300+0703	 &	T4.5 &    T4.5  &    $0.316 \pm  0.008$ (T4/5) & $0.504 \pm  0.009$ (T4)   & $0.583 \pm  0.006$ ($<$T7) &        -           &        -           &        -           & \\
ULAS~J2315+0344	 &	T7   &	  T7	&    $0.093 \pm  0.017$ (T7)   & $0.270 \pm  0.023$ (T6/7) & $0.382 \pm  0.018$ (T6/7) & $0.184 \pm  0.024$  (T7/8) & $0.193 \pm  0.016$  (T7) & $0.254 \pm  0.046$  (T4/5) & \\
ULAS~J2318+0433	 &	T7.5 & 	  T7.5  &    $0.054 \pm  0.007$ ($>$T7)  & $0.248 \pm  0.005$ (T7)   & $0.343 \pm  0.006$ (T8) & $0.257 \pm  0.015$  (T6/7) & $0.182 \pm  0.011$  (T7) & $0.077 \pm  0.019$  ($>$T6) & \\
ULAS~J2326+0201	 &	T8   &	  T8	&    $0.052 \pm  0.006$ ($>$T7)  & $0.169 \pm  0.012$ ($>$T7)  & $0.304 \pm  0.006$ (T8) & $0.203 \pm  0.011$  (T7/8) & $0.097 \pm  0.006$  ($>$T7) & $0.046 \pm  0.010$  ($>$T6) & \\
ULAS~J2331+0426	 &	T4   &	  T4	&    $0.388 \pm  0.013$ (T3/4) & $0.489 \pm  0.010$ (T4)   & $0.693 \pm  0.013$ ($<$T7) & $0.466 \pm  0.019$  (T2/3) & $0.654 \pm  0.016$  (T3) & $0.429 \pm  0.017$  (T3) & \\
ULAS~J2342+0856	 &	T6   &	  T6	&    $0.156 \pm  0.003$ (T6)   & $0.340 \pm  0.003$ (T6)   & $0.437 \pm  0.002$ ($<$T7) &        -           &        -           &        -           &  T7 (photometric) S10\\
ULAS~J2352+1244  &	T6.5 &    T6.5  &    $0.152 \pm  0.008$ (T6)   & $0.312 \pm  0.006$ (T6)   & $0.449 \pm  0.009$ ($<$T7) & $0.276 \pm  0.014$  (T6) & $0.206 \pm  0.014$  (T7) & $0.091 \pm  0.021$  ($>$T6) & \\
ULAS~J2357+0132	 &	T5.5p &	  T5.5p &    $0.173 \pm  0.003$ (T6)   & $0.362 \pm  0.004$ (T5/6) & $0.484 \pm  0.005$ ($<$T7) & $0.313 \pm  0.007$  (T6) & $0.420 \pm  0.007$  (T5) & $0.186 \pm  0.004$  (T5) & \\
\hline
\end{tabular}
}
\caption{(Continued) Spectral typing ratios for the confirmed T~dwarfs as set out
  by \citet{burgasser06,ben09}, along with the types from by-eye comparison to template spectral standards
  and the final adopted types. The notes column indicates spectral types determined by authors where: K11 = \citet{kirkpatrick2011}; S12 = \citet{scholz2012}; M13 = \citet{mace2013}. In the case of T8+ objects, the notes column indicates the spectral type using the \citet{cushing2011} system.
\label{tab:types}
}

\end{table}
\end{landscape}

\section{Space-based mid-IR photometry}
\label{sec:midIR}

\subsection{WISE cross-matched photometry}
\label{sec:wise}

We cross-matched our full list of 171 spectroscopically confirmed T~dwarfs within the UKIDSS LAS (which includes some objects that were confirmed in the literature rather than by our follow-up) against the WISE all sky release catalogue, with matching radius of 6\arcsec. The typical epoch difference between the UKIDSS and WISE observations is less than 3 years, so this ensured that high proper motion targets would still be matched. All apparent matches were visually inspected to remove spurious correlations. Sixty seven T~dwarfs were found with WISE photometry, of which six were affected by blending with another source.  Twenty five of the 67 T~dwarfs with WISE photometry are confirmed here for the first time, and their details are given in Table~\ref{tab:wise}.

\begin{table*}
\begin{tabular}{ l >{$}c<{$} >{$}c<{$}  >{$}c<{$}  >{$}c<{$} c }
\hline
Name & W1 & W2 & W3 & W4 & WISE blend? \\ 
\hline
CFBDS~J0133+0231 & 17.78 \pm 0.27 & 15.10 \pm 0.09 & >12.91 & >9.26  & N \\
ULAS~J0139+1503 & 17.86 \pm 0.26 & 15.94 \pm 0.16 & >12.93 & >9.31 & N\\
ULAS~J0200+0908 & 16.16 \pm 0.07 & 15.70 \pm 0.14 & >12.88 & >9.24 & Y\\
ULAS~J0329+0430 & 17.52 \pm 0.25 & 15.35 \pm 0.14 & >12.20 & >8.53 & N\\
ULAS~J0745+2332 & 14.79 \pm 0.04 & 14.47 \pm 0.07 & 12.51 \pm 0.50 & >8.41 & Y\\
ULAS~J0758+2225 & 16.95 \pm 0.16 & 15.23 \pm 0.12 & >12.65 & >9.03 & N\\
ULAS~J0819+2103 & 17.16 \pm 0.16 & 15.28 \pm 0.10 & >12.17 & >8.81 & N\\
ULAS~J0821+2509 & 17.40 \pm 0.22 & 15.48 \pm 0.14 & >12.67 & >9.17 & N\\
ULAS~J0950+0117 & 18.05 \pm 0.34 & 14.48 \pm 0.06 & >12.85 & >9.20 & N\\
ULAS~J0954+0623 & 16.67 \pm 0.13 & 14.66 \pm 0.08 & >12.64 & >8.70 & N\\
ULAS~J1021+0544 & 16.63 \pm 0.14 & 15.31 \pm 0.28 & >11.94 & >8.14 & N\\
ULAS~J1029+0935 & 16.84 \pm 0.13 & 14.29 \pm 0.08 & 11.58 \pm 0.33 & >8.58 & N\\
ULAS~J1043+1048 & >18.28 & 15.66 \pm 0.18 & 12.10 \pm 0.32 & >9.07 & N\\
ULAS~J1152+0359 & 16.97 \pm 0.15 & 15.34 \pm 0.13 & 12.29 \pm 0.38 & >9.03 & N\\
ULAS~J1152+1134 & 16.89 \pm 0.15 & 14.66 \pm 0.08 & 12.37 \pm 0.41 & >8.57 & N\\
ULAS~J1204-0150 & 16.66 \pm 0.11 & 14.70 \pm 0.08 & 12.48 \pm 0.42 & >8.60 & N\\
ULAS~J1206+1018 & 17.55 \pm 0.24 & 15.83 \pm 0.19 & >12.36 & >8.86 & N\\
ULAS~J1338-0142 & >18.22 & 16.12 \pm 0.19 & >13.01 & >9.34 & N\\
ULAS~J1417+1330 & 16.67 \pm 0.08 & 14.70 \pm 0.06 & 12.51 \pm 0.29 & >9.49 & N\\
ULAS~J1449+1147 & 17.39 \pm 0.16 & 14.84 \pm 0.07 & >12.40 & >9.44 & N\\
ULAS~J1517+0529 & >18.10 & 15.13 \pm 0.08 & >13.13 & >9.52 & N\\
ULAS~J1549+2621 & 17.13 \pm 0.10 & 16.09 \pm 0.13 & 13.42 \pm 0.5 & >9.89 & N\\
ULAS~J1639+3232 & 16.62 \pm 0.10 & 15.02 \pm 0.08 & 12.28 \pm 0.28 & >9.05 & N\\
ULAS~J2326+0201 & 18.03 \pm 0.42 & 15.45 \pm 0.16 & >12.53 & >8.96 & N\\
ULAS~J2342+0856 & 16.07 \pm 0.08 & 13.97 \pm 0.05 & 12.63 \pm 0.53 & 9.08 \pm 0.54 & N\\
\hline\end{tabular}
\caption{WISE photometry for 25 of the spectroscopically confirmed T~dwarfs presented here for the first time.
\label{tab:wise}
}
\end{table*}

In Figures~\ref{fig:brighthist} and~\ref{fig:fainthist} we compare the numbers of T~dwarfs detected by UKIDSS with those detected by both WISE and UKIDSS for objects brighter than $J = 18.3$ and for objects in the $J = 18.3 - 18.8$ range (the faintest 0.5 magnitude bin of our complete sample). It can be seen that UKIDSS is considerably more sensitive to earlier type objects than WISE, with roughly twice as many T4 -- T6.5 dwarfs identified in the $J < 18.3$ regime. In the fainter $18.3 < J < 18.8$ regime this effect is even more pronounced, and extends to the T7 -- T7.5 bin. This highlights that although WISE now dominates the search for cool and faint T8+ dwarfs, wide and deep near-infrared surveys such as UKIDSS, and the (wider) VISTA VHS and (deeper) VIKING surveys continue represent an important resource for exploring the L and T dwarf sequences. 

The faint nature of the large number of L and T~dwarfs that will be revealed by VHS and VIKING, in particular, will be extremely challenging to confirm spectroscopically. To take advantage of the opportunity these offer for robust statistical studies of the substellar component of the Galaxy, it will be essential to develop methods for determining their properties from the photometric data that will be supplied by the surveys.

\begin{figure}
\includegraphics[width=220pt, angle=0]{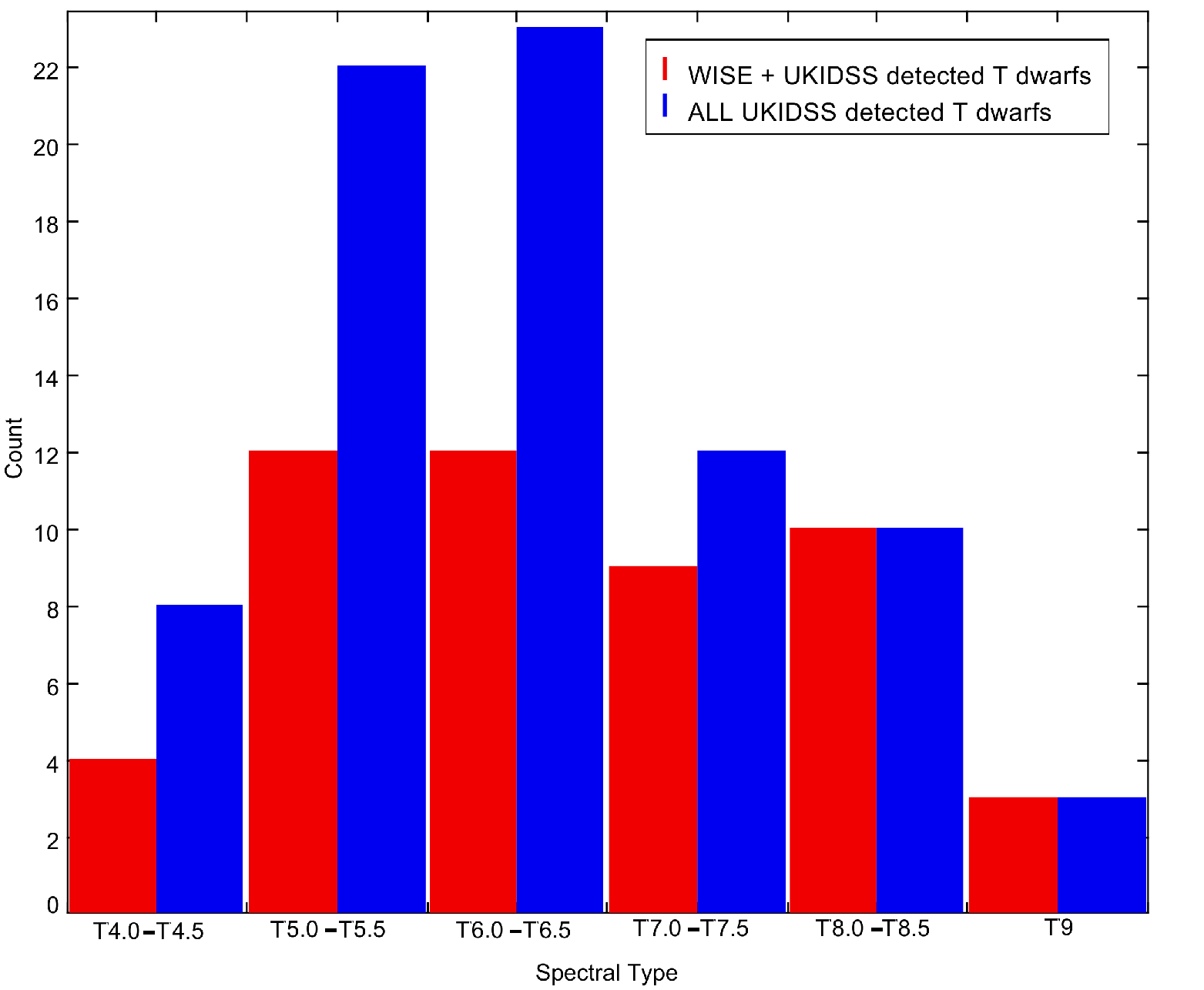}
\caption{A histogram of UKIDSS and WISE  + UKIDSS detected T~dwarfs in our sample for objects with $J < 18.3$.}
\label{fig:brighthist}
\end{figure}

\begin{figure}
\includegraphics[width=220pt, angle=0]{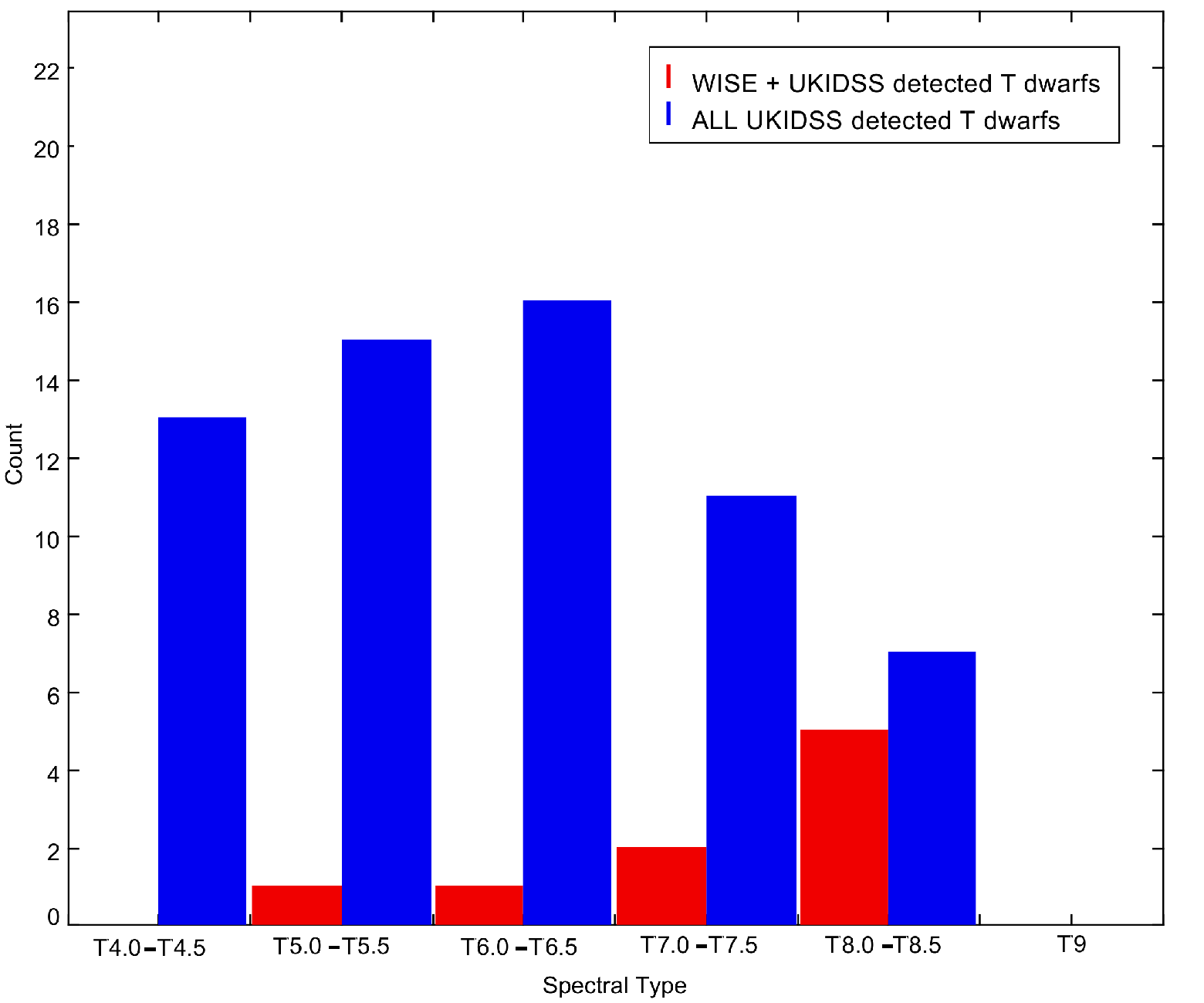}
\caption{A histogram of UKIDSS and WISE  + UKIDSS detected T~dwarfs in our sample for objects with $18.3 < J < 18.8$.}
\label{fig:fainthist}
\end{figure}

%\begin{figure}
%\includegraphics[width=220pt, angle=0]{YJ_JW2_colplot_bytype.ps}
%\caption{A $J - W2$ vs $Y - J$ colour-colour diagram for 61 spectroscopically confirmed T~dwarfs in the UKIDSS LAS with W2 photometry. Spectral types are indicated by coloured symbols.}
%\label{fig:YJW2}
%\end{figure}

\subsection{Warm-{\it Spitzer} photometry}
\label{sec:spitz}

Warm-{\it Spitzer}  IRAC photometry was obtained for some of the T~dwarfs presented in this work via Cycles 6, 7 and 8 GO programs   60093, 70058 and  80077 (PI~Leggett). The observations were carried out in both the [3.6] and [4.5] bands (hereafter Ch1 and Ch2 respectively), with typically a 30s frame time, repeated 3 to 6 times per pointing, and dither patterns consisting of 12 or 16 positions. In all cases the post-basic-calibrated-data mosaics generated by the Spitzer pipeline were used to obtain aperture photometry. Generally the sky levels were determined from annular regions. Aperture corrections were taken from the IRAC handbook.

Table~\ref{tab:spitz}  lists the targets observed, the associated program ID, the date of the observation, the pipeline version used,  the frame time, the total integration time, the aperture size, the derived photometry and uncertainties, and any notes on the dataset.   The Table includes known sources not presented in this work, for which the IRAC data has not been previously published. We include it here so that it is available to the community.

\begin{table*}
\begin{tabular}{ c c c c c c c c c c c }
\hline
Name      &   Spectral  &    Program  &   Obs date &  Pipeline &  Frame  &   Integration & Aperture  &    [3.6] &  [4.5]  & Notes \\
	       & type & number & (UT) & version & time (s) & (min) & (arcsec) & (mag) & (mag) & \\
\hline
CFBDS~J005910.90-011401.3 & T8.5$^{1}$ & 50667 & 2009-01-26 & S18.7.0 & 12 & 7.2 & 7.2 & $15.71\pm 0.01$ & $13.66 \pm 0.01$ & A \\
CFBDS~J030135-161418  &  T7.5$^{2}$ & 60093 & 2009-09-02 & S18.12.0 & 30 & 48.0 & 7.2 & $16.95 \pm 0.01$ & $15.42 \pm 0.01$ & \\
2MASS~J07290002-3954043& T8$^{3}$ & 60093 & 2009-12-04 & S18.13.0 & 12 & 1.0 & 4.8 & $14.47 \pm 0.01$ & $12.95 \pm 0.01$ & \\
ULAS~J0809+2126  & T8 & 70058 & 2011-05-31 & S18.18.0 & 30 & 48.0 & 4.8 & $17.74 \pm 0.05$ & $16.03 \pm 0.01$ & B \\
CFBDS~J092250+152741 &  & 60093 &  2010-05-29 &  S18.18.0 &  30 & 48.0 & 7.2 &  $17.45 \pm 0.04$ & $16.11 \pm 0.01$ & \\
ULAS~J0950+0117 & T8p & 60093 &  2010-01-06 &  S18.13.0 & 30 & 48.0 & 7.2 & $16.28 \pm 0.01$ & $14.35 \pm 0.02$ & \\
ULAS~J1043+1048 & T8 & 70058 & 2012-02-01 & S19.1.0 & 30 & 48.0 & 7.2 & $16.94 \pm 0.01$ & $15.34 \pm 0.03$ & \\
ULAS~J123327.45+121952.2 & T3.5p$^{4}$ & 80077 & 2012-09-03 & S19.1.0 & 30 &  48.0 &  7.2 & $16.83 \pm 0.01$ &  $15.61 \pm 0.01$ &  B \\
ULAS~J1339+0104 & T5 & 80077 & 2012-09-18  & S19.1.0 & 30 &  48.0 & 7.2  & $16.93 \pm 0.01$ & $16.08 \pm 0.02$ & \\
ULAS~J2237+0642 & T6.5p & 80077 & 2012-09-08 & S19.1.0 & 30 & 48.0 & 7.2 & $17.87 \pm 0.01$ & $15.48 \pm 0.02$ &  \\
ULAS~J2326+0201  & T8 & 80077 & 2012-01-31  & S19.1.0 & 30 & 24.0 & 7.2 & $16.84 \pm 0.03$ & $15.37 \pm 0.01$ & \\
\hline
\multicolumn{11}{l}{$^{1}$ \citet{cushing2011}}\\
\multicolumn{11}{l}{$^{2}$ \citet{albert2011}}\\
\multicolumn{11}{l}{$^{3}$ \citet{looper07}}\\
\multicolumn{11}{l}{$^{4}$ \citet{ben10b}}\\
\end{tabular}
\caption{Spitzer photometry for a subset of T~dwarfs, selected either as examples of late-T types, peculiar spectra or as benchmark objects. Full designations are given for those objects whose discoveries are reported in other publications, whilst abbreviated names are given for those objects whose discovery is presented in this work. Notes:
{\bf A:} The data were taken when Spitzer was cold, and longer wavelength photometry was also obtained: ${\rm [5.8]} =14.24 \pm 0.03$,  ${\rm [8.0]} = 13.31 \pm 0.03$.
{\bf B:} Separate, non-annular, skies used due to background structure.
\label{tab:spitz}
}
\end{table*}

\section{Wide common proper motion binary systems}
\label{sec:wcpm}

\subsection{Identifying companions}
\label{sec:compsel}

In Table~\ref{tab:pms} we present proper motions for the targets
identified in this paper, along with those for late-T~dwarfs found
within the UKIDSS LAS sky in previous works. The vast majority
of these proper motions have been drawn from the catalogue of Smith et
al (in prep) which presents proper motions calculated from 2 epochs of
$J$~band UKIDSS LAS observations.  Here we present only the absolute proper motions, and refer the reader to the Smith et al catalogue for further astrometric parameters.
A small number of additional proper
motions have been calculated using an identical method to that used in
Smith et al (in prep), but using our follow-up WFCAM $J$ band
observations for the second epoch, instead of UKIDSS survey data.

We have cross matched our full catalogue of T~dwarfs
identified within the UKIDSS LAS sky that have proper motions (128
targets) against
several astrometric catalogues. We crossmatched our targets against
the Hipparcos \citep{hipparcos,newhip}, LSPM-NORTH \citep{lspm} and
NOMAD \citep{nomad} catalogues. We searched projected separations up to
20000AU, assuming a minimum likely distance for each source. Minimum
and maximum likely distances for each source were determined by
considering the  $\pm 0.5$~subtype spectral type uncertainty, the
mean M$_J$ for each spectral subtype and the scatter about M$_J$ as presented in
\citet{dupuy2012}. Since a significant fraction of wide-common proper motion companions are themselves multiple systems \citep[e.g.][]{faherty2010}, we also assigned an upper limit to the distance based on the target being an unresolved equal luminosity binary system. These distances are presented in Table~\ref{tab:pms}.

%\begin{landscape}
\begin{table*}
{\scriptsize
\begin{tabular}{ c  c c c c  >{$}c<{$}  >{$}c<{$}  c  >{$}c<{$}  >{$}c<{$}  >{$}c<{$}  }
\hline
Name & $\alpha$ & $\delta$ & SpType & Ref &  \mu_{\alpha \cos \delta} & \mu_{\delta} & Baseline & D_{min} & D_{max} & D{max}(binary) \\
& J2000 & J2000 & & & mas/yr & mas/yr & years & pc & pc & pc \\
\hline
ULAS~J013017.79+080453.9  &  01:30:17.79  &  +08:04:53.90  &  T6    &    			 &  12.98  \pm  24.45  &  -43.98  \pm  59.09  &  1.29$^{W}$  &30.1&51.3&  72.4 \\
ULAS~J020013.18+090835.2  &  02:00:13.18  &  +09:08:35.20  &  T6    &    			 &  7.16  \pm  27.97  &  -40.39  \pm  28.01  &1.28&28.4&48.4&  68.4 \\
ULAS~J022603.18+070231.4  &  02:26:03.18  &  +07:02:31.40  &  T7    &    			 &  54.55  \pm  16.02  &  59.24  \pm  16.38  &  2.02 $^{W}$  &22.6&53.0&  74.8 \\
ULAS~J024557.88+065359.4  &  02:45:57.88  &  +06:53:59.40  &  T7    &    			 &  -101.13  \pm  32.52  &  -112.72  \pm  29.84  &  1.05 $^{W}$  & 21.0 &49.2&  69.5 \\
SDSS~J032553.17+042540.1  &  03:25:53.13  &  +04:25:40.10  &  T5.5  &  1 	 &  -194.65  \pm  15.04  &  -102.68  \pm  14.86  &  2.09 $^{W}$  &13.4& 24.1&  34.0 \\
ULAS~J032920.22+043024.5  &  03:29:20.22  &  +04:30:24.50  &  T5    &    			 &  225.25  \pm  15.8  &  -59.84  \pm  15.51  &  2.09 $^{W}$  &31.5&53.2&  75.2 \\
SDSS~J074149.14+235127.3  &  07:41:49.01  &  +23:51:25.90  &  T5    &  2	 &  -259.64  \pm  11  &  -216.24  \pm  11.02  &2.03&14.6&24.7&  34.9 \\
ULAS~J074503.03+233240.3  &  07:45:03.03  &  +23:32:40.30  &  T9    &    			 &  -251.7  \pm  21.34  &  -287.98  \pm  21.1  &2.03&8.7&19.4&  27.4 \\
ULAS~J074617.17+235532.2  &  07:46:17.17  &  +23:55:32.20  &  T7    &    			 &  -3.75  \pm  24.51  &  -120.29  \pm  24.53  &2.03&28.2&66.1&  93.3 \\
ULAS~J074502.79+245516.3  &  07:45:02.79  &  +24:55:16.30  &  T6.5  &    			 &  -84.88  \pm  12.55  &  -60.28  \pm  12.52  &2.03&31.6&45.1&  63.7 \\
WISEP~J075003.84+272544.8  &  07:50:03.84  &  +27:25:44.80  &  T8.5  &  3   &  -734.03  \pm  14.2  &  -195.27  \pm  14.49  & 2.00 &12.9&36.4&  51.4 \\
2MASS~J07554795+2212169  &  07:55:47.95  &  +22:12:14.50  &  T5   &  4   &  -4.2  \pm  10.28  &  -252.3  \pm  10.65  &2.91&12.1&20.4&  28.9 \\
ULAS~J075829.83+222526.7  &  07:58:29.83  &  +22:25:26.70  &  T6.5  &    			 &  -105.34  \pm  10.37  &  -57.49  \pm  10.84  &2.91&24.5&35.0&  49.4 \\
ULAS~J075937.75+185555.0  &  07:59:37.75  &  +18:55:55.00  &  T6    &    			 &  -48.17  \pm  15.64  &  -81.2  \pm  15.91  &2.9&42.9&73.1&  103.3 \\
ULAS~J080918.41+212615.2  &  08:09:18.41  &  +21:26:15.20  &  T8    &    			 &  -152.69  \pm  15.28  &  -154.7  \pm  16.28  &1.86& 12.0 &44.3&  62.5 \\
ULAS~J081110.86+252931.8  &  08:11:10.86  &  +25:29:31.80  &  T7    &    			 &  45.25  \pm  11.99  &  -231.89  \pm  12.05  &1.97&14.6&34.2&  48.3 \\
ULAS~J081407.51+245200.9  &  08:14:07.51  &  +24:52:00.90  &  T5p   &    			 &  -51.29  \pm  14.3  &  -9.48  \pm  14.11  &2.03&49.7&83.9&  118.6 \\
ULAS~J081507.26+271119.2  &  08:15:07.26  &  +27:11:19.20  &  T7p   &    			 &  -50.15  \pm  37.06  &  -79.91  \pm  36.98  &  0.90 $^{W}$  &20.5&48.1&  67.9 \\
ULAS~J081918.58+210310.4  &  08:19:18.58  &  +21:03:10.40  &  T6    &    			 &  -57.72  \pm  11.49  &  -181.4  \pm  10.67  &1.86&19.2&32.7&  46.2 \\
ULAS~J081948.08+073323.3  &  08:19:48.08  &  +07:33:23.30  &  T6p   & 5  	 &  13.17  \pm  9.41  &  -68.5  \pm  9.4  &5.09&34.7&59.2&  83.6 \\
ULAS~J082155.49+250939.6  &  08:21:55.49  &  +25:09:39.60  &  T4.5  &    			 &  -449.31  \pm  14.02  &  -56.45  \pm  10.26  &2.03&31.3&45.9&  64.9 \\
ULAS~J082327.46+002424.4  &  08:23:27.46  &  +00:24:24.40  &  T4.0  & 6  &  -35.13  \pm  10.06  &  -221.22  \pm  10.11  &5.04&55.4&83.9&  118.5 \\
ULAS~J082707.67-020408.2  &  08:27:07.67  &  -02:04:08.20  &  T5.5  & 7       &  23.21  \pm  8.01  &  -111.73  \pm  7.97  &6.07&22.7&40.8&  57.6 \\
SDSS~J083048.81+012831.0  &  08:30:48.89  &  +01:28:28.90  &  T4.5  &  4  &  186.16  \pm  9.08  &  -361.52  \pm  8.96  &5.07&18.2&26.7&  37.7 \\
ULAS~J083756.19-004156.0  &  08:37:56.19  &  -00:41:56.00  &  T3.0  &  6  &  -13.13  \pm  10.51  &  -94.16  \pm  10.53  &6.06&48.8&78.0&  110.2 \\
ULAS~J084036.72+075933.6  &  08:40:36.72  &  +07:59:33.60  &  T4.5  &  5  	 &  -270.68  \pm  12.98  &  -82.66  \pm  12.93  &5.28&72.1&105.7&  149.3 \\
ULAS~J084211.68+093611.9  &  08:42:11.68  &  +09:36:11.90  &  T5.5  &  5  	 &  -201.73  \pm  12.49  &  -53.25  \pm  12.62  &5.20&39.8&71.4&  100.9 \\
ULAS~J085139.03+005340.9  &  08:51:39.03  &  +00:53:40.90  &  T4.0  &  5  	 &  -61.66  \pm  11.07  &  -39.68  \pm  10.86  &5.18&62.5&94.6&  133.7 \\
ULAS~J085342.94+000651.8  &  08:53:42.94  &  +00:06:51.80  &  T6p   &  5  	 &  -43.79  \pm  9.51  &  120.23  \pm  9.62  &5.19&41.5&70.8&  100 \\
ULAS~J090116.23-030635.0  &  09:01:16.23  &  -03:06:35.00  &  T7.5  &  7  	 &  -56.37  \pm  8.62  &  -253.84  \pm  10.39  &6.18&16.1&31.3&  44.2 \\
ULAS~J091309.55-003136.9  &  09:13:09.55  &  -00:31:36.90  &  T6    &    			 &  72.67  \pm  11.54  &  -51.54  \pm  11.36  &5.28&50.4&85.9&  121.3 \\
ULAS~J092624.76+071140.7  &  09:26:24.76  &  +07:11:40.70  &  T3.5  &  5  	 &  -51.48  \pm  10.67  &  -420.41  \pm  12.3  &5.12&34.5&47.7&  67.4 \\
WISEP~J092906.77+040957.9 &  09:29:06.75  &  +04:09:57.70  &  T7    &   3 	 &  526.12  \pm  32.38  &  -438.45  \pm  30.96  &  0.97 $^{W}$  &10.6&24.8&  35.0 \\
ULAS~J092926.44+110547.3  &  09:29:26.44  &  +11:05:47.30  &  T2.0  &  5  	 &  -41.48  \pm  11.88  &  9.95  \pm  11.8  &5.12&63.1&100.5&  141.9 \\
ULAS~J093245.48+310206.4  &  09:32:45.48  &  +31:02:06.40  &  T2    &    			 &  -44.82  \pm  14.75  &  -6.29  \pm  14.51  &2.02&53.7&85.6&  120.9 \\
ULAS~J093829.28-001112.6  &  09:38:29.28  &  -00:11:12.60  &  T4.5  &  6  &  -255.55  \pm  12.59  &  -81.28  \pm  11.48  &5.32&57.3&83.9&  118.6 \\
ULAS~J093951.0400:6:3.60  &  09:39:51.04 &  +00:16:53.60  &  T5.5  &  6  &  159.08  \pm  11.15  &  -299.34  \pm  11.03  &5.32&31.6&56.8&  80.2 \\
ULAS~J094331.49+085849.2  &  09:43:31.49  &  +08:58:49.20  &  T5p   &  5  	 &  -83.54  \pm  11.48  &  -78.78  \pm  11.5  &5.21&51.1&86.3&  121.9 \\
ULAS~J094349.60+094203.4  &  09:43:49.60  &  +09:42:03.40  &  T4.5p &  5  	 &  44.98  \pm  13.17  &  -123.92  \pm  12.72  &4.80&65.8&96.4&  136.1 \\
ULAS~J094516.39+075545.6  &  09:45:16.39  &  +07:55:45.60  &  T5.0  &  5  	 &  -129.17  \pm  11.35  &  -41.17  \pm  10.61  &5.33&30.6&51.8&  73.1 \\
ULAS~J094806.06+064805.0  &  09:48:06.06  &  +06:48:05.00  &  T7.0  &  7  	 &  238.8  \pm  12.16  &  -273.6  \pm  11.96  &6.06&26.0&61.0&  86.2 \\
ULAS~J095047.28+011734.3  &  09:50:47.28  &  +01:17:34.30  &  T8p    &    			 &  242.68  \pm  11.79  &  -386.56  \pm  11.71  &3.97&9.3&34.2&  48.3 \\
ULAS~J095429.90+062309.6  &  09:54:29.90  &  +06:23:09.60  &  T5    &   			 &  -494.26  \pm  9.92  &  -436.26  \pm  10.79  &2.75&20.3&34.4&  48.5 \\
ULAS~J095829.86-003932.0  &  09:58:29.86  &  -00:39:32.00  &  T5.5  &  6  &  -59.18  \pm  12.49  &  1.73  \pm  12.33  &6.24&51.8&92.9&  131.2 \\
CFBDS~J100113+022622   &  10:01:13.04  &  +02:26:22.40  &  T5   &  8  &  -90.21  \pm  13.7  &  49.41  \pm  14.23  &3.99&59.1&99.8&  141.0 \\
ULAS~J100759.90-010031.1  &  10:07:59.90  &  -01:00:31.10  &  T5.5  &  7  	 &  -226.6  \pm  11.08  &  145.69  \pm  11.19  &6.08&45.7&82.1&  115.9 \\
ULAS~J101243.54+102101.7  &  10:12:43.54  &  +10:21:01.70  &  T5.5  &  5  	 &  -390.6  \pm  11.94  &  -555  \pm  15.99  &4.74&19.9&35.6&  50.4 \\
ULAS~J101721.40+011817.9  &  10:17:21.40  &  +01:18:17.90  &  T8p   &  9  	 &  -83  \pm  12.46  &  -15.14  \pm  11.74  &5.05&11.7&43.3&  61.1 \\
ULAS~J101821.78+072547.1  &  10:18:21.78  &  +07:25:47.10  &  T5.0  &  7  	 &  -168.38  \pm  12.88  &  -14.81  \pm  10.33  &6.00&33.8&57.2&  80.7 \\
ULAS~J102144.87+054446.1  &  10:21:44.87  &  +05:44:46.10  &  T6    &    			 &  -29.92  \pm  36.78  &  30.79  \pm  32.49  &  0.90 $^{W}$  &26.5&45.3& 64.0\\
ULAS~J102305.44+044739.2  &  10:23:05.44  &  +04:47:39.20  &  T6.5  &    			 &  15.49  \pm  36.22  &  -83.08  \pm  37.91  &  0.91 $^{W}$  &35.0&49.9&  70.5 \\
ULAS~J102940.52+093514.6  &  10:29:40.52  &  +09:35:14.60  &  T8    &    			 &  -407.57  \pm  37.19  &  -145.97  \pm  26.38  &  0.91 $^{W}$  &6.6&24.3&  34.4 \\
ULAS~J103434.52-001553.0  &  10:34:34.52  &  -00:15:53.00  &  T6.5p &  5  	 &  -100.45  \pm  14.95  &  -30.41  \pm  14.54  &4.01&43.5&61.9&  87.5 \\
ULAS~J104355.37+104803.4  &  10:43:55.37  &  +10:48:03.40  &  T8    &    			 &  96.75  \pm  37.4  &  -73.04  \pm  37.95  &  0.91 $^{W}$  &10.2&37.7&  53.2 \\
ULAS~J105134.32-015449.8  &  10:51:34.32  &  -01:54:49.80  &  T6    &    			 &  -65.76  \pm  32.13  &  -40.03  \pm  32.4  &  0.99 $^{W}$  &27.7&47.2&  66.7 \\
ULAS~J105235.42+001632.7  &  10:52:35.42  &  +00:16:32.70  &  T5    &   5 			 &  -30.88  \pm  18.54  &  -140.23  \pm  18.4  &2.39&56.8&95.9&  135.5 \\
SDSS~J111010.01+011613.1  &  11:10:09.85 &  +01:16:10.50  &  T5.5  &  4  &  -235.42  \pm  34.43  &  -273.03  \pm  31.09  &  0.95$^{W}$  &14.2&25.5&  36.0 \\
ULAS~J114925.58-014343.2  &  11:49:25.58  &  -01:43:43.20  &  T5    &  5  	 &  -112.51  \pm  13.11  &  11  \pm  12.55  &4.05&40.7&68.9&  97.3 \\
ULAS~J115038.79+094942.9  &  11:50:38.79  &  +09:49:42.90  &  T6.5  &  6  &  -94.21  \pm  13.94  &  -19.38  \pm  14.08  &5.59 &40.0&57.0&  80.5 \\
ULAS~J115338.74-014724.1  &  11:53:38.74  &  -01:47:24.10  &  T6    &  5  	 &  -568.28  \pm  16.79  &  -333.86  \pm  13.12  &3.90 &25.7&43.9&  61.9 \\
ULAS~J115508.39+044502.3  &  11:55:08.39  &  +04:45:02.30  &  T7    &    			 &  483.25  \pm  14.18  &  -533.47  \pm  13.2  &2.96&20.7&48.5&  68.6 \\
ULAS~J115718.02-013923.9  &  11:57:18.02  &  -01:39:23.90  &  T5    &  5  	 &  108.24  \pm  13.62  &  -20.85  \pm  12.8  &3.90&42.1&71.1&  100.5 \\
ULAS~J115759.04+092200.7  &  11:57:59.04  &  +09:22:00.70  &  T2.5  &  6  &  -90.15  \pm  11.89  &  30.25  \pm  11.08  &4.79&25.4&34.4&  48.5 \\
\hline
\multicolumn{11}{l}{ 1)\citet{chiu06}; 2)\citet{knapp04}; 3)\citet{kirkpatrick2011};  4)\citet{burgasser06}; 5)\citet{ben10b};  6) \citet{pinfield08}; } \\
\multicolumn{11}{l}{7) \citet{lod07}; 8) \citet{albert2011}; 9) \citet{ben08} ; 10) \citet{ben2011b};  11) \citet{tsvetanov2000}; 12)  \citet{ben10a}; }\\
\multicolumn{11}{l}{ 13)  \citet{kendall07}; 14) \citet{murray2011}}\\
\end{tabular}
}
\caption{Proper motions for T~dwarfs within UKIDSS LAS sky. Unless otherwise stated proper motions are drawn from the catalogue of Smith et al (in prep). Epoch baselines denoted with $^{W}$ indicate that the proper motion has been calculated from our own WFCAM follow-up following the same method as used for the Smith et al catalogue. Spectral types are on the system of \citet{ben08}. Maximum and minimum plausible distances have been calculated using the mean magnitudes for each spectral subtype from \citet{dupuy2012} and assuming $\pm 0.5$ subtype precision on types. An additional maximum distance to account for possible unresolved binaries is given in the final column.
\label{tab:pms}
}
\end{table*}
%\end{landscape}

%\begin{landscape}
\begin{table*}
\addtocounter{table}{-1}
{\scriptsize
\begin{tabular}{ c c c c c  >{$}c<{$}  >{$}c<{$}  c  >{$}c<{$}  >{$}c<{$}  >{$}c<{$}  }
\hline
Name& $\alpha$ & $\delta$ & SpType & Ref &  \mu_{\alpha \cos \delta} & \mu_{\delta} & Baseline & D_{min} & D_{max} & D{max}(binary) \\
& J2000 & J2000 & & & mas/yr & mas/yr & years & pc & pc & pc \\
\hline

ULAS~J120257.05+090158.8  &  12:02:57.05  &  +09:01:58.80  &  T5.0  &  5  	 &  -42.73  \pm  12.09  &  -62.13  \pm  10.63  &4.79&21.4&36.1&  51.1 \\
ULAS~J120444.67-015034.9  &  12:04:44.67  &  -01:50:34.90  &  T4.5  &    			 &  -402.6  \pm  15.73  &  132.24  \pm  11.38  &3.64&25&36.6&  51.8 \\
ULAS~J120621.03+101802.9  &  12:06:21.03  &  +10:18:02.90  &  T5    &    			 &  -400.83  \pm  19.5  &  -87.42  \pm  18.75  &3.85&64.6&109.1&  154.2 \\
ULAS~J120744.65+133902.7  &  12:07:44.65  &  +13:39:02.70  &  T6.0  &  5  	 &  -154.53  \pm  12.49  &  0.84  \pm  12.29  &4.79&35.3&60.3&  85.1 \\
ULAS~J121226.80+101007.4  &  12:12:26.80  &  +10:10:07.40  &  T5    &    			 &  -181.77  \pm  16.23  &  -123.53  \pm  16.5  &3.86&53.2&89.9&  127.1 \\
ULAS~J122343.35-013100.7  &  12:23:43.35  &  -01:31:00.70  &  T6    &    			 &  -168.28  	 \pm  15.18  &  65.79  \pm  14.32  &3.64&42.8&73.0&  103.1 \\
ULAS~J123153.60+091205.4  & 12:31:53.60   &  +09:12:05.40  &  T4.5p &  5  	 &  82.62  \pm  15.08  &  -80.16  \pm  13.8  &4.96&71.4&104.7&  147.9 \\
ULAS~J123327.45+121952.2 &  12:33:27.45  &  +12:19:52.20  &  T4p   &  5  	 &  35.8  \pm  10.74  &  92.94  \pm  11.15  &4.89&40.7&61.7&  87.1 \\
ULAS~J123828.51+095351.3 &  12:38:28.51  &  +09:53:51.30  &  T8.5  &  9  	 &  -450.71  \pm  15.46  &  41.72  \pm  14.17  &4.96&12.8&36.1&  51.1 \\
ULAS~J123903.75+102518.6 &  12:39:03.75  &  +10:25:18.60  &  T0    &  5  	 &  -209.49  \pm  13.53  &  92.23  \pm  13.62  &4.96&65.2&86.3&  121.9 \\
ULAS~J124804.56+075904.0 &  12:48:04.56  &  +07:59:04.00  &  T7.0  &  5  	 &  -230.78  \pm  12.95  &  -145.35  \pm  11.95  &5.03&15.6&36.6&  51.7 \\
ULAS~J125708.07+110850.4 &  12:57:08.07  &  +11:08:50.40  &  T4.5  &  5  			 &  28.65  \pm  14.42  &  -57.13  \pm  13.94  &4.81&53.5&78.3&  110.7 \\
ULAS~J125835.97+030736.1 &  12:58:35.97  &  +03:07:36.10  &  T5    &    			 &  -170.86  \pm  14.19  &  -30.75  \pm  12.34  &2.83&46.1&78.0&  110.2 \\
ULAS~J130041.73+122114.7 &  13:00:41.73  &  +12:21:14.70  &  T8.5  &  10 	 &  -635.95  \pm  14.43  &  -27.9  \pm  12.29  &3.83&5.1&14.3&  20.2 \\
ULAS~J130217.21+130851.2 &  13:02:17.21  &  +13:08:51.20  &  T8.5  &  5  	 &  -427.96  \pm  13.41  &  -9.34  \pm  12.92  &4.88&9.7&27.3&  38.5 \\
ULAS~J130227.54+143428.0 &  13:02:27.54  &  +14:34:28.00  &  T4.5  &    			 &  -42.14  \pm  16.29  &  -14.66  \pm  15.2  &3.71&58.9&86.3&  121.9 \\
ULAS~J130303.54+001627.7 &  13:03:03.54  &  +00:16:27.70  &  T5.5  &  6  &  12.3  \pm  20.99  &  -274.16  \pm  19.97  &2.64&53.5&95.9&  135.5 \\
ULAS~J131508.42+082627.4 &  13:15:08.42  &  +08:26:27.40  &  T7.5  &  6  &  -36.52  \pm  12.73  &  -103.85  \pm  12.22  &5.67&25.2&49.0&  69.2 \\
ULAS~J131943.77+120900.2 &  13:19:43.77  &  +12:09:00.20  &  T5    &  5  	 &  -121.9  \pm  16.04  &  -22.9  \pm  14.59  &4.87&58.6&99.1&  140.0 \\
ULAS~J132048.12+102910.6 &  13:20:48.12  &  +10:29:10.60  &  T5    &  5  	 &  102.21  \pm  11.03  &  -56.56  \pm  11.44  &4.94&35.6&60.3&  85.1 \\
ULAS~J132605.18+120009.9 &  13:26:05.18  &  +12:00:09.90  &  T6p   &  5  	 &  70.63  \pm  11.04  &  -29.76  \pm  12.02  &4.87&24.7&42.1&  59.4 \\
ULAS~J133502.11+150653.5 &  13:35:02.11  &  +15:06:53.50  &  T6    &   			 &  3.02  \pm  14.16  &  -106.03  \pm  12.56  &3.85&30.6&52.2&  73.8 \\
ULAS~J133553.45+113005.2 &  13:35:53.45  &  +11:30:05.20  &  T9    &  9  	 &  -183.38  \pm  13.1  &  -214.17  \pm  10.92  &4.86&5.5&12.1&  17.1 \\
ULAS~J133828.69-014245.4 &  13:38:28.69  &  -01:42:45.40  &  T7.5  &    			 &  200.75  \pm  21.02  &  -109.89  \pm  20.2  &2.56&23.3&45.3& 64.0\\
ULAS~J133933.64-005621.1 &  13:39:33.64  &  -00:56:21.10  &  T7    &    			 &  72.34  \pm  17.15  &  -14.95  \pm  15.8  &2.53&19.9&46.6&  65.8 \\
ULAS~J133943.79+010436.4 &  13:39:43.79  &  +01:04:36.40  &  T5    &    			 &  -130.23  \pm  14.58  &  -23.76  \pm  14.72  &2.72&40.1&67.9&  95.9 \\
SDSS~J134646.43-003150.3 &  13:46:46.10  &  -00:31:51.40  &  T6.5  & 11 &  -512.71  \pm  12.41  &  -112.02  \pm  12.16  &3.02&9.9&14.1&  19.9 \\
ULAS~J134940.81+091833.3 &  13:49:40.81  &  +09:18:33.30  &  T7    &  5  	 &  -154.6  \pm  13.16  &  -73.34  \pm  12.13  &5.05&30.3&71.1&  100.5 \\
ULAS~J135607.41+085345.2 &  13:56:07.41  &  +08:53:45.20  &  T5    &  5  	 &  -67.84  \pm  12.12  &  -2.95  \pm  11.24  &5.05&39.4&66.7&  94.2 \\
ULAS~J141623.94+134836.3 &  14:16:23.94  &  +13:48:36.30  &  T7.5p &  12 	 &  86.22  \pm  12.43  &  128.58  \pm  12.53  &3.75&12.6&24.4&  34.5 \\
ULAS~J141756.2213:0:5.80 &  14:17:56.22 &  +13:30:45.80  &  T5    &    			 &  -123.05  \pm  12.48  &  45.93  \pm  10.32  &4.88&22&37.2&  52.5 \\
ULAS~J142145.63+013619.0 &  14:21:45.63  &  +01:36:19.00  &  T4.5  &    			 &  -213.24  \pm  17.99  &  57.23  \pm  17.19  &3.84&56.8&83.2&  117.5 \\
ULAS~J142320.79+011638.2 &  14:23:20.79  &  +01:16:38.20  &  T8p   &    			 &  281.48  	 \pm  19.56  &  -492.01  \pm  17.36  &3.84&13.1&48.1&  67.9 \\
ULAS~J142536.35+045132.3 &  14:25:36.35  &  +04:51:32.30  &  T6.5  &    			 &  137.3  	 \pm  15.42  &  -44.87  \pm  14.93  &3.84&40.4&57.5&  81.3 \\
ULAS~J144458.87+105531.1 &  14:44:58.87  &  +10:55:31.10  &  T5    &  5  	 &  -185.13  \pm  12.32  &  -137.62  \pm  11.88  &4.98&56.5&95.5&  134.9 \\
ULAS~J144555.24+125735.1 &  14:45:55.24  &  +12:57:35.10  &  T6.5  &  5  	 &  -369.47  \pm  15.11  &  108.78  \pm  14.93  &4.92&37.8&54.0&  76.2 \\
ULAS~J144901.91+114711.4 &  14:49:01.91  &  +11:47:11.40  &  T5.5  &    			 &  -248.94  \pm  10.8  &  -252.19  \pm  11.94  &4.96&24.8&44.6&  63.0 \\
ULAS~J145243.59+065542.9 &  14:52:43.59  &  +06:55:42.90  &  T4.5  &  13 &  57.71  \pm  13.11  &  -154.94  \pm  12.25  &6.79&59.7&87.4&  123.5 \\
ULAS~J145935.25+085751.2 &  14:59:35.25  &  +08:57:51.20  &  T4.5  &  5  	 &  -174.75  \pm  12.06  &  -78.04  \pm  10.63  &5.05&44.3&64.9&  91.6 \\
ULAS~J150135.33+082215.2 &  15:01:35.33  &  +08:22:15.20  &  T4.5  &  6  &  93.61  \pm  13.5  &  -190.91  \pm  13.57  &5.70&51.8&75.9&  107.2 \\
SDSS~J150411.63+102718.4   &  15:04:11.73  &  +10:27:16.90  &  T7    &  1  	 &  379.37  \pm  10.5  &  -382.55  \pm  10.84  &5.01&8.9&20.9&  29.6 \\
ULAS~J150457.66+053800.8 &  15:04:57.66  &  +05:38:00.80  &  T6p   & 14  &  -609.61  \pm  12.68  &  -514.86  \pm  11.82  &4.02&16.2&27.7&  39.1 \\
ULAS~J150547.89+070316.6 &  15:05:47.89  &  +07:03:16.60  &  T4.0  &  6  &  37.29  \pm  13.24  &  -115.07  \pm  12.79  &5.79&67.3&101.9&  143.9 \\
ULAS~J151637.89+011050.1 &  15:16:37.89  &  +01:10:50.10  &  T6.5  &    			 &  -110.69  \pm  16.45  &  -80.97  \pm  15.7  &2.96&35.3&50.4&  71.1 \\
WISE~J151721.13+052929.3 &  15:17:21.12  &  +05:29:29.03  &  T8p    &    			 &  -78.89  \pm  15.18  &  221.33  \pm  14.49  &3.18&11.8&43.5&  61.5 \\
ULAS~J152526.25+095814.3 &  15:25:26.25  &  +09:58:14.30  &  T6.5  &    5			 &  -83.02  \pm  12.05  &  128.17  \pm  12.14  &5.01&37.5&53.5&  75.5 \\
CFBDS~J152655.78+034534.8 &  15:26:55.78  &  +03:45:34.80  &  T4   &  8  &  -85.78  \pm  12.84  &  -7.49  \pm  11.88  &3.96&43.4&65.7&  92.8 \\
ULAS~J152912.23+092228.5 &  15:29:12.23  &  +09:22:28.50  &  T6    &  5  	 &  -125.96  \pm  11.49  &  41.27  \pm  11.43  &5.03&41.1&70.1&  99.1 \\
ULAS~J153406.06+055643.9 &  15:34:06.06  &  +05:56:43.90  &  T5    &    			 &  -14.9  \pm  12.06  &  -107.46  \pm  11.32  &6.68&61.9&104.7&  147.9 \\
ULAS~J153653.8001:5:0.60 &  15:36:53.80 &  +01:55:40.60  &  T5    &    			 &  -205.16  \pm  12.8  &  47.85  \pm  14  &2.97&37.6&63.5&  89.7 \\
ULAS~J154427.34+081926.6 &  15:44:27.34  &  +08:19:26.60  &  T3.5  &  6  &  -57.78  \pm  11.17  &  0.08  \pm  11.71  &5.79&56&77.3&  109.1 \\
ULAS~J154701.84+005320.3 &  15:47:01.84  &  +00:53:20.30  &  T5.5  &  6  &  -76.28  \pm  10.79  &  6.92  \pm  10.37  &5.68&38.7&69.5&  98.2 \\
ULAS~J154914.45+262145.6 &  15:49:14.45  &  +26:21:45.60  &  T5    &    			 &  -151.56  \pm  12.61  &  208.94  \pm  13.13  &3.00&39.6&67&  94.6 \\
ULAS~J160143.75+264623.4 &  16:01:43.75  &  +26:46:23.40  &  T6.5  &    			 &  -41.6  \pm  17  &  -42.18  \pm  17.11  &2.99&35.6&50.8&  71.8 \\
ULAS~J161436.96+244230.1 &  16:14:36.96  &  +24:42:30.10  &  T7    &    			 &  -122.92  \pm  11.46  &  32.01  \pm  11.64  &3.99&22.6&53.0&  74.8 \\
ULAS~J161710.39+235031.4 &  16:17:10.39  &  +23:50:31.40  &  T6    &    			 &  -146.63  \pm  10.05  &  48.19  \pm  9.97  &3.99&27.3&46.6&  65.8 \\
ULAS~J161934.78+235829.3 &  16:19:34.78  &  +23:58:29.30  &  T6    &    			 &  -84.43  \pm  12.4  &  11.54  \pm  13.15  &3.99&41.3&70.5&  99.5 \\
ULAS~J161938.12+300756.4 &  16:19:38.12  &  +30:07:56.40  &  T5    &   			 &  -5.54  \pm  14.78  &  -237.43  \pm  15.47  &3.03&51.3&86.7&  122.5 \\
ULAS~J162655.04+252446.8 &  16:26:55.04  &  +25:24:46.80  &  T5    &    			 &  -30.06  \pm  13.66  &  -44.36  \pm  14.39  &3.99&46.6&78.7&  111.2 \\
ULAS~J163931.52+323212.7 &  16:39:31.52  &  +32:32:12.70  &  T3    &    			 &  129.07  \pm  9.63  &  -119.83  \pm  10.96  &3.06&21.2&34&  48.0 \\
ULAS~J223728.91+064220.1 &  22:37:28.91  &  +06:42:20.10  &  T6.5p &    			 &  347.86  \pm  16.86  &  252.29  \pm  15.73  &  2.14$^{W}$  &41.9&59.7&  84.3 \\
ULAS~J230049.08+070338.0 &  23:00:49.08  &  +07:03:38.00  &  T4.5  &    			 &  128.88  \pm  41.56  &  -152.06  \pm  29.72  &  0.78$^{W}$  &38.4&56.2&  79.4 \\
ULAS~J234228.97+085620.1 &  23:42:28.97  &  +08:56:20.10  &  T6    &    			 &  242.18  \pm  13.57  &  -62.65  \pm  21.61  &  2.17$^{W}$  &14.8&25.2&  35.6 \\
ULAS~J235715.98+013240.3 &  23:57:15.98  &  +01:32:40.30  &  T5.5p &    			 &  47.32  \pm  21.43  &  1.3  \pm  25.73  &  1.28$^{W}$  &42.1&75.5&  106.7 \\

\hline
\multicolumn{11}{l}{ 1)\citet{chiu06}; 2)\citet{knapp04}; 3)\citet{kirkpatrick2011};  4)\citet{burgasser06}; 5)\citet{ben10b};  6) \citet{pinfield08}; } \\
\multicolumn{11}{l}{7) \citet{lod07}; 8) \citet{albert2011}; 9) \citet{ben08} ; 10) \citet{ben2011b};  11) \citet{tsvetanov2000}; 12)  \citet{ben10a}; }\\
\multicolumn{11}{l}{ 13)  \citet{kendall07}; 14) \citet{murray2011}}\\
\end{tabular}
}
\caption{Proper motions for T~dwarfs within UKIDSS LAS sky. Unless otherwise stated proper motions are drawn from the catalogue of Smith et al (in prep). Epoch baselines denoted with $^{W}$ indicate that the proper motion has been calculated from our own WFCAM follow-up following the same method as used for the Smith et al catalogue. Maximum and minimum plausible distances have been calculated using the mean magnitudes for each spectral subtype from \citet{dupuy2012} and assuming $\pm 0.5$ subtype precision on types. An additional maximum distance to account for possible unresolved binaries is given in the final column.
\label{tab:pms}
}

\end{table*}
%\end{landscape}

To identify apparent pairs with common proper motion we selected only
objects with total proper motions that are more than 3$\sigma$
significant, and greater than 100 mas/yr (92 objects). To be considered common
proper motion pairs we required 4$\sigma$ matches in both
$\mu_{\alpha \cos \delta}$ and $\mu_{\delta}$.  

To assess if possible pairs share a common distance (in the absence of trigonometric parallax), we estimated the maximum and minimum plausible absolute magnitudes of the candidate primary stars based on the hypothesis that they lie at the same distance as their candidate companions.  The candidate primary stars are compared to Hipparcos stars with $VJ$ photometry in Figure~\ref{fig:hipcmd}. Those targets whose maximum and minimum hypothesised $M_J$ bracket the main sequence (or white dwarf or giant branches) were accepted as candidate common proper motion binary pairs to our T~dwarfs. Of our 9 candidate primaries with 4$\sigma$ matched proper motions and $VJ$ colours, 5 appear very likely to have common distance to our T~dwarfs. One more has a minimum hypothetical $M_J$ value that lies on the periphery of the main sequence and is thus consistent with the target sharing a common distance to the T~dwarf, if the T~dwarf is itself an unresolved binary. Three pairs are ruled out by the common distance test.  The initial characterisation of these pairs is given given in Table~\ref{tab:pmmatches}.

\begin{table*}
\begin{tabular}{ c c c c c c c c }
\hline
T dwarf & Primary name & Separation (arcsec) & $\Delta \mu_{\alpha \cos \delta}$ & $\Delta \mu_{\delta}$ & $V-J$ & Distance match?$^{a}$ & Notes \\
\hline
  ULAS~J0950+0117 & LHS~6176 & 52 &  0.5$\sigma$ &  3.5$\sigma$ &  4.11 & Y & 0.0002\% \\
  ULAS~J1043+1048 & NOMAD 1006-0190624 & 1021 & 3.9$\sigma$ &  0.7$\sigma$ & 1.2 & N & \\
  ULAS~J1300+1221 & Ross 458 & 105 & 0.3$\sigma$ & 0.2$\sigma$ & 3.28 & Y & 1 \\
  ULAS~J1315+0826 & NOMAD 0983-0263649 & 382 & 1.6$\sigma$ & 2.0$\sigma$ &  1.3 & B & 2\%\\
  ULAS~J1335+1130 & LSPM J1334+1123 & 1625 &  1.0$\sigma$ &  2.5$\sigma$ &  3.11 & N & \\ 
  ULAS~J1339+0104 & HD118865 & 148 & 2.3$\sigma$ &  1.6$\sigma$ & 1.0 & Y & 0.3\%\\
  ULAS~J1423+0116 & BD+01 2920  & 153 & 2.9$\sigma$ &  0.8$\sigma$ &  1.21& Y &  2 \\
  ULAS~J1459+0857 & LSPM J1459+0851 & 386 &  0.6$\sigma$ & 1.6$\sigma$ &  -  & - & 3 \\ 
  ULAS~J1504+0538 & HIP 73786 & 64 &  0.1$\sigma$ &  0.8$\sigma$ & 2.64 & Y &  4,5 \\
\hline
\multicolumn{8}{l}{$^a$ Y = good match; N = bad match; B  = requires binarity.} \\
\multicolumn{8}{l}{ (1) \citet{goldman2010}, (2) \citet{pinfield2012}, (3) \citet{adj2011}, (4) \citet{scholz2010b}, (5) \citet{murray2011}} \\

\end{tabular}
\caption{Initial characterisation of candidate wide binary pairs selected by our crossmatches against the LSPM, NOMAD and Hipparcos catalogues. The ``notes" column includes previous discovery references and chance alignment probabilities for new candidates that passed the common distance test.
\label{tab:pmmatches}
}
\end{table*}

It is worth highlighting that this method restricts us to investigating stars with $V$ and $J$ band photometry in the NOMAD and LSPM catalogues, and we have likely thus excluded a number of genuine binary companions.  
For example, we also recovered the white dwarf - T~dwarf pair LSPM 1459+0851AB \citep{adj2011} as a proper motion match, however the lack of 2MASS photometry for the WD primary excluded it from our analysis at this stage. This was the only LSPM candidate with proper motion agreement that lacked $VJ$ photometry. However, a large number of NOMAD candidate primaries with proper motion agreement lacked appropriate photometry for the common distance distance check. It is thus likely that a number of additional binary companions may remain unidentified in our sample, particularly for more distant red primaries that lack Tycho photometry (e.g. M dwarfs).   

Of the 5 strong candidates, three are previously identified binary systems in our sample: Ross 458ABC \citep{goldman2010}, BD+01~2930AB \citep{pinfield2012} and Hip~73786AB \citep[LHS~3020AB][]{scholz2010b,murray2011}, and two are new candidates: ULAS~J0950+0117 (T8) + LHS~6176 (estimated M4); ULAS~J1339+0104 (T5) + HD118865 (F8). The latter of these was also identified in our crossmatch against  Hipparcos, and the parallax for the primary is consistent with our estimated distance to the T~dwarf secondary. The former has been independently identified as a candidate proper motion pair by \citet{luhman2012} since our detailed study of it had already commenced.

To assess the probability of chance alignment for our new candidate binary pairs, we followed the method described in \citet{dhital2010}, which calculates the frequency of unrelated pairings using a Galactic model that is parameterised by empirically measured stellar number density \citep{juric2008,bochanski2010} and space velocity \citep{bochanski2007} distributions. All stars in the model are single (and hence unrelated); therefore any stars within the 5D ellipsoid defined by the binary's position, angular separation, distance, and proper motions is a chance alignment. We performed 10$^6$ Monte Carlo realisations to calculate the probability of chance alignment.  The chance-alignment probabilities for the three new candidates are given in the ``notes" column of Table~\ref{tab:pmmatches}. The weaker candidate has a correspondingly higher probability of chance alignment, and in the absence of further data on the primary star, or improved distance estimates, it is not reasonable to pursue further analysis of this candidate system. The two strongest new candidates, however, are likely to be bona fide common proper motion systems and we proceed on this basis.

\begin{figure}
\includegraphics[width=220pt, angle=90]{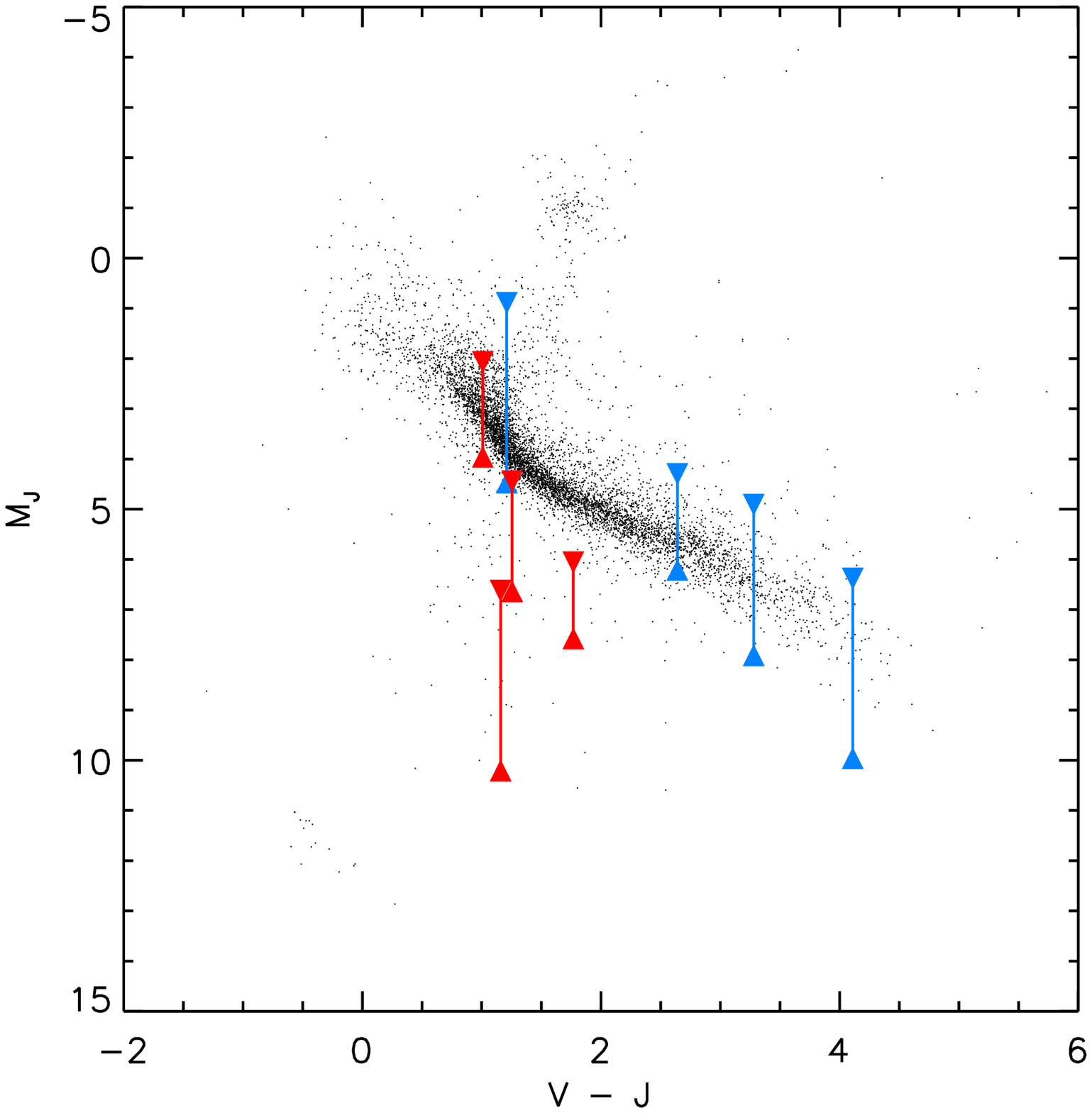}
\caption{A $V-J$ versus $M_J$ colour magnitude diagram showing the Hipparcos stars with $V$ and $J$ photometry (black dots) and our candidate primary stars. Blue symbols indicate the hypothetical range of $M_J$ for primaries selected from LSPM, whilst the red symbols indicate the same for candidate primaries selected from NOMAD. One candidate primary lies beyond the plotted range.}
\label{fig:hipcmd}
\end{figure}

\subsection{LHS 6176AB} 
\label{sec:6176AB}

\subsubsection{Distance to LHS 6176AB}
\label{sec:6176pi}

As part of our wider campaign for determining accurate distances to late-T~dwarfs in our sample \citep[e.g. ][]{smart2010,marocco2010}, we have obtained a trigonometric parallax measurement for ULAS~J0950+0117 (LHS 6176B), and also for the proposed primary LHS~6176. 
The astrometric observations and image reduction procedures were
identical to those described in \citet{smart2010}.
Since the observing strategy was optimised for measuring the distance to the T dwarf, LHS 6176A was often close to 
saturation on the image and, as a result,  the centroiding precision is reduced. For this reason,  the
original survey image from 2008 could not be used at all in the solution so the final astrometric 
parameter precision  for LHS~6176 is much lower than for the T~dwarf. 
In total 35 observations with 78 reference stars over a baseline of 4.22 years were used for LHS~6176B, compared to 33 observations with 525 reference stars over 2.16 years for LHS~6176A. 
The greater number of reference stars for LHS~6176A is due to their being drawn from the entire WFCAM chip, rather than from just the immediate vicinity of the target. 
All astrometric parameters 
indicate a common distance and common motion for the two objects, supporting our interpretation of the pair as a binary system.  
 The proper motions are also consistent with those found for LHS 6176
in \citet{lspm} (249, $-346$ mas/yr). 
The resulting astrometric parameters for the pair are given in Table~\ref{tab:6176Aprop}.
 When we re-run our chance alignment estimate calculation using the new trigonometric distance estimates, the probability of chance alignment is smaller than one part in $10^7$.
Assuming the more precise distance to LHS~6176B as the distance to the system, we find that the projected separation of the pair is thus 970~AU

\subsubsection{Spectroscopy of LHS 6176A}
\label{sec:6176spec}

Optical spectroscopy of LHS 6176A was obtained on the night of 5$^{th}$ May 2012 with the SuperNova Integral Field Spectrograph \citep[SNIFS; ][]{Lantz2004} on the University of Hawaii 2.2m telescope on Mauna Kea. SNIFS was operated with a dichroic mirror that separated the incoming light into blue (3200\AA\ to 5200\AA) and red (5100\AA\ to 9700\AA) spectrograph channels as well as an imaging channel that was used for guiding. The observations yielded a resolution of $R\approx 1000$ for the blue channel and $R\approx 1300$ for the red. Integration time was 210s, which was sufficient for high S/N ($\approx$80 per \AA) in the red channel, which is our region of interest for spectral typing an object as red as LHS~6176A. Basic reduction was performed automatically by the SNIFS processing pipeline, which included dark, bias, flat-field corrections, and cleaning of bad pixels and cosmic rays. The clean data were then assembled into blue and red data cubes. Wavelengths were calibrated with arc lamp exposures taken at the same telescope pointing as the science data (to correct for flexures).

The calibrated spectrum was then sky-subtracted, and a 1D spectrum was extracted from the image cube using a point-spread function model. Observations of the Feige~66, BD+75325, and G191B2B spectrophotometric standards \citep{Oke1990} taken over the course of the night were used to correct each spectrum for instrument response and remove telluric lines. Spectra were then shifted in wavelength to the rest frames of their source stars by cross correlating each spectrum with similar spectral type templates from the Sloan Digital Sky Survey \citep{Stoughton2002, bochanski2007}. More details on the SNIFS data processing pipeline can be found in \citet{Bacon2001} and \citet{Aldering2006}, and more information on additional data processing can be found in \citet{Lepine2012}.

Figure~\ref{fig:lhs6176Aspec} shows the extracted red channel. Overplotted is an M4 template spectrum based on mean spectra of inactive M dwarfs from \citet{bochanski2007}, their extremely close agreement across nearly the entire wavelength range covered by our data leads us to adopt a spectral type of M4V for LHS 6176. Their excellent agreement, and the lack of H$\alpha$ in emission also suggests a lack of activity for this object. The absence of H$\alpha$ in emission has been found to be typical of M4 dwarfs with ages of $>$3.5 Gyrs \citep{west08}.

\begin{figure*}
\includegraphics[width=200pt, angle=90]{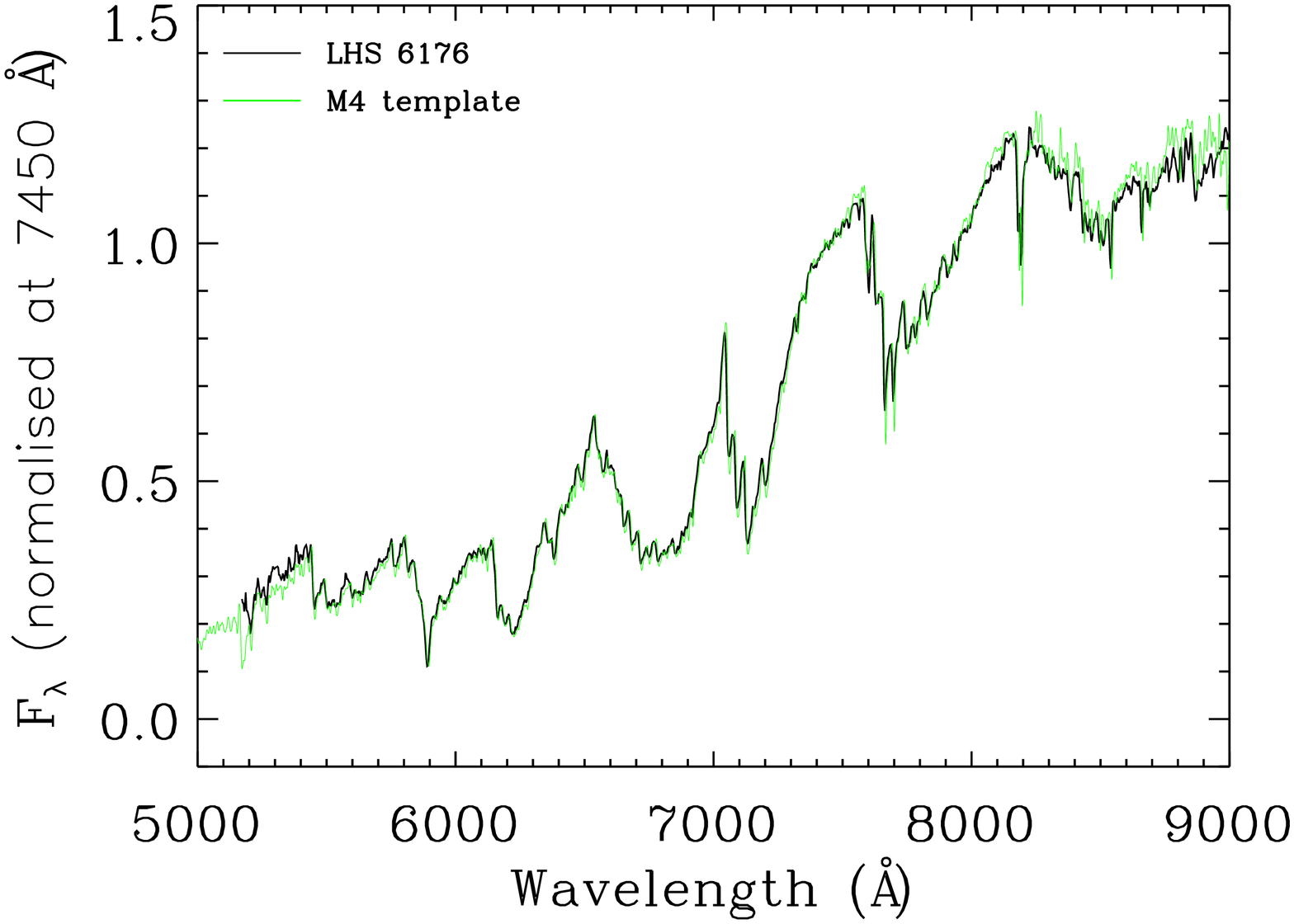}
\caption{Our SNIFS spectrum of LHS~6176A compared with the non-active M4 template spectrum from \citet{bochanski2007}.}
\label{fig:lhs6176Aspec}
\end{figure*}

We have also obtained near-infrared spectroscopy of LHS~6176A  on 2012~April~30~UT at the NASA Infrared Telescope Facilty on the summit of Mauna Kea, Hawaii. We used
the facility spectrograph SpeX \citep{vacca2003} in short-wavelength cross-dispersed mode with the 0.3\arcsec\ slit, which
provided an average spectral resolution ($R\equiv\Delta\lambda/\lambda$) of $\approx$2000. We obtained 6~exposures, each with a 120-second
integration time and dithered in an ABBA pattern, for a total of 12~minutes on-source. We observed the A0~V star HD~92245 contemporaneously for
telluric calibration. All spectra were reduced using version 3.4 of the SpeXtool software package \citep{vacca2003,cushing2004}. The resulting $JHK$ spectrum is shown in Figure~\ref{fig:lhs6176spex}.

\begin{figure*}
\includegraphics[width=200pt, angle=90]{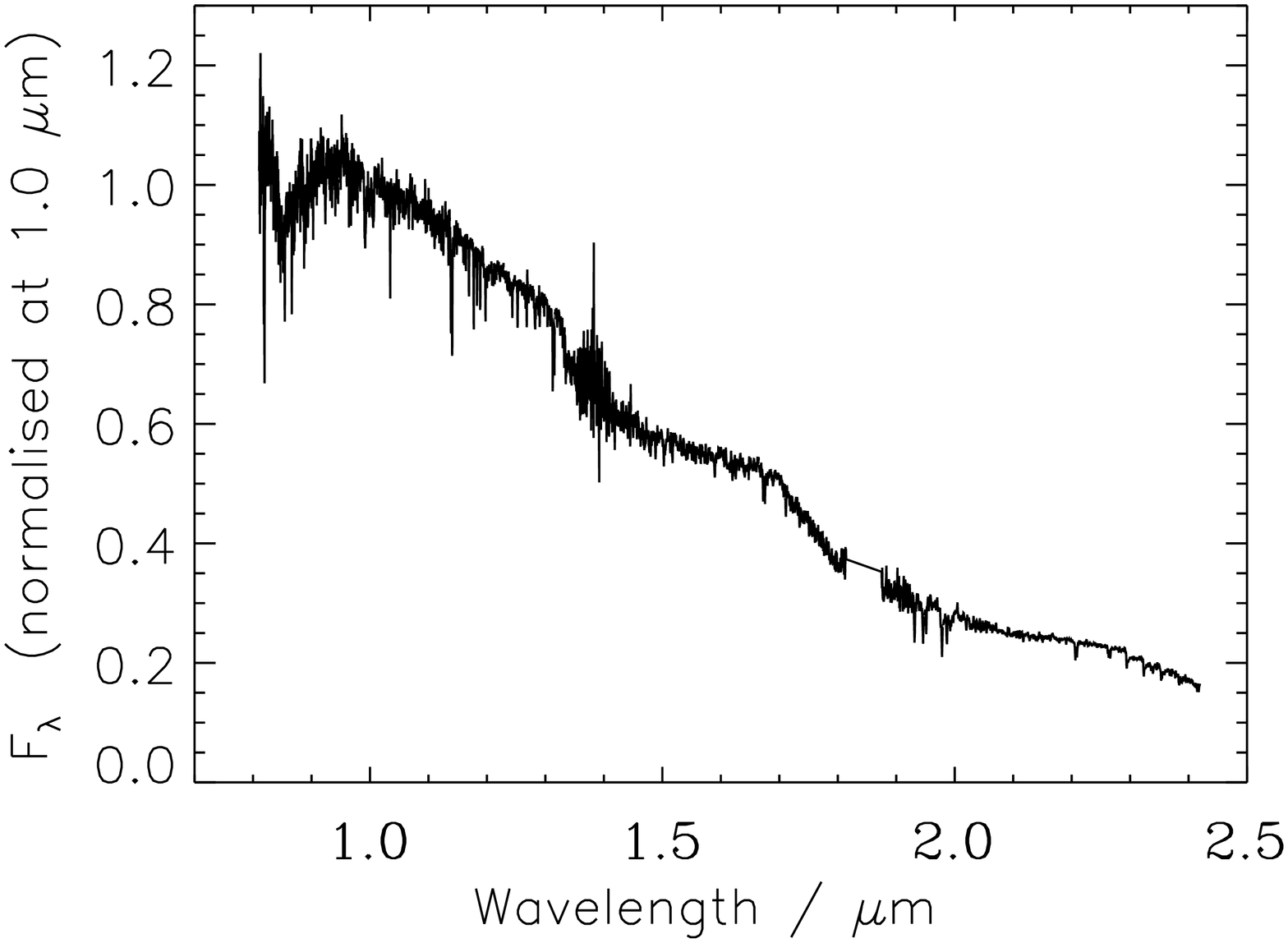}
\caption{Our $JHK$ SpeX spectrum of LHS~6176A. }
\label{fig:lhs6176spex}
\end{figure*}

We have taken three approaches to estimating the metallicity of LHS~6176A: \\

{\bf Method 1:} Our parallax for LHS~6176 also allows us to estimate its metallicity using the improved [Fe/H] vs. $M_{K_s}/V - K_s$ calibrations of \citet{schlaufman2010} and \citet{neves2012}. Since the uncertainty on the parallax of the T~dwarf component is considerably smaller than that of LHS~6176A, we have adopted the former's distance for the system. To maximise the precision of our metallicity estimate we have obtained new $V$ band photometry of LHS~6176A.
Johnson $V$ band data were obtained using the 50cm pt5m telescope on La Palma on the night of 17$^{th}$ December 2012. Five sixty second exposures were obtained. These exposures had the dark current and bias levels subtracted and were flat fielded using twilight sky frames. Objects were detected and instrumental magnitudes calculated using SExtractor \citep{sex}. The instrumental magnitudes were calibrated against $V$ band magnitudes from the AAVSO Photometric All Sky Survey (APASS\footnote{\url{http://www.aavso.org/apass}}). The resulting transformation was $V = 0.9774 \times I + 0.1010$, where V is the calibrated Johnson V band magnitude and I is the instrumental magnitude. The photon noise in our measurement is 0.004 magnitudes, but this is outweighed by approximately 0.03 magnitudes of calibration error, which we instead quote as our uncertainty (see Table~\ref{tab:6176Aprop}). 

We thus estimate a moderately-low metallicity of ${\rm [Fe/H]} =  -0.43 \pm 0.19$ for LHS~6176A from the \citet{schlaufman2010} calibration and ${\rm [Fe/H]} =  -0.36 \pm 0.17$~dex using the \citet{neves2012} calibration. The quoted error reflects the dispersion about the metallicity relations which dominates over our photometric uncertainties. 

{\bf Method 2:} We have applied the method of  \citet{mann2012} to our SNIFS optical and our $JHK$ SpeX near-infrared spectra.  From the optical regions we estimate ${\rm [Fe/H]} = -0.31 \pm 0.08$~dex, whilst the near-infrared regions yield an estimate of ${\rm [Fe/H]} = -0.25 \pm 0.05$~dex. These are consistent with one another, and the optically based estimate is consistent with the photometric estimate based on the \citet{neves2012} relations.

{\bf Method 3:}  Using the strengths of metal sensitive $K$ band features and the calibration described by \citet{babs2012} provides an estimate of ${\rm [Fe/H]} = -0.26 \pm 0.14$~dex. This is consistent with all the other estimates, and matches the other spectroscopic estimates particularly well. 

Although estimating M~dwarf metallicities is challenging, the good agreement of the different estimates that we have obtained for LHS~6176A highlights the excellent progress that has been made in this field in recent years. We adopt the mean of the estimates from \citet{neves2012,mann2012} and \citet{babs2012}, and assign a metallicity of ${\rm [Fe/H]} = -0.3 \pm 0.1$~dex for LHS~6176A. Table~\ref{tab:6176Aprop} summarises our determined properties for LHS~6176A.

\begin{table*}
\begin{tabular}{l c c}
\hline
 & LHS~6176A & LHS~6176B \\
\hline
R.A. (ep=2011.2579 eq=2000) &  09:50:49.8 & 09:50:47.3 \\
Dec (ep=2011.2579 eq=2000)  & +01:18:09.4 & +01:17:33.0 \\
PM$_{\alpha \cos{\delta}}$ (mas/yr) &   $242.42 \pm  19.0$ & $237.18 \pm 2.84$ \\
PM$_{\delta}$ (mas/yr) & $ -351.46 \pm  4.50$&  $-360.03 \pm 3.13$ \\
Spectral type & M4 & T8 \\
$V$ & $13.88 \pm 0.03$ & - \\
$B_J$ & 15.2 $^{a}$ & - \\
$J$(2MASS) & $9.80 \pm 0.02 ^{b}$ & -  \\
$J-H$ (2MASS) & $0.57 \pm 0.04 ^{b}$ & -  \\
$H-K_s$ (2MASS) & $0.28 \pm 0.04 ^{b}$ & -  \\
$V-K_s$ & $4.93 \pm  0.04$ & -  \\
$J$(UKIDSS) & - & $18.02 \pm 0.03$ \\
$Y-J$(UKIDSS) & - & $0.88 \pm 0.04$\\
$J-H$(UKIDSS) & - & $-0.38 \pm 0.04$ \\
$H-K$(UKIDSS) & - & $-0.45 \pm 0.08$\\
$W1$ & $8.77 \pm 0.02$ & $18.05 \pm 0.34$ \\
$W2$ & $8.60 \pm 0.02$ & $14.48 \pm 0.06$\\
$W3$ & $8.50 \pm 0.02$ & $>12.85$ \\
$W4$ & $> 7.98$ & $>9.20$\\
${\rm [3.6]}$ & - & $16.28 \pm 0.01$ \\
${\rm [4.5]}$ & - & $14.35 \pm 0.02$ \\
$\pi$ & $46.14 \pm  10.7$  & $53.40 \pm 3.51$ \\
Distance & $21.7^{+6.5}_{-4.1}$ pc  & $18.73^{1.32}_{-1.15}$\\
$[{\rm Fe/H}]$ & $-0.30 \pm 0.1$ & -  \\
H$_{\alpha}$EW & $-0.29 \pm 0.23$ \AA & - \\
Age & $>3.5$ Gyr $^{c}$ & - \\
$ \log (L_{*} / \Lsun)$ & - & $ -5.63 \pm 0.07$\\
Projected separation & \multicolumn{2}{c}{52 \arcsec, $\sim 970$~AU} \\
\hline
\multicolumn{3}{l}{$^a$ \citet{lspm}} \\
\multicolumn{3}{l}{$^b$ From 2MASS All-Sky Point Source Catalog}\\
\multicolumn{3}{l}{$^c$ Derived from activity life-time information presented}\\
\multicolumn{3}{l}{in \citet{west08}}\\
\hline
\end{tabular}
\caption{Properties of LHS~6176AB.}
\label{tab:6176Aprop}
\end{table*}

\subsubsection{The properties of LHS 6176B}
\label{sec:6176Bprop}

We have used our derived properties for LHS~6176A to constrain the metallicity and age of LHS~6176B, and determine more precise properties than would otherwise have been possible.  Since our warm-{\it Spitzer} photometry is more precise than the WISE survey photometry we have used it in preference in our calculations. We followed the same method as that described in \citet{ben2011b} and \citet{pinfield2012}. 

Briefly, we have calculated the bolometric flux using our $YJHK$ spectroscopy flux calibrated to our $J$ band photometry, and warm-{\it Spitzer} photometry. We have filled the gaps in our wavelength coverage by scaling the latest BT Settl model spectra \citep{allard2010} to match our $Y$ band spectrum for wavelengths blueward of our GNIRS spectrum, and to match our Ch1 and Ch2 photometry for wavelengths redward of it.  To avoid biasing our derived flux by assumptions regarding the $T_{\rm eff}$ of the target, we initially calculated the flux using a wide range of models covering $T_{\rm eff} = 500 - 1000K$, $\log g = 4.5 - 5.5$ and metallicity, ${\rm [M/H]} = 0.0$ and~$-0.3$. The resulting flux ruled out high and low $T_{\rm eff}$ extremes and lowest-gravity cases.  We thus  recalculated the flux using $T_{\rm eff} = 600 - 900$~K models with $\log g = 5.0 - 5.5$.

To account for random uncertainties in our flux calibration, we calculated each flux estimate (for each model spectrum) as the mean of a set of 100 different scalings, each offset by a random value drawn from the uncertainty in the photometry. 
Our final flux estimate is the mean of the estimates made using the range of models and different scalings, and our error is taken as the standard deviation on this value.  Thus our uncertainty implicitly includes both random and (identified) systematic elements. We thus calculate the flux from LHS~6176B as $F_{bol} = 2.13 \pm 0.15 \times 10^{-16} {\rm Wm^{-2}}$, and its luminosity as $8.94 \pm 1.30 \times 10^{20} {\rm W}$ or $ \log (L_{*} / \Lsun) = -5.63 \pm 0.07$.  This is approximately 60\% higher than the luminosity of the similarly metal poor benchmark T8 dwarf BD+01~2920B which was calculated using the same method \citep{pinfield2012}.

We can use the measured luminosity for LHS~6176B, in combination with the evolutionary models of \citet{baraffe03}, to estimate its radius and mass, assuming our estimated age for LHS~6176A.  The COND models assume solar metallicity, and the effect of low-metallicity on the luminosity-radius relation of brown dwarfs is not well constrained by observations. However, theoretical correlations between metallicity and radius derived by \citet{burrows2011} suggest radii are reduced by less than 5\% for a $-0.5$~dex shift from solar metallicity for objects at the stellar/substellar boundary, with considerably smaller shifts at lower masses. The \citet{sm08} evolutionary sequences suggest that shifts in radius of less than 1\% can be expected as a result of decreasing metallicity by 0.3~dex.

 The derived properties also depend strongly on the assumed multiplicity (or otherwise) of LHS~6176B. If we assume that LHS~6176B is a single object, with an age in excess of 3.5~Gyr (based on the lack of H$\alpha$ emission seen in LHS~6176A), we estimate its radius to be $0.078 \Rsun < R < 0.094 \Rsun$ mass to be $0.055 \Msun > M  > 0.030 \Msun$,  with a corresponding temperature of $850 > T_{\rm eff}  > 710$~K and gravity $5.3 > \log g > 5.0$ (respectively). 
If, on the other hand, we assume that LHS~6176B is an equal luminosity binary system, we find that the components would have radius $0.081 \Rsun < R < 0.096 \Rsun$ and mass $0.045 \Msun > M > 0.022 \Msun$, and with a corresponding temperature of $700 > T_{\rm eff}  > 590$~K and gravity $5.30 > \log g > 4.8$. 
Although LHS~6176B has a $J$ band magnitude ($M_J = 16.65$) at the faint end of the scatter about the T8 mean magnitude in \citet[][$M_J = 16.39 \pm 0.35$]{dupuy2012}, we cannot rule out binarity.  
It should be borne in mind that the mean T8 M$_{J}$ is calculated from a sample that specifically excludes peculiar objects, and so is dominated by objects with higher metallicity than LHS~6176B. 
A comparison with other low-metallicity objects would thus be more relevant. 
BD+01~2920B \citep{pinfield2012} provides just such a comparison. LHS~6176B is approximately 0.85 mag brighter than this object the $J$ band. 
It is thus quite possible that LHS~6176B is an unresolved binary system, and we refrain from adopting a single set of properties for this object.

\subsection{HD118865 AB}
\label{sec:hd118865AB}

The probability of chance alignment for HD~118865A and ULAS~J1339+0104 is $0.2\%$, which is significantly higher than has been found previously for wide binary systems in our searches \citep[e.g.][]{ben09,pinfield2012}. This is due to the combination of a large range of plausible distances for the T5 dwarf (40 -- 95~pc, allowing for the possibility that it is a binary) and the relatively low proper motions of the proposed components. However, if our range of plausible distances is reduced to account only for the spread in T5 absolute magnitudes (rather than T4.5 -- T5.5 and binarity; since the former arguably already incorporates these effects), then we find the probability of chance alignment is reduced to $0.001\%$. On balance, it is reasonable to proceed with analysis of this system as a bona fide common proper motion binary system, although we caution that a parallax for the T~dwarf is required to confirm beyond doubt that this pair is associated.

In many respects HD~118865A represents an ideal benchmark primary star. With a Hipparcos parallax \citep{hipparcos,newhip}, kinematic and model-based age estimates and well measured metallicity \citep{casagrande2011}, it avoids many of the pitfalls associated with less massive primary stars.  Indeed, the metallicity diagnostics for the M dwarf primaries that dominate the substellar wide binary sample have been benchmarked against nearby binary systems containing FGK primary stars for these reasons.  The properties of HD118865A are summarised in Table~\ref{tab:hd118865ab}.

To determine the properties of HD118865B we have followed an identical method to that applied for LHS~6176B (see Section~\ref{sec:6176Bprop}). In this case we have used models spanning the $T_{\rm eff} = 1000 - 1400$~K range, $\log g = 4.5 - 5.5$ and solar metallicity. After a first round of calculations we found that the lowest gravity models were inconsistent with our estimated luminosity and age for the system, and so were excluded from the next iteration. We find the flux from HD118865B to be $F_{bol} = 1.47 \pm 0.14 \times 10^{-16} {\rm Wm^{-2}}$ and the luminosity to be $L_{bol} = 6.85 \pm 1.00 \times 10^{21}$W or $\log (L_{*} / \Lsun) = -4.75 \pm 0.07$. 

If we assume that HD118865B is a single object  with an age of between 1.5~and 4.9~Gyr, then we find that its luminosity corresponds to a radius of $0.080 \Rsun < R <  0.091 \Rsun$ and a mass of $0.065 \Msun > M > 0.040 \Msun$, with corresponding temperature $1320 K > T_{\rm eff} > 1240 K$ and gravity $5.4 > \log g > 5.0$,    according to the COND evolutionary models. If HD~118865B is an equal mass binary, these properties should be revised to $0.079 \Rsun < R < 0.095 \Rsun$, $0.060 \Msun > M > 0.030 \Msun$, $1120 K > T_{\rm eff} > 1020 K$ and $5.4 > \log g > 5.0$.

\begin{table*}
\begin{tabular}{l c c}
\hline
 & HD~118865A & HD~118865B \\
\hline
R.A. (J2000) &  13 39 34.33$^{a}$ & 13:39:43.79 \\
Dec (J2000)  &  +01 05 18.12$^{a}$ & +01:04:36.40  \\
PM$_{\alpha \cos{\delta}}$ (mas/yr) &   $-95.75 \pm  0.80^{a}$ & $-130.23 \pm 14.58$ \\
PM$_{\delta}$ (mas/yr) & $ -48.81 \pm  0.54^{a}$ &  $-23.76 \pm 14.72$ \\
Spectral type & F5 & T5 \\
$B_T$ & $8.52 \pm 0.02^{b}$ & - \\
$V_T$ & $7.98 \pm 0.01^{b}$ & - \\
$J$(2MASS) & $6.98 \pm 0.02 ^{c}$ & -  \\
$J-H$ (2MASS) & $0.25 \pm 0.05^{c}$ & -  \\
$H-K_s$ (2MASS) & $0.06 \pm 0.05^{c}$ & -  \\
$J$(UKIDSS) & - & $18.08 \pm 0.04$ \\
$Y-J$(UKIDSS) & - & $1.07 \pm 0.06$\\
$J-H$(UKIDSS) & - & $-0.31 \pm 0.14$ \\
$H-K$(UKIDSS) & - & $0.0 \pm 0.13$\\
${\rm [3.6]}$ & - & $16.93 \pm 0.01$ \\
${\rm [4.5]}$ & - & $16.08 \pm 0.02$ \\ 
$\pi$ & $16.02 \pm  0.86^{a}$  & - \\
Distance & $62.4^{+3.2}_{-3.6}$ pc  &  - \\
$[{\rm Fe/H}]$ & $0.09 \pm 0.09^{d}$ & -  \\
Age & \multicolumn{2}{c}{$1.5 - 4.9$ Gyr (1$\sigma$ interval)$^{d}$}\\
$ \log (L_{*} / \Lsun)$ & - & $-5.24 \pm 0.04$ \\
Projected separation & \multicolumn{2}{c}{148 \arcsec, $\sim 9200$~AU} \\
\hline
\multicolumn{3}{l}{$^a$ \citet{newhip}} \\
\multicolumn{3}{l}{$^b$ \citet{tycho}}\\
\multicolumn{3}{l}{$^c$ from 2MASS All-Sky Point Source Catalog}\\
\multicolumn{3}{l}{$^d$ \citet{casagrande2011}}\\
\hline
\end{tabular}
\caption{Properties of HD118865AB.}
\label{tab:hd118865ab}
\end{table*}

 \subsection{HIP 73786B} 
 \label{sec:hip73786}
 
 \citet{scholz2010b} and \citet{murray2011} independently identified HIP~73786AB as a wide binary system consisting of a metal poor K5 dwarf moving in common motion with a T6p dwarf, and we recovered it with our search. At the time of its discovery longer wavelength photometry for the T~dwarf  was not available, and so no bolometric flux estimate was made. The advent of the WISE all-sky release, however, allows such an estimate to made relatively conveniently and for the purposes of including this object in subsequent discussion we have estimated the properties for HIP~73786B following the same method as described in the previous sections. 
 
In this case we have used BT Settl models spanning the $T_{\rm eff} = 800 - 1200$~K range, $\log g = 4.5 - 5.5$ and solar metallicity, scaling the longer wavelength portions to match the W1 and W2 survey photometry, and stitching them to our flux-calibrated near-infrared spectrum at 2.4$\mu$m. As before, the regions blueward of 1$\mu$m have been filled using models scaled to match the $Y$ band flux in our near-infrared spectrum.  The current lack of low-metallicity BT Settl models in this temperature range has prevented us from using models covering the entire expected parameter range for this object, but since we scale the models to match the long wavelength photometry, we do not expect this to have a significant impact on our flux estimate.   We find the flux from HIP~73786B to be $F_{bol} = 5.31 \pm 0.30 \times 10^{-16} {\rm Wm^{-2}}$ and the luminosity to be $L_{bol} = 2.20 \pm 0.12 \times 10^{21}$W or $\log (L_{*} / \Lsun) = -5.24 \pm 0.04$.

\citet{murray2011} loosely constrained the age of the HIP~73786AB system to be 1.6--10Gyrs. If we accept this age for HIP~73786B, then from its luminosity and the COND evolutionary models we infer a radius of $0.076 \Rsun < R <  0.096 \Rsun$ and a mass of $0.063 \Msun > M > 0.028 \Msun$, with corresponding temperature $1020 K > T_{\rm eff} > 910 K$ and gravity $5.5 > \log g > 4.9$ under the assumption that it is a single object. If HIP~73786B is an equal luminosity binary then we derive a radius $0.076 \Rsun < R < 0.098 \Rsun$, $0.058 \Msun > M > 0.022 \Msun$, $860 K > T_{\rm eff} > 760 K$ and $5.4 > \log g > 4.8$.

\subsection{Properties of the UKIDSS wide binary sample}
\label{sec:propbins}

The relatively uniform manner in which our wide binary systems have been identified allows us to draw some preliminary conclusions about the properties of the late-T dwarf wide binary companion  population. Firstly, it is apparent that with 7 wide binary companions out of a total of 92 T~dwarfs within our proper motion $>100$~mas/yr selection, we can place a minimum value of $~8\%$ on the wide binary companion fraction. This is consistent with the minimum value of 5\% found by Gomes et al (submitted) for L dwarfs. 
This represents a lower limit, as our selection of candidate primary stars is limited by available photometry for faint red primary star candidates, as highlighted in Section~\ref{sec:compsel}.  This source of incompleteness also explains another feature of our binary sample. That is, that the latest type T~dwarfs are significantly over-represented, with 25\% of T8 and later objects (3 out of 12 in our selection) appearing as wide binary companions, compared to 5\% for T4.5 to T6.5 dwarfs. Whether this suggests that the true wide binary companion fraction for T~dwarfs is nearer the 25\% value seen for the latest type objects is an open issue that will require significantly improved selection of earlier type binary systems. 

The scarcity of earlier type wide-binary companions can be understood not in terms of bias within our proper motion sample, but rather as a reflection of the relative space densities of late and mid-type T~dwarfs, and the bias against fainter red candidate primary stars with Tycho photometry. The space density of the T8 and later dwarfs is significantly (e.g. $4 \times$) higher than that of mid-type T~dwarfs. This means that a larger proportion of the latest type objects will be found at close distances, where the LSPM and NOMAD catalogues are relatively complete for primary stars and proper motions are larger. The earlier type objects, however, will be less numerous in the nearby volume, and so larger fraction will be found at larger distances, where the catalogues of potential primary stars are most incomplete for the most common M dwarf type of primaries, and proper motions are smaller making reliable identification more problematic.   It thus appears, at this stage, that we can (weakly) conclude that the spectral type distribution of T~dwarfs as wide binary companions does not appear to be drastically different from that of isolated (i.e. single) T~dwarfs.

\subsection{The colours of benchmark T~dwarfs and the latest model atmospheres}
\label{sec:benchcols}

Figure~\ref{fig:hs2plots} shows $H-K$ vs $H-{\rm [4.5]}$ colour-colour plots for the compendium of MKO and Spitzer photometry of late-T~dwarf benchmark systems including that presented by \citet{sandy10} and updated with additional photometry presented here.  Figure~\ref{fig:jw2plots} shows similar $Y-J$ vs $J - W2$ colour-colour diagrams for benchmark T~dwarfs with WISE photometry. The properties of the benchmarks are summarised in Table~\ref{tab:bmprops}. 

\begin{table}
\begin{tabular}{l c c c c}
\hline
Name & $T_{\rm eff}$ / K & $\log g$  & [Fe / H] \\
\hline
Wolf~940B & $605 \pm 20^a$ & $5.0 \pm 0.2^a$ & $+0.02 \pm 0.05^b$  \\
BD+01 2920B & $680 \pm 55^c$ & $5.0 \pm 0.3^c$ & $-0.36 \pm 0.06^c$  \\
Ross~458C & $695 \pm 60^d$ & $4.35 \pm 0.35^d$ & $+0.09 \pm 0.05^b$ \\
LHS~6176B & $780 \pm 70^e$ & $5.15 \pm 0.15^e$ & $-0.30 \pm 0.10^e$ \\ 
Gl~570D  & $800 \pm 20^f$ & $5.1 \pm 0.15^f$ & $+0.09 \pm 0.04^f$ \\
HD3651B & $810 \pm 30^g$ & $5.3 \pm 0.2^g$ & $+0.12 \pm 0.04^g$  \\
HIP~73786B & $965 \pm 55^e$ & $5.2 \pm 0.3^e$ & $-0.30 \pm 0.1^h$\\
LHS~2803B & $1120 \pm 80^i$ & $5.4 \pm 0.1^i$ & $\sim 0.0^i$\\
HD~118865B & $1280 \pm 40^e$ & $5.2 \pm 0.2^e$ & $0.09 \pm 0.10^e$ \\
\hline
\multicolumn{4}{l}{$^a$ \citet{sandy2010a}} \\
\multicolumn{4}{l}{$^b$ from the $V-K$ calibration of \citet{neves2012}. } \\
\multicolumn{4}{l}{$^c$ \citet{pinfield2012}}\\
\multicolumn{4}{l}{$^d$ \citet{ben2011b}}\\
\multicolumn{4}{l}{$^e$ this work}\\
\multicolumn{4}{l}{$^f$ \citet{saumon06}}\\
\multicolumn{4}{l}{$^g$ \citet{liu07}}\\ 
\multicolumn{4}{l}{$^h$ \citet{murray2011}}\\
\multicolumn{4}{l}{$^i$ \citet{deacon2012b}}\\
\hline
\end{tabular}
\caption{Summary of the properties of the benchmark systems shown in Figures~\ref{fig:hs2plots} and~\ref{fig:jw2plots}.}
\label{tab:bmprops}
\end{table}
 
\begin{figure*}
\includegraphics[width=400pt, angle=00]{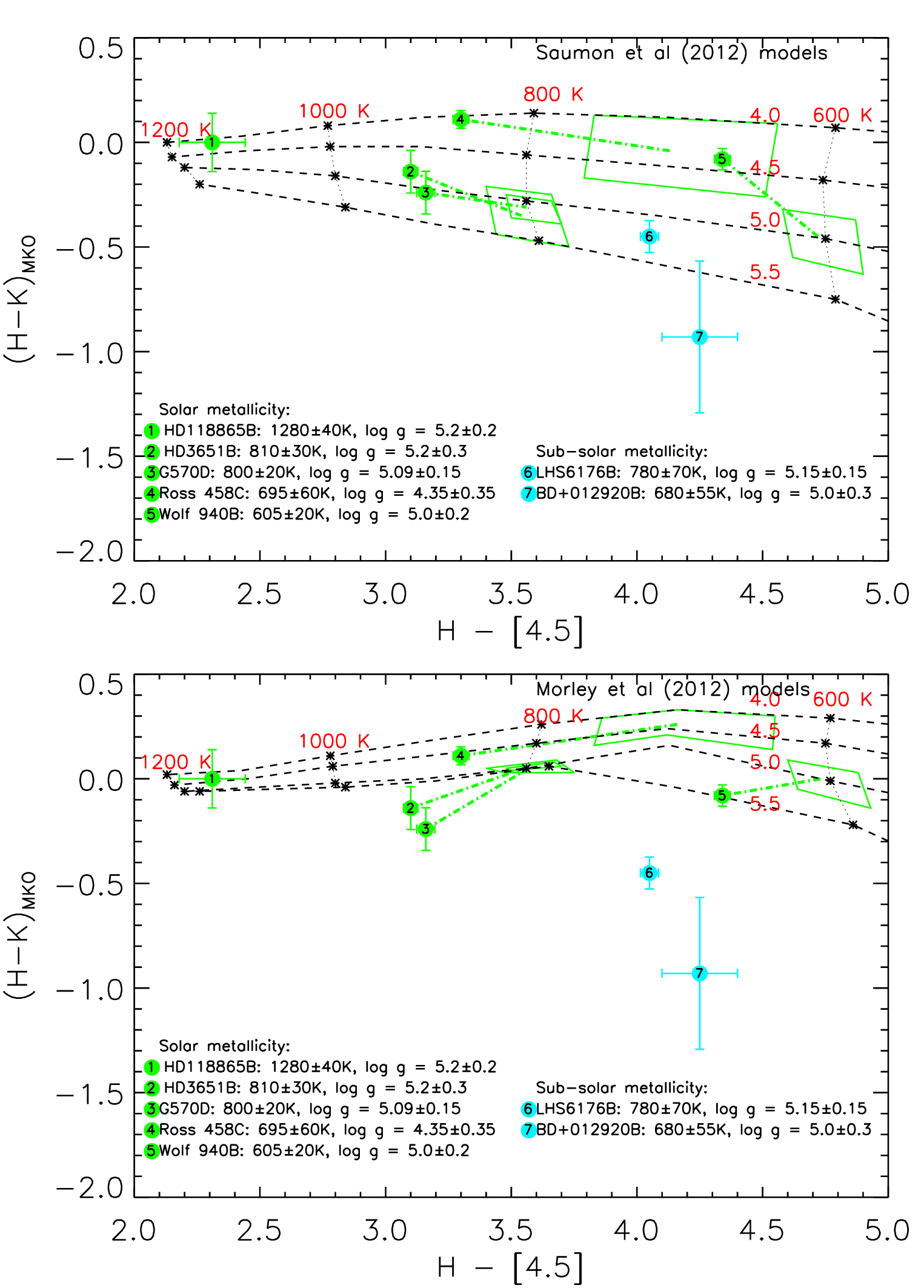}
\caption{$H$ - [4.5] vs $H-K$ colour-colour plots for the compilation of photometry for T~dwarfs from \citet{sandy10}  and benchmark systems from the literature, along with new photometry and benchmarks presented in this paper. Benchmark systems with roughly Solar metallicity are shown in green, whilst those with sub-Solar metallicity are shown in blue. Model colour tracks for \citet{saumon2012}, top panel, and \citet{morley2012}, bottom panel, are shown for comparison. The latter assumed $f_{sed} = 5$. The models shown are all solar metallicity, and the $T_{\rm eff}$ and $\log g$ values are indicated on the colour sequences. Each benchmark object is linked to a 1$\sigma$ box indicating the model prediction for its colours.}
\label{fig:hs2plots}
\end{figure*}

\begin{figure*}
\includegraphics[width=400pt, angle=0]{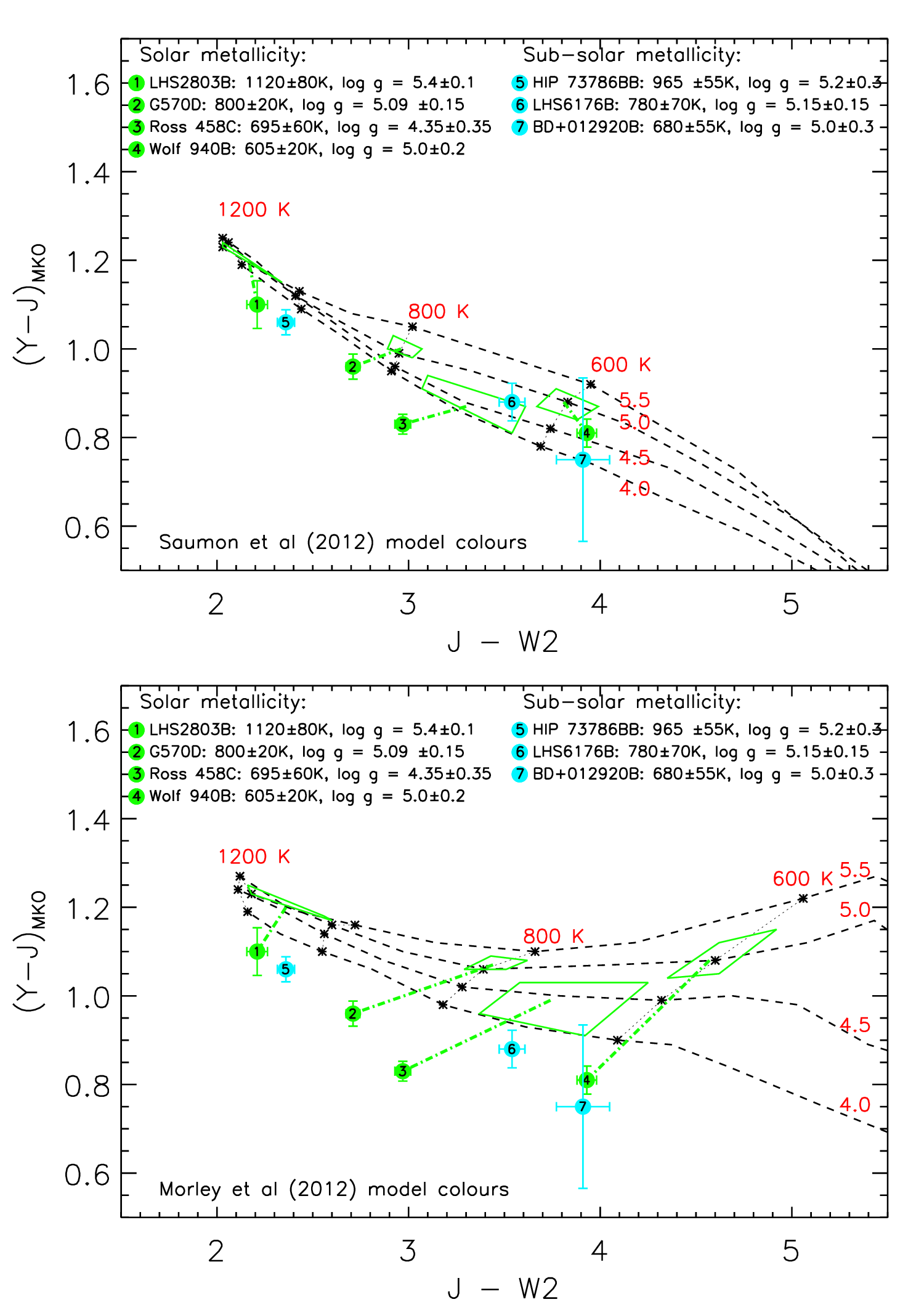}
\caption{$J - W2$ vs $Y-J$ colour-colour plots for the UKIDSS T~dwarfs with $YJ$ and WISE photometry and benchmark systems from the literature, along with new photometry and benchmarks presented in this paper. Benchmark systems with roughly Solar metallicity are shown in green, whilst those with sub-Solar metallicity are shown in blue. Model colour tracks for \citet{saumon2012}, top panel, and \citet{morley2012}, bottom panel, are shown for comparison. The latter assumed $f_{sed} = 5$. The models shown are all solar metallicity, and the $T_{\rm eff}$ and $\log g$ values are indicated on the colour sequences. Each benchmark object is linked to a 1$\sigma$ box indicating the model prediction for its colours.}
\label{fig:jw2plots}
\end{figure*}

Overlaid on the two panels of Figures~\ref{fig:hs2plots} and~\ref{fig:jw2plots} are colour tracks for the models of \citet{saumon2012} and \citet{morley2012}, and a 1$\sigma$ box for the model-predicted colours of each benchmark is connected to each objects observed colours (where a model prediction is available on these grids). These models incorporate the latest NH$_{3}$ opacities from \citet{yurchenko2011}, and a new treatment of collisionally induced H$_{2}$ absorption \citep[CIA H$_{2}$; ][]{abel2011,abel2012}. The \citet{morley2012} models also include the effects of proposed sulphide and alkali condensates, which may become important in atmospheres below $T_{\rm eff} \approx 800$~K.  Low-metallicity versions of these models are not yet available, so it is not possible to assess if these recent developments have impacted the predictions for low-metallicity atmospheres.

It is difficult to draw firm conclusions about the success or otherwise of the different models by comparing the colour predictions to the observed colours of the benchmark systems as deficiencies in the models can be both masked and amplified by their differing impacts at the wavelengths of the two photometric bands being compared in any particular case. None-the-less some of the principal differences between the models and the observations can be attributed to known deficiencies in the model grids.   For example, the predicted $H-{\rm [4.5]}$ and $J-W2$ colours for the solar-metallicity benchmarks are almost universally too red in both Figures~\ref{fig:hs2plots} and~\ref{fig:jw2plots}. This is due to the fact that neither model set includes non-equilibrium chemistry, which has been demonstrated to be important for determining the emergent flux in the $4-5 \micron$ region \citep[e.g. ][]{saumon06}.

The most obvious conclusion that can be drawn from examining the colours of the benchmark systems is that metallicity and gravity both have a significant impact on the near to mid-infrared colours of late-T~dwarfs, along with the (generally considered) dominant influence of temperature. For example, the low-gravity benchmark Ross~458C appears only 0.15--0.2 mag redder in $H-{\rm [4.5]}$ than Gl~570D and HD3651B,  despite being approximately 100K cooler. Comparing the colours of Wolf~940B with Gl~570D and HD3651B suggests that $\Delta T_{\rm eff} = -100$~K should (approximately) correspond to $\Delta(H-{\rm [4.5]}) = +0.6 $~mag. This implies that the low-gravity of Ross~458C imposes a blueward shift of $\sim 0.4$~mag for $\Delta \log g \approx -0.7$~dex, or roughly $+0.1$~mag per $+0.18$~dex in $\log g$.

The impact of metallicity appears to be even more significant, with  lower metallicity objects appearing redder (in e.g. $H - {\rm [4.5]}$ or $J - W2$) at a given $T_{\rm eff}$ than objects with near-Solar metallicity. 
For example, comparing the colours of LHS~6176B and BD+01~2920B with the warmer solar metallicity objects suggest that a shift of $\Delta {\rm [Fe/H]} \approx -0.3$~dex can give rise to a shift of $\Delta (H-{\rm [4.5]}) \approx +0.6$, if we assume that LHS~6176B is an unresolved binary with a $T_{\rm eff} = 645 \pm 55$~K and $\log g = 5.05 \pm 0.25$.
If LHS~6176B is a single object then the apparent shift in colour due to its metallicity would more like 0.9~mag. This is somewhat higher than for BD+01~2920B, and further argues for a binary interpretation for this object.

This effect has been suggested previously, based on comparisons of photometric colours and atmospheric models \citep{sandy09,sandy10}, and can largely be attributed to increased flux in the 4.5$\micron$ region due to reduced CO opacity. 
Comparison of Solar and low-metallicity cases for the BT~ Settl \citep{btsettlCS16} models that were first shown in \citet{pinfield2012} and the \citet{sm08} models indicates that although both predict a colour shift due to lower metallicity, both under-predict its magnitude.
The \citet{sm08} models predict $\Delta (H- {\rm [4.5]}) \approx +0.2$ for $\Delta {\rm [M/H]} = -0.3$, whilst the BT~Settl grid predicts  $\Delta (H- {\rm [4.5]}) \approx +0.3$ for the same change at $T_{\rm eff} = 700$~K, compared with the $\Delta (H-{\rm [4.5]}) \approx +0.6$ shift seen in our benchmarks.

The slightly stronger metallicity dependence that the BT~Settl model colours exhibit can likely be attributed to the combination of two factors. Firstly, the BT~Settl models include non-equilibrium chemistry for CO$_2$, which will result in a greater relative increase in flux in the ${\rm [4.5]}$ band in response to reducing the metallicity. Secondly, the BT~Settl models include additional methane opacity in the $H$~band, where the methane line lists are very incomplete,  based on a statistical estimate of the contributions from the hot vibrational bands.  With decreased metallicity these hot bands are likely to become more important as the chemistry shifts in favour of CH$_4$ over CO and CO$_2$ \citep[e.g. ][]{lodders2002}. 
Indeed, it is likely that the failure of the models to accurately reproduce the strong metallicity dependence of the $H - {\rm [4.5]}$ colour can be partially attributed to the incomplete nature of the methane line lists.

%In Figure~\ref{fig:metplot} we have compared the solar and low-metallicity cases for just the $\log g = 4.5$ case of the BT~ Settl \citep{btsettlCS16} models that were first shown in \citet{pinfield2012} and the \citet{sm08} models that were used for the colour-colour plots in \citet{sandy10}. 

%\begin{figure*}
%\includegraphics[width=400pt, angle=0]{metplot.ps}
%\caption{$H-K$ vs $H-{\rm [4.5]}$ colour-colour plot showing the same T dwarf set as Figure~\ref{fig:hs2plots}, but also comparing solar-metallicity (black lines) and ${\rm [M/H]} = -0.3$ (cyan lines) model colours for $\log g = 4.5$ for BT-Settl models \citep{btsettlCS16} and the models of \citet{sm08}. The latter are shown for $K_{zz} = 10^{4}$, as in the similar plot presented by \citet{sandy10}.}
%\label{fig:metplot}
%\end{figure*}

The strong dependence on metallicity of the $H-{\rm [4.5]}$ and $J-W2$ colours should be considered carefully when using these colours to estimate the properties of cool brown dwarfs. 
 Two specific examples worth highlighting in this context are: SDSS~1416+1338B, the second reddest known T dwarf in $H-{\rm [4.5]}$ despite a spectral type of T7.5 \citep{ben10a}; and WISEPC~J1828+2650, which is the reddest known Y~dwarf in the same colours \citep{cushing2011,leggett2013}. In the case of the former, its anomalously red colour was initially interpreted as being indicative of $T_{\rm eff} \approx 500$~K \citep{ben10a}. However, its parallax has since been measured, and its luminosity appears to rule out such a low temperature, and its extreme colours are now attributed to some combination of low-metallicity and/or high gravity with a significantly higher $T_{\rm eff}$ (Murray et al in prep). In the case of WISEPC~J1828+2650, \citet{leggett2013} argues that its extremely red $H-{\rm [4.5]}$ can only be consistent with its luminosity (which is higher than for some earlier and bluer Y dwarfs) if it is either younger than 50~Myr (with a mass $< 1 M_{Jup}$) or an unresolved binary system. Given the example of SDSS~J1416+1348B and the colours of benchmark systems in Figure~\ref{fig:hs2plots} (albeit at higher $T_{\rm eff}$), it seems reasonable to also offer a third (somewhat speculative)  interpretation e.g.  this object may in fact be somewhat warmer and more massive, but with its colours reddened by low-metallicity, high-gravity and/or their combined impact on cloud properties.

 It is important to emphasise that what follows is based on a highly simplistic extrapolation of the trends we have identified at cool T~dwarf temperatures ($T_{\rm eff} = 600-800$K) to Y~dwarf temperatures some $200-500$K cooler.
We note that at these low-temperatures the gravity range is limited by the long cooling times for the most massive objects \citep[e.g. ][]{sm08}, so the bulk of any colour shift at a given $T_{\rm eff}$ must be driven by metallicity. If we apply the shift that we have identified at higher-$T_{\rm eff}$ of $\Delta H-{\rm [4.5]} =+0.6$~mag for roughly $\Delta {\rm [M/H]}  = -0.3$~dex, and assume that the shift is entirely driven by extra flux in [4.5], we find that a halo-like low-metallicity of ${\rm [M/H]} \approx -1$ could provide the additional $1.5 - 2.0$~magnitudes of flux seen in the 4.5$\micron$ region from WISEPC~J1828+2650 \citep{leggett2013,beichman2013}. Although the tangential velocity of WISEPC~J1828+2650 has been measured as $51 \pm 5$~km~s$^{-1}$ \citep{beichman2013}, which is most consistent with thin-disk membership, this does not rule out a metallicity as low as ${\rm [M/H]} \approx -1$. Its near-infrared colours may also lend weight to the low-metallicity interpretation. \citet{leggett2013} report $H-K = -0.63 \pm 0.43 $ for this target, making it the bluest known $Y$ dwarf in these colours, which at higher temperatures would be associated with low-metallicity and/or high-gravity. Its red $J-H = +0.63 \pm 0.33$ colour is also not at odds with low-metallicity. For example, in the T~dwarf regime, metallicity has little impact on the $J-H$ colours of T~dwarfs (see Figure~\ref{fig:jhsplot}). Interestingly, the BT~Settl models predict redder $J-H$ colours for lower metallicity T~dwarfs, with a more pronounced effect at lower~$T_{\rm eff}$.  It will be interesting to see how the picture of the impact of varying metallicity develops as new models grids are calculated and new benchmark systems are discovered over the coming years.

\begin{figure*}
\includegraphics[width=430pt, angle=00]{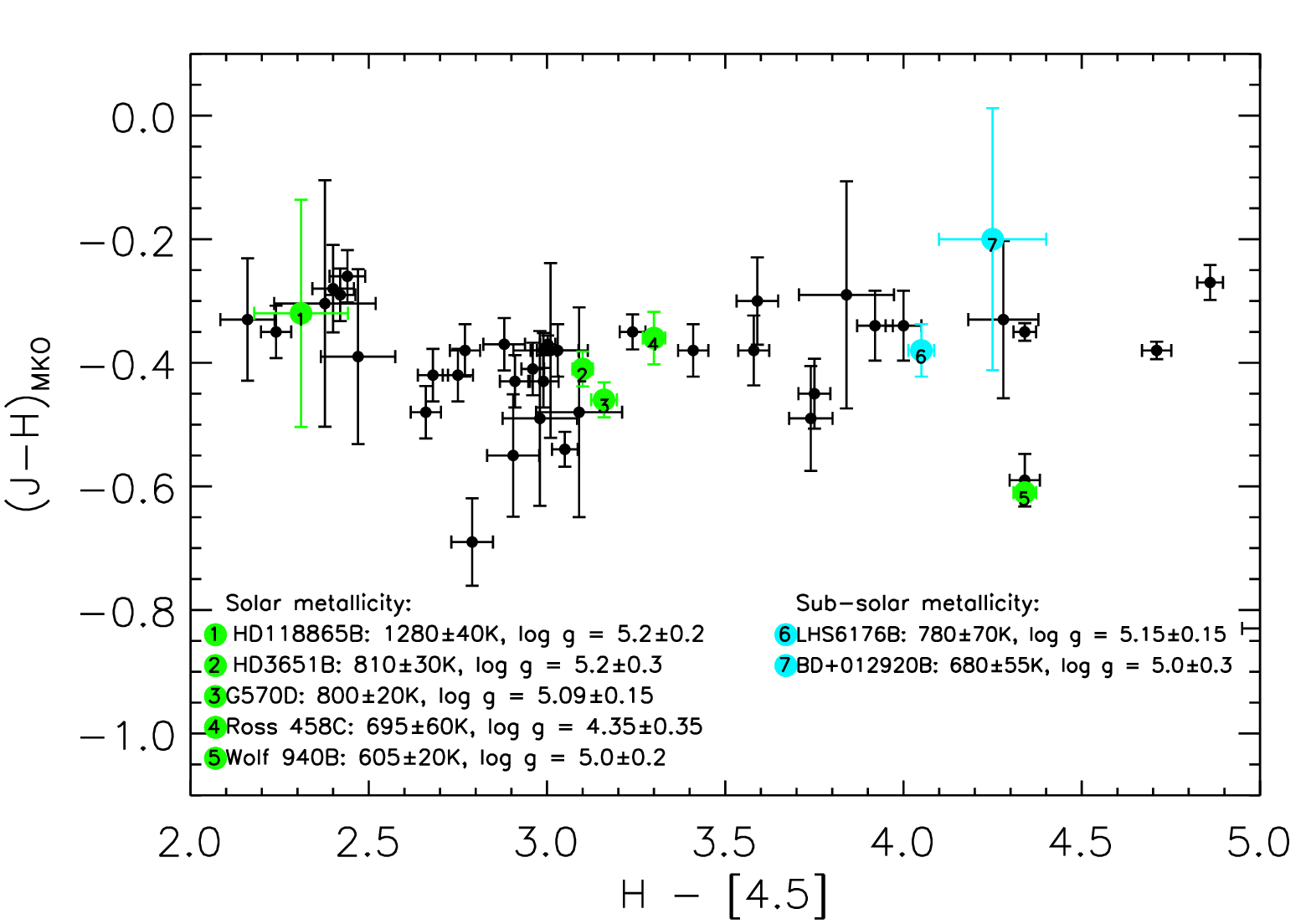}
\caption{$H$ - [4.5] vs $J-H$ colour-colour plots for the compilation of photometry for T~dwarfs from \citet{sandy10}, along with new photometry presented in this paper. Benchmark systems are indicated and numbered as for Figure~\ref{fig:hs2plots}.}
\label{fig:jhsplot}
\end{figure*}

\section{The distribution of T dwarf colours}
\label{sec:scat}

Beyond comparisons to our benchmark sample, it is also interesting to make more qualitative comparisons between the spread in colours predicted by the atmospheric models and that seen for the wider T~dwarf population. Such comparisons provide insight as to whether the impact of varying parameters such as gravity in models results in colour shifts of the similar proportions seen in the data. 
Figure~\ref{fig:hs2plots} shows $H-K$ vs $H-{\rm [4.5]}$ colour-colour plots for the compendium of MKO and Spitzer photometry of late-T~dwarfs presented by \citet{sandy10}, updated with additional photometry presented here.  The benchmark T~dwarfs are labelled as for Figure~\ref{fig:hs2plots}.  Figure~\ref{fig:jw2plots} shows similar $Y-J$ vs $J - W2$ colour-colour diagrams for the UKIDSS T~dwarfs with WISE photometry. 
These plots reveal that the two model sets are each able to reproduce the spread in late-T dwarf colours in one of $H-K$ or $Y-J$, but not both.   We note that neither set of models include refractory condensate clouds that are important at warmer $T_{\rm eff}$, and so they are not expected to match the data for earlier type T dwarfs, which they do not.

The upper panel of Figure~\ref{fig:hs2scat} suggests that the revised CIA~H$_{2}$ opacity included in the \citet{saumon2012} models has significantly improved the match to the spread and pattern seen in $H-K$ colours over previous model grids, which tended to predict $H-K$ that was too blue, and descending to more negative values rapidly with $T_{\rm eff}$ \citep[e.g. ][]{ben09, pinfield2012}.
The lower panel of Figure~\ref{fig:hs2scat} shows the colour tracks for the \citet{morley2012} models, which, in addition to using the \citet{saumon2012} CIA H$_{2}$ opacity also include alkali and sulphide clouds that may become important at $T_{\rm eff} \ltsimeq 800$~K. These models show slightly redder $H-K$, with a smaller spread in $H-K$ colour as a function of gravity. 
The reduced spread in $H-K$ arises as higher gravity leads to thicker cloud layers, pushing the colours redward and partially counteracting the blueward trend that would otherwise result from the increased CIA  H$_{2}$ opacity. 

As can be seen in Figure~\ref{fig:jw2scat}, the introduction of the low-$T_{\rm eff}$ clouds tends to increase the $Y-J$ spread due to gravity, as the effect of increasing gravity on pressure sensitive gas-phase opacities and the impact of thickening cloud layers with increasing gravity both tend to give redder $Y-J$ colours, and here the cloudy models better reproduce the observed spread in colours than those without clouds. 
Variations in metallicity may have a significant impact on cloud properties. Since only solar metallicity realisations are available for the new model grids, it is thus impossible to determine whether their failure to reproduce the observed colour spread in both Figures~\ref{fig:hs2scat} and~\ref{fig:jw2scat} merely reflects this fact, or if it suggests some other as-yet unidentified shortcoming in the theoretical approach.

\begin{figure*}
\includegraphics[width=430pt, angle=00]{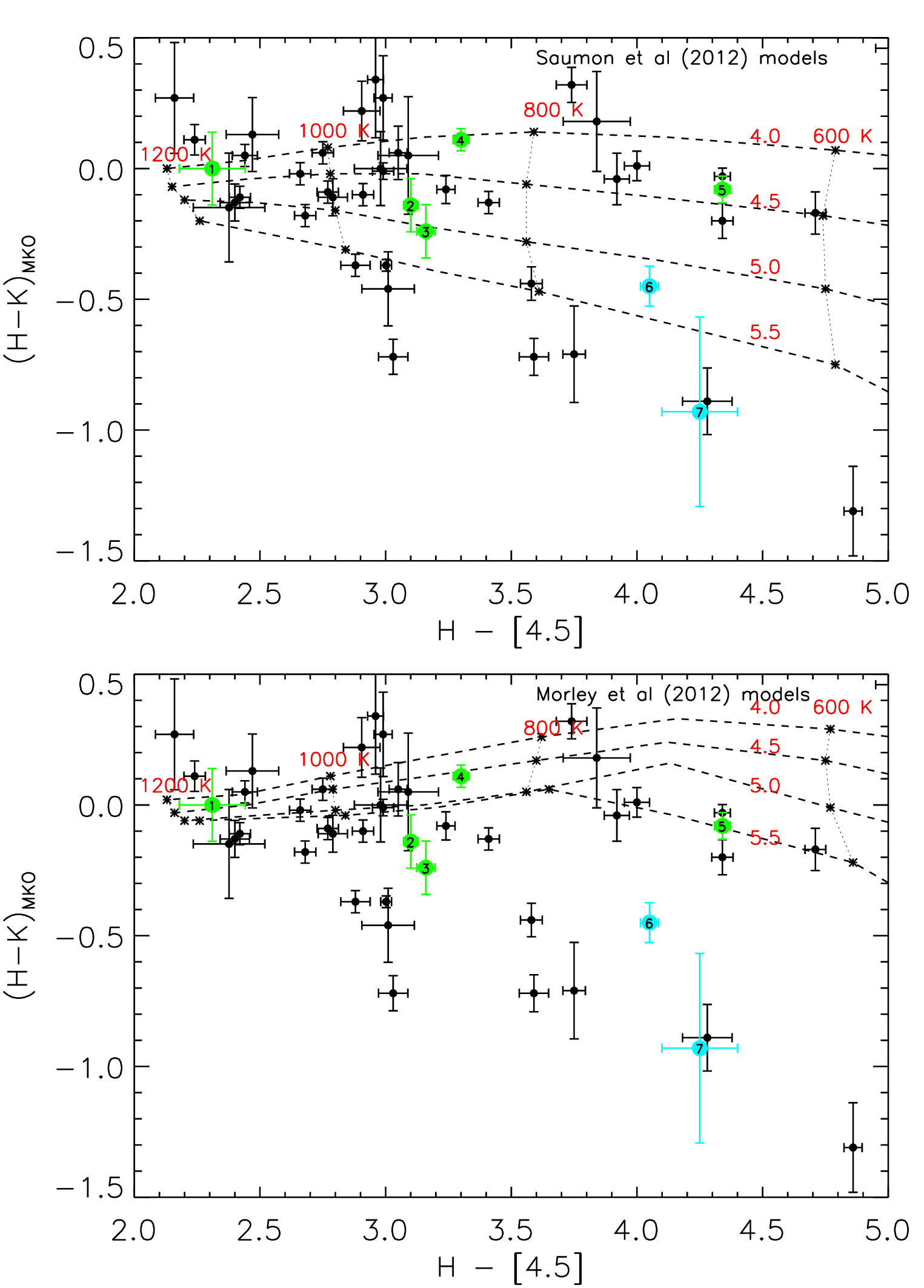}
\caption{$H$ - [4.5] vs $H-K$ colour-colour plots for the compilation of photometry for T~dwarfs from \citet{sandy10}, along with new photometry presented in this paper. Model colour tracks for \citet{saumon2012}, top panel, and \citet{morley2012}, bottom panel, are shown for comparison, the latter assumed $f_{sed} = 5$. The models shown are all solar metallicity, and the $T_{\rm eff}$ and $\log g$ values are indicated on the colour sequences. Benchmark systems are indicated and numbered as for Figure~\ref{fig:hs2plots}.}
\label{fig:hs2scat}
\end{figure*}

\begin{figure*}
\includegraphics[width=430pt, angle=0]{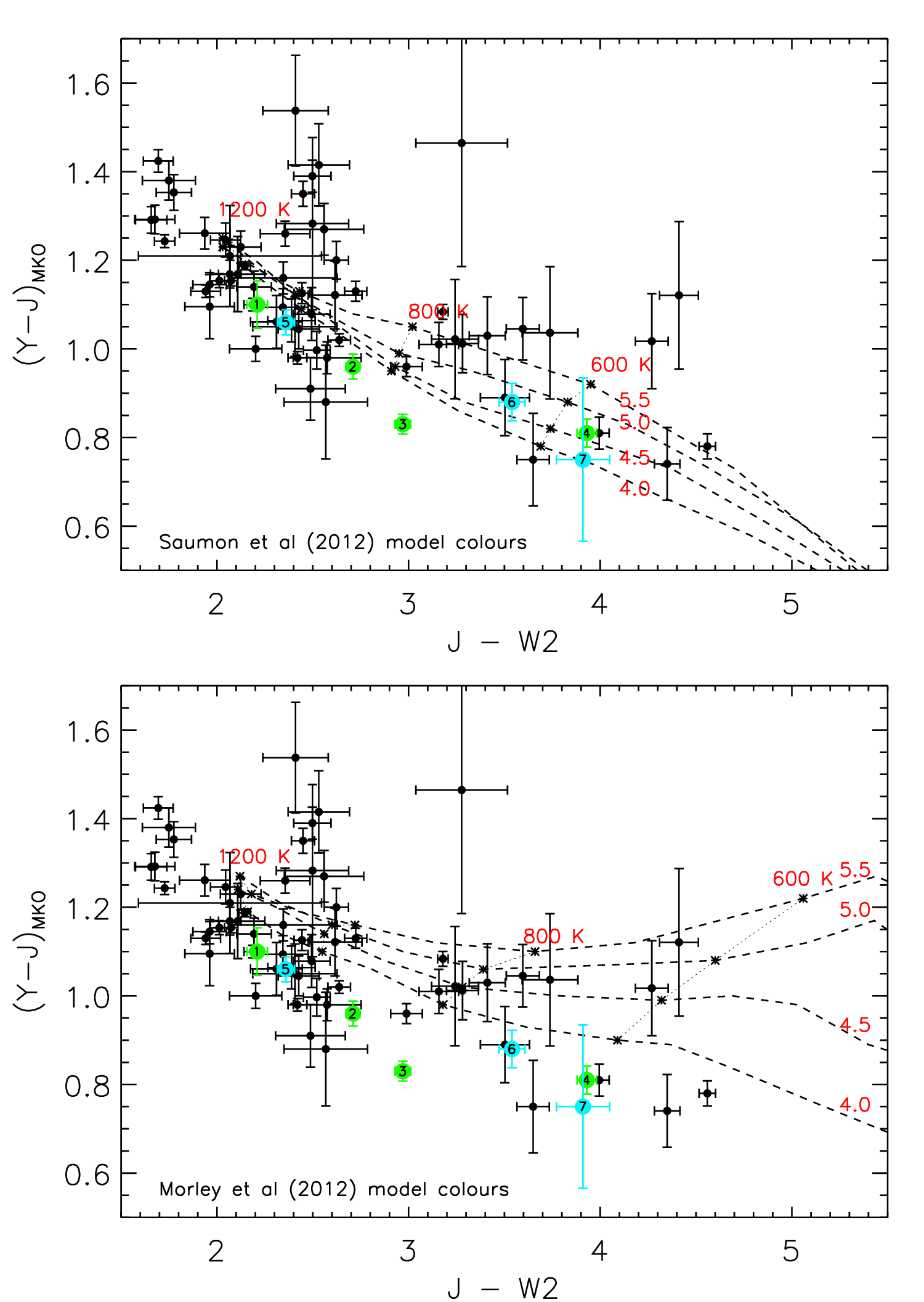}
\caption{$J - W2$ vs $Y-J$ colour-colour plots for the UKIDSS T~dwarfs with $YJ$ and WISE photometry. Model colour tracks for \citet{saumon2012}, top panel, and \citet{morley2012}, bottom panel, are shown for comparison, the latter assumed $f_{sed} = 5$. The models shown are all solar metallicity, and the $T_{\rm eff}$ and $\log g$ values are indicated on the colour sequences. Benchmark systems are indicated and numbered as for Figure~\ref{fig:jw2plots}.}
\label{fig:jw2scat}
\end{figure*}

We have presented plots similar to Figure~\ref{fig:hs2scat} in previous work \citep[e.g. ][]{sandy10,ben2011b,pinfield2012} comparing the colours of the late-T~dwarf sample with the previous generation of models that did not include the improved CIA H$_{2}$ treatment, nor the low-temperature sulphide and alkali condensates proposed by \citet{morley2012}. 
The previous generations of models typically predicted significantly bluer colours in $H-K$, with $H-K$ becoming increasingly blue with decreasing $T_{\rm eff}$. The preference for redder $H-K$ colours amongst the late-T sample led us to conclude that there was some bias in favour of young, low-gravity objects either within our selection or in the population itself.  The model colour tracks in the top panel of Figure~\ref{fig:hs2scat} match the observed spread in $H-K$ colours very well, with no need for such a young-age bias in the sample. Indeed, when both Figures~\ref{fig:hs2scat} and~\ref{fig:jw2scat} are considered, the strong model dependence of previous conclusions drawn from such plots is apparent.

\section{Kinematics of the UKIDSS late-T dwarf sample}
\label{sec:kin}

The implication from Figures~\ref{fig:hs2plots} to~\ref{fig:jw2scat} that the late-T dwarf sample is not dominated by young low-mass objects as had been previously suggested \citep{sandy10,ben2011b,pinfield2012} highlights the inherent model dependency of determining the properties of cool brown dwarfs from colour-colour diagrams. To shed further light on this issue we have constructed a $J$ band reduced proper motion diagram for our sample (see Figure~\ref{fig:rpmj}). It is apparent that there are no obvious features which distinguish
the distribution of kinematic properties between the latest type T dwarfs and the earlier type objects in UKIDSS.
A more robust statistical treatment to assess this will be presented by Smith et al (in prep.).  Nonetheless, it appears that the late-T dwarf sample is kinematically indistinct from the mid-T~dwarfs, which themselves have been shown to match the kinematics of the earlier type L and M dwarfs, reflecting the typical Galactic disk age distribution \citep{faherty2009}.  

We note that our earliest type objects, in the T4 class, appear to be preferentially distributed to higher speeds. Previous detailed studies of T dwarf kinematics \citep[e.g.]{faherty2009} argue this is not a real effect. It likely arises as a result of the combination of our $J-H < 0.1$ and $J-K < 0.1$ colour requirements, which may exclude T4 dwarfs with redder $H-K$ colours, and $J-H$ colours near our colour cut offs. As such, we are likely to preferentially select objects that are bluer in $H-K$, due to high-gravity and/or low-metallicity, in this region.  Such objects can be expected to be typically older, and thus will exhibit higher velocities.

\begin{figure}
\includegraphics[width=250pt, angle=0]{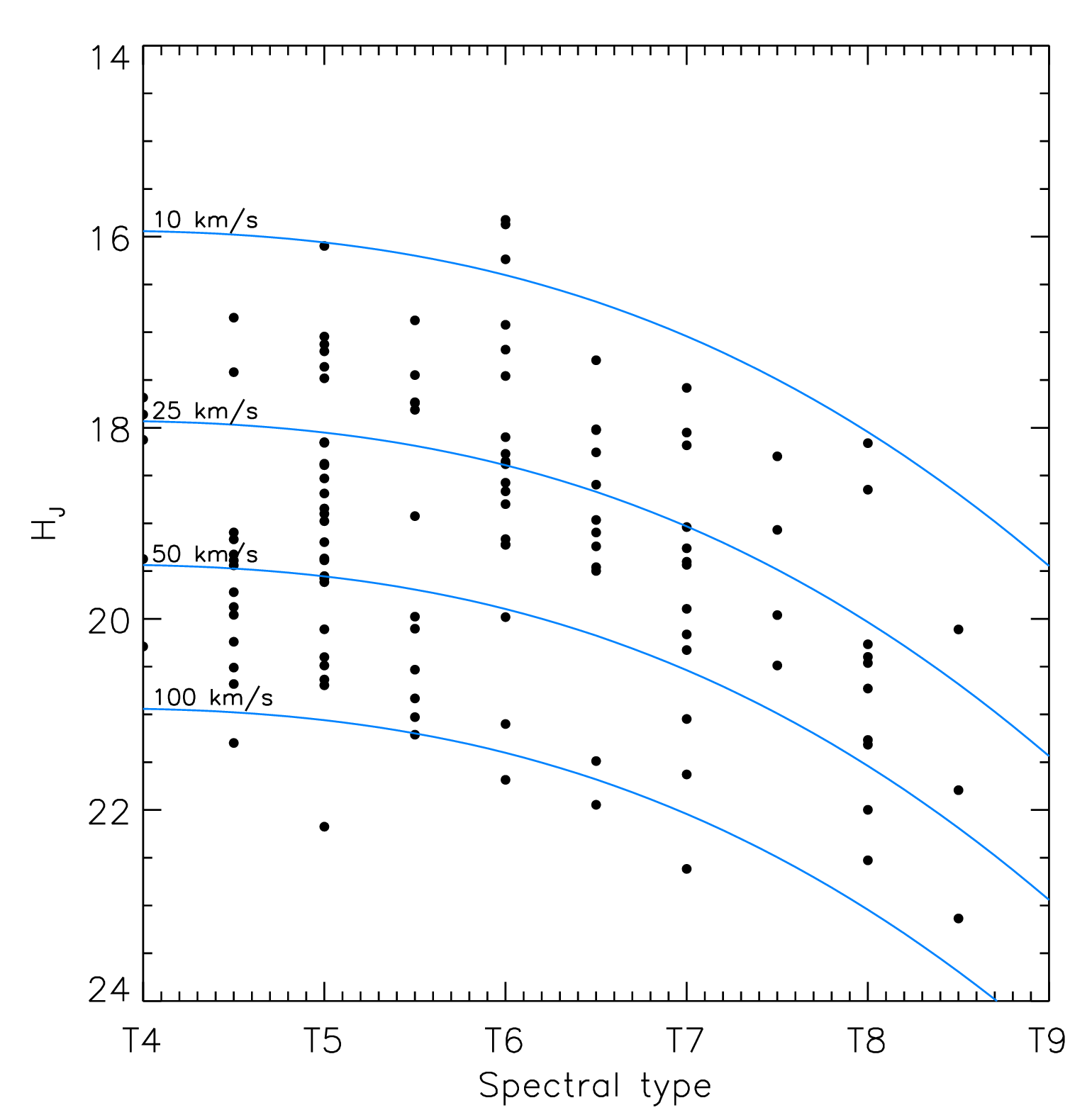}
\caption{A $J$ band reduced proper motion versus spectral type diagram for the UKIDSS late-T~dwarfs with well measured proper motions. Iso-$V_{tan}$ contours have been over plotted using by applying the \citet{dupuy2012} $M_{J}$-spectral type relations. To apply these relations, we have adopted \citet{cushing2011} spectral types for the latest type objects.}
\label{fig:rpmj}
\end{figure}

\section{Updated space density estimate}
\label{sec:density}

We now have near-complete follow-up of all candidates with $J < 18.8$ in UKIDSS LAS Data Release 8 (DR8). This covers 2270 square degrees of sky within the SDSS DR8 footprint. 
Our significantly increased sample of $\geq$T6 dwarfs allows us to improve on the space density estimate derived in \citet{ben10b}. However, the optimisation of our selection method since that work (to include CH$_{4}$ imaging) somewhat complicates our completeness correction and bias selection. Unifying a sample selected in a slightly inhomogeneous manner, such as ours,  to derive a space density is best achieved via a Bayesian parameter estimate method.  For this reason we defer this more rigorous derivation of the space density to a future work, since it is beyond the scope of this discovery paper. Here, instead, we follow the method of \citet{pinfield08} and \citet{ben10b}, and derive approximate correction factors for the incompleteness introduced by our different photometric cuts in the same manner as we previously did for correcting for our $J-H$ selection cut, and simply scale these by the proportion of the sample to which they were applied.

There are 76 dwarfs with measured spectral types of T6 or later with $J < 18.8$ in the region of sky covered by UKIDSS DR8 and SDSS DR8 probed by our searches. Of these, 39 have already had spectra published \citep{tsvetanov2000,burgasser04a,chiu06,lod07,chiu08,pinfield08,ben08,delorme08,ben09,ben10b,kirkpatrick2011,scholz2012}, and 37 have been presented for the first time here. In addition, we have four targets with ${\rm CH_{4}}s - {\rm CH_{4}}l < -0.5$ which have not not yet been followed up spectroscopically (see Table~\ref{tab:photom}). The ${\rm CH}_{4}$ types for two of these objects are constrained to earlier than T5.5, whilst two are consistent with types as late as T6.  We thus include the later two of these four in our sample, bringing it to a total of 78 T6 and later dwarfs. 

We also have one target for which we do not have well calibrated methane photometry and no spectroscopic observations:  ULAS~J~0800+1908 has a LIRIS methane colour of $H - {\rm CH}_{4}l  = -0.73$. This is suggestive of a type intermediate between that of the T7 dwarf ULAS~J0746+2355 which has LIRIS $H - {\rm CH}_{4}l  = -0.87$ and the  T6.5 dwarf ULAS~J1023+0447. However, the lack of a spectral type calibration for this methane colour prevents us from assigning this object to a specific spectral type bin in our sample. Instead, we incorporate it into the uncertainty in the T6 and T7 bins.

Two thirds of our sample were selected using the method presented in \citet{pinfield08} and \citet{ben10b}. For this method we found that a correction factor of 1.03 should be applied to all spectral type bins, to account for the biases introduced by the following effects: scatter out of our $J-H < 0.1$ selection due to photometric error; mis-matching of bona fide T~dwarfs with background stars in our SDSS crossmatch; wide binary companions to stars (which should be excluded from a T dwarf primary space density).  We do not correct for the possibility of excluding objects with the $Y-J < 0.5$ that we applied to $YJ$-only selected targets fainter than $J = 18.5$ since no T~dwarfs have yet been found with such colours. Although $Y$ dwarfs are now known with such blue $Y-J$ colours \citep{leggett2013}, we are not sensitive to such objects due to their inherently faint nature.

One third of our sample was selected using the new follow-up method described in Section~\ref{sec:phot}. Since this method uses the same starting point of $YJH$ and $YJ$-only colour selections, it is subject to the same biases. In addition to those, however, it also includes bias due to the $z'$ band and ${\rm CH}_{4}$ follow-up criteria. The follow-up $z'$ band observations delivered typically uncertainties of $\pm 0.2$ magnitudes for faint  red T~dwarfs, and considerably smaller uncertainties for bluer M dwarfs. We ruled out from further investigation targets that had $z' - J < 2.5$ after $z'$~band follow-up. This is roughly 3$\sigma$ bluer than the bluest measured $z' - J$ for late-T~dwarfs. As such, we estimate that less than 1\% of bona fide late-T~dwarfs would be scattered out of our selection during the $z'$ band follow-up step. 

Our ${\rm CH}_{4}$ photometry step may also exclude some objects since we only obtained spectroscopy for those objects with ${\rm CH_{4}}s-{\rm CH_{4}}l < -0.5$. The strong dependence of ${\rm CH}_{4}$ colour on spectral type means that the fractional bias must be determined for each spectral type bin.
We have estimated the number of objects that we would expect to scatter out of our ${\rm CH_{4}}s-{\rm CH_{4}}l < -0.5$ cut by summing the probabilities of our confirmed dwarfs in each bin being scattered beyond the cut based on their measured ${\rm CH}_{4}$ colours and 1$\sigma$ uncertainties. We find that we would expect 8\% of T6--T6.5 dwarfs to be scattered out of our selection at this stage of follow-up. Following the same method for T7--T7.5 dwarfs we find that less the 0.1\% will be excluded, and similarly negligible fractions for the later type bins also.

We thus find that our bias corrections factors for our full sample due to our selection method are 1.03 for the T7--T7.5 and later bins, and 1.05 for our T6--T6.5 bin. We must also correct for biases inherent to any magnitude limited survey due to preferential inclusion of unresolved binary systems and the Malmquist bias. In \citet{ben10b} we derived binary correction factors of 0.97 and 0.55 based on the broadest range of likely binary fractions reported in the literature (5\% -- 45\%). In \citet{pinfield08}, the Malmquist correction factor was found to be between 0.84 and 0.86. We apply these same correction factors here.  

Calculating a space density from our corrected sample requires the application of some $M_J$-spectral type conversion to determine the volume probed for each subtype by our magnitude limited sample. The choice of relation has a significant impact on the resulting space densities, and differences in approach to this issue can contribute significantly to differences in measured space density between different projects. In \citet{ben10b} we used the $M_J$-spectral type relations of \citet{liu06}. However, to take advantage of the improved sample of T~dwarfs with measured parallaxes we will instead use the polynomial relations presented by \citet{dupuy2012} here. Since these relations are only valid for the \citet{cushing2011} system for the latest spectral types we have included our T9 dwarfs in the T8--T8.5 bin.  In Table~\ref{tab:density}  we provide a summary of our update to this calculation. By this method we find that the space density of T6--T8.5 dwarfs \citep[on the ][ system]{cushing2011} is $3.00 \pm 1.3  - 5.52 \pm 2.4 \times 10^{-3} pc^{-3}$, depending on the underlying binary fraction.  

\begin{table*}
\begin{tabular}{ c c c c c c c c c c }
  \hline
Type & $T_{\rm eff}$ range & N & N$_{c}$(min) & N$_{c}$(max) & $M_J$(MKO) & Range (pc) & Volume (pc$^3$) & $\rho_{min}$ ($10^{-3} pc^{-3}$) & $\rho_{max}$ ($10^{-3} pc^{-3}$) \\
\hline
T6-6.5 & 900--1050K & 40 & $19.4 \pm 3.1$ & $35.6 \pm 5.7$ & $14.90 \pm 0.39$ & $60 \pm 11$ & $50400 \pm 27100$ & $0.39 \pm 0.22$ & $0.71 \pm 0.40$ \\ 
T7-7.5 & 800--900K & 21 & $10.0 \pm 2.2$ & $18.5 \pm 4.0$  & $15.65 \pm 0.39$ & $43 \pm 8$ & $18000 \pm 9700$ & $0.56 \pm 0.32$ & $1.02 \pm 0.64$ \\
T8-8.5 & 500--800K & 17  & $6.7 \pm 1.8$ & $12.3 \pm 3.3$ & $16.75 \pm 0.39$ & $26 \pm 5$ & $3900 \pm 2100$ & $2.05 \pm 1.21$ & $3.79 \pm 2.24$ \\ 

\hline
\end{tabular}
\caption{Summary of the updated space space density calculation following the method of \citet{ben10b} for our $J < 18.8$
  sample of $\geq$T6 dwarfs. N$_{c}$ refers to corrected numbers based
  on the sample corrections described in the text, with maximum and
  minimum values arising from the different possible binary
  corrections.
The values of $M_{J}$ used to calculate the
  distance limit and volume probed for each type were calculated using
  the polynomial relations in $M_{J}$ versus spectral type derived by
  \citet{dupuy2012}. The uncertainties in $M_{J}$ reflect the RMS
  scatter about the \citet{dupuy2012} polynomials.
The uncertainties in the computed space
  densities reflect the volume uncertainty that arises from the
  uncertainty in $M_J$ and Poisson noise in our sample. The minimum
  and maximum space densities reflect the range encompassed by likely
  binary fractions \citep[see text and ][]{ben10b}.   The latest spectral types are on the \citet{cushing2011} system.
\label{tab:density}}

\end{table*}

\citet{kirkpatrick2012} also used polynomial relations to estimate the magnitudes of objects for which no trigonometric parallax is available and they applied a single 30\% unresolved binary correction. To allow comparison between these samples we have carried out an additional space density calculation assuming the same binary fraction. Table~\ref{tab:denscomp} summarises this calculation, from which we find a space density of $3.9\pm 1.7 \times 10^{-3} pc^{-3}$ for T6--T8.5 dwarfs on the \citet{cushing2011} system. This is very close to the value of $3.43 \pm 0.32 \times 10^{-3} pc^{-3}$ found by \citet{kirkpatrick2012}. The error we have assigned to Kirkpatrick value has been calculated assuming only Poisson noise in their count of each spectral subtype.

\begin{table*}
\begin{tabular}{ c c c c c c c c  c}
  \hline
Type & $T_{\rm eff}$ range & N & N$_{c}$& $M_J$(MKO) & Range (pc) & Volume (pc$^3$) & $\rho$ ($10^{-3} pc^{-3}$) & $\rho$ ($10^{-3} pc^{-3}$) \\
&&&&&&&&  \citet{kirkpatrick2012} \\
\hline
T6-6.5 & 900--1050K & 40 & $25.3 \pm 4.0$ & $14.90 \pm 0.39$ & $60 \pm 11$ & $50400 \pm 27100$ &  $0.50 \pm 0.28$ & $1.10 \pm 0.18$\\ 
T7-7.5 & 800--900K & 21 & $13.0 \pm 2.8$ & $15.65 \pm 0.39$ & $43 \pm 8$ & $18000 \pm 9700$ &  $0.73 \pm 0.42$ & $0.93 \pm 0.17$ \\
T8-8.5 & 500--800K & 17  & $10.5 \pm 2.6$ & $16.75 \pm 0.39$ & $26 \pm 5$ & $3900 \pm 2100$ & $2.67 \pm 1.58$  & $1.40 \pm 0.21$  \\ 

\hline
\end{tabular}
\caption{Summary of the updated space space density calculation following the method of \citet{ben10b} for our $J < 18.8$
  sample of $\geq$T6 dwarfs. N$_{c}$ refers to corrected numbers based
  on the sample corrections described in the text, and 30\% binary correction as applied by \citet{kirkpatrick2012}. The values of $M_{J}$ and the uncertainties are as for Table~\ref{tab:density}.
The latest spectral types are on the \citet{cushing2011} system. The uncertainties for the \citet{kirkpatrick2012} densities estimated purely on the basis of Poisson noise on their number counts. 
\label{tab:denscomp}}

\end{table*}

Using polynomial relations is not necessarily, however, the most appropriate way to estimate absolute magnitudes for targets of a given subtype. 
 This is because the subtypes are not a continuous variable, but rather discrete and based on an inherently subjective spectral typing scheme. As such, polynomial fits (and their associated residuals) can mask true scatter in the $M_J$ values of certain subtypes. We have thus performed an additional density estimate using the mean magnitudes for each spectral subtype from \citet{dupuy2012}, and using the \citet{cushing2011} spectral typing scheme. For the purposes of this calculation we have also used a single binary correction factor of 30\%. Table~\ref{tab:denscomp2} summarises this calculation that we include here for illustrative purposes. 
 
 This method leads us to an estimate of  $6.2 \pm 1.9 \times 10^{-3} pc^{-3}$ for the space density of T6--T8.5 dwarfs \citep[on the system of ][]{cushing2011}. This is somewhat higher than the other methods determined (for the same binary fraction), and it is driven by the fainter absolute magnitude assigned to the coolest spectral type bin.  \citet{metchev08} also used a mean magnitude approach for estimating their survey volumes, applying a mean $M_J(2MASS) = 15.75 \pm 0.50$ to the T6--T8 bin, and deriving a space density of $4.3^{+2.9}_{-2.6} \times 10^{-3} pc^{-3}$ (for a binary fraction of $\sim 26\%$). This is consistent with our value of $2.8 \pm 0.8 \times 10^{-3} pc^{-3}$ across the same spectral type range. Their slightly higher value can largely be attributed to fact that they use a single mean for the entire T6--T8 bin, which is fainter than the value we use for the most numerous T6 dwarfs. This will tend lead to an overestimate of the space density across the whole bin. This illustrates  the significant impact that the assumed $M_J$-spectral-type relation can have on derived space densities where trigonometric parallaxes are not available for the entire sample. The uncertainties introduced by the scatter about the assumed $M_J$, and the systematics associated with the choice of $M_J$ estimate relation, now significantly outweigh the random error due to our sample size.

\begin{table*}
\begin{tabular}{ c c c c c c c}
  \hline
Type & N & N$_{c}$& $M_J$(MKO) & Range (pc) & Volume (pc$^3$) & $\rho$ ($10^{-3} pc^{-3}$) \\
\hline
T6  &  24 & $15.2 \pm 3.0$ & $15.22 \pm 0.15$ & $52 \pm 4$ & $32400 \pm 6700$ &  $0.47 \pm 0.12$ \\ 
T6.5 & 16 & $10.1 \pm 2.5$ & $15.22 \pm 0.31$ & $52 \pm 7$ & $32400 \pm 13900$ &  $0.31 \pm 0.14$ \\  
T6 - T6.5 & 40 & &  & & &  $0.78 \pm 0.19$ \\ 
T7 & 14 & $8.7 \pm 2.3$ & $15.54 \pm 0.25$ & $45 \pm 5$ & $20800 \pm 7200$ &  $0.42 \pm 0.17$ \\
T7.5 & 7 &$4.3 \pm 1.6$ & $16.05 \pm 0.65$ & $35 \pm 11$ & $10300 \pm 9200$ &  $0.42 \pm 0.39$  \\
T7 - T7.5 & 21 & &  & & &  $0.84 \pm 0.43$ \\ 
T8 & 12  & $7.4 \pm 2.1$ &  $16.39 \pm 0.35$ & $30 \pm 5$ & $6400 \pm 3100$ & $1.16 \pm 0.63$  \\ 
T8.5 & 5 & $3.1 \pm 1.4$ &  $17.81 \pm 0.33$ & $16 \pm 2$ & $900 \pm 400$ & $3.43 \pm 1.77$   \\ 
T8 - T8.5 & 17 & &  & & &  $4.58 \pm 1.88$ \\ 
\hline
\end{tabular}
\caption{Summary of our space space density calculation following the method of \citet{ben10b} to correct biases in our $J < 18.8$
  sample of $\geq$T6 dwarfs.
  N$_{c}$ refers to corrected numbers based
  on the sample corrections described in the text, and a 30\% binary correction as applied by \citet{kirkpatrick2012}, and the latest spectral type bin is on the system of \citet{cushing2011}.
  In this case we have used the mean absolute magnitudes of \citet{dupuy2012} for our $M_J$-spectral type conversion.  We have applied these to each half subtype before combing them to arrive at space densities for full subtype bins.
  The uncertainties in the computed space
  densities reflect the volume uncertainty that arises from the
  uncertainty in $M_J$ and Poisson noise in our sample. 
\label{tab:denscomp2}}

\end{table*}

In Figure~\ref{fig:imf} we compare our space densities from Table~\ref{tab:denscomp} and that of \citet{kirkpatrick2012}, to the predictions from Monte Carlo simulations of the Galactic field population of T~dwarfs assuming different forms of the IMF and brown dwarf formation history \citep[see ][ for full descriptions]{ben10b,dh06}. 
As in \citet{ben10b} we have used the system mass function normalisation of 0.0024~pc$^{-3}$ for objects in the $0.09 - 0.10~\Msun$ range, taken from \citet{deacon2008}.
For the purposes of this comparison we have transformed the space densities which are supplied from the simulations as a function of $T_{\rm eff}$, to densities as a function of spectral type using the $T_{\rm eff}$ ranges given in Table~\ref{tab:denscomp}. These conversions are uncertain and function as a very ``broad brush" to facilitate the comparison between simulations and observation.
  We have also included predictions for the \citet{chabrier2005} log-normal system mass function, since this now appears to be the preferred function fitted to young clusters across the low-mass stellar/substellar/planetary mass regime \citep[e.g.][ and references therein]{bastian2010,lod12,alves2012}.

\begin{figure}
\includegraphics[height=200pt, angle=90]{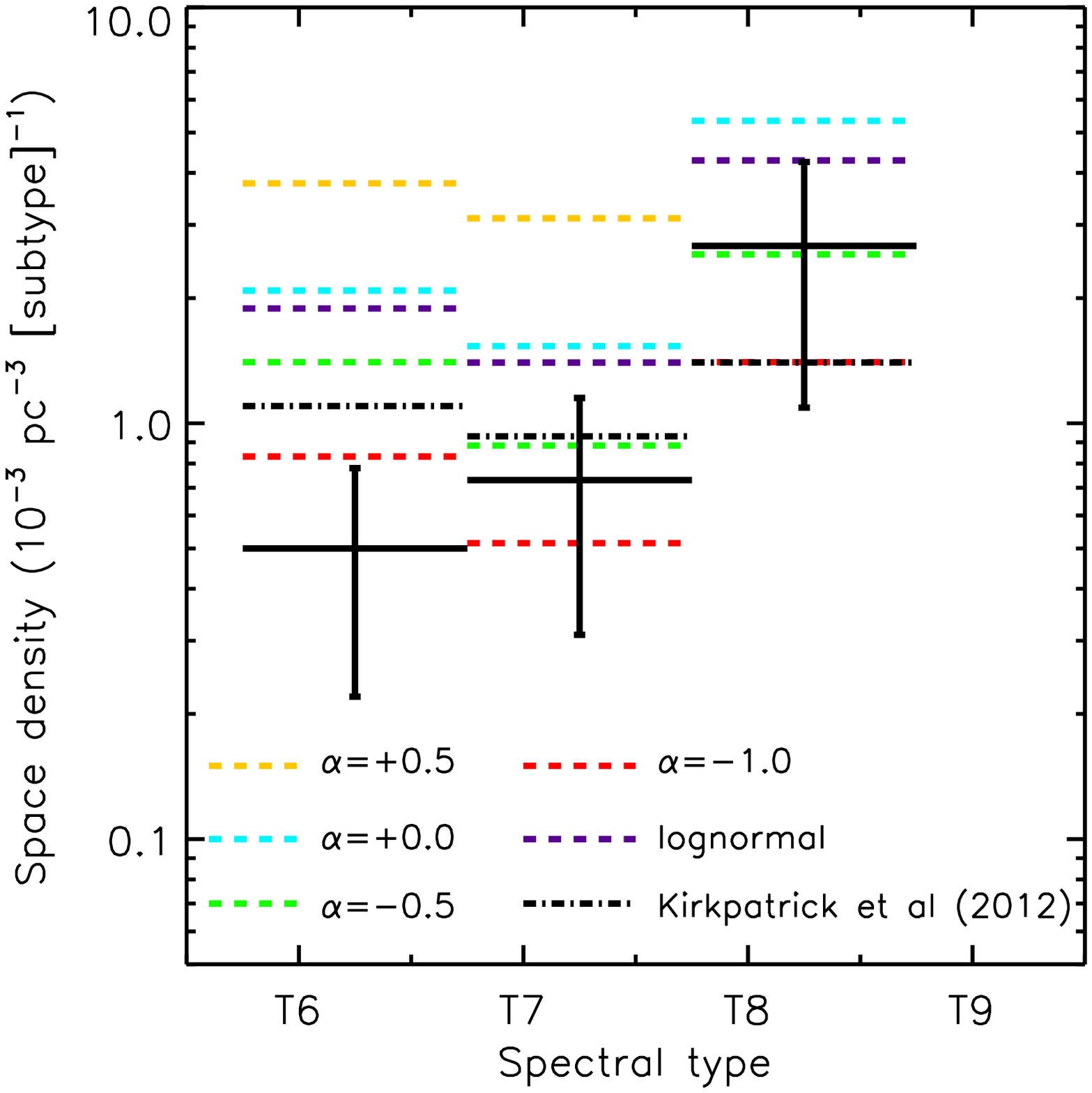}
\caption{Computed space densities for different spectral types  from
  Monte Carlo simulations of the  field population of T~dwarfs for a
  uniform birthrate \citep[i.e. $\beta~=~0.0$, see ][ for full definition]{dh06,ben10b} and various underlying mass
  functions. The power-law mass functions are of the form $\psi(M)~\propto~M^{-\alpha}$~pc$^{-3}$ \Msun$^{-1}$, and the log-normal mass function is the \citet{chabrier2005} system mass function.  Our observed space density is shown as solid black lines, \citet{kirkpatrick2012} as dash-dotted black lines.
   Uncertainties are indicated with bars at the midpoint of each
  spectral type bin, and reflect volume uncertainties and Poisson
  counting uncertainties.
}
\label{fig:imf}
\end{figure}

It is clear that the estimated space densities from both UKIDSS and WISE are significantly lower than predicted for an $\alpha = 0$ or log-normal mass function, coinciding best with values of $-1.0 < \alpha < -0.5$. As has been noted before \citep[e.g.][]{pinfield08,ben10b}, this represents a significant discrepancy with the IMF measured in many young clusters where the mass function has been fitted with $\alpha \approx +0.6$ power laws \citep[e.g.][]{barrado02,lod07a,lodieu09} or the log-normal mass function with a characteristic mass in the region of $0.2 \Msun$ that has been measured by studies using the UKIDSS Galactic Clusters Survey, amongst others \citep[e.g.][]{lod07b,lodieu09,lod11,lod11b,lod12,lod12a,alves2012,alves2012b,boudreault2012}.   \citet{kirkpatrick2012} note that the ratio of stars to brown dwarfs in the field is consistent with that seen in young clusters when a universal substellar IMF is assumed \citep{andersen2008}, and thus conclude that the underlying mass functions are the same. However, the discrepancy between the predictions based on the functional form that has been fitted to the IMF in young clusters, and observed field population should not be ignored. Assuming a 30\% binary fraction, both our UKIDSS T6--T8.5 space density and that of \citet{kirkpatrick2012} are approximately half that predicted by the log-normal mass function, and corresponding to a significance of 2$\sigma$ and 13$\sigma$ respectively.

The kinematics of our sample (see Section~\ref{sec:kin}) imply that this discrepancy cannot be attributed to late-T~dwarfs having a significantly different formation history to the rest of the Galactic disk, and we are left with two plausible options. Firstly,  errors in the evolutionary models used to estimate masses in young clusters and/or in the cooling times used to produce the Monte Carlo simulations in field could give rise to such a discrepancy. These options can be investigated by employing shifted cooling curves in the Monte Carlo simulations to investigate their impact on the predicted space densities and age profiles, and we defer further discussion of this possibility to a future work.

The second plausible origin for such a discrepancy is if the bulk of the field population formed in different environments than the nearby young clusters that have been subject to detailed study. For example,  \citet{bressert2010} have found that only 26\% of low-mass stars form in high density regions, which they define as those with surface densities of young stellar objects (YSOs) $> 200 pc^{-2}$, corresponding roughly to the most dense parts of the Taurus star forming complex.  \citet{luhman2004} found a deficit of brown dwarfs in Taurus relative to the high densities of the Trapezium cluster, and more recent studies have further hinted at similar enrichment of the substellar component at higher densities \citep{andersen2011}. 
A combination of low-density dominated low-mass star and brown dwarf formation with a substellar IMF that declines steeply in the lowest density regions, but has the log-normal form currently preferred in higher density regions, would naturally reconcile the two conflicting views of form for the substellar IMF that we are currently presented. 
Unfortunately, obtaining a statistically meaningful census of the young substellar population in the lowest density regions where \citet{bressert2010} argue that most low-mass stars are formed (e.g. $< 50$~YSOs~pc$^{-2}$) is extremely challenging due to the large areas that must be surveyed and the inherently faint nature of the targets.  Given the significant investment of observing time that would be required to test this hypothesis, it would make sense to first investigate the potential impact of systematic effects in the evolutionary models for both young and old objects.

\section{Summary}
\label{sec:summ}

The expanded T dwarf sample presented here, along with access to the new 2 epoch LAS catalogue of Smith et al (in prep) has allowed the discovery of two new benchmark systems along with the first estimate for the wide-binary companion fraction amongst late-T~dwarfs. By examining the colours of  the current census of late-type benchmarks we have identified the $H-{\rm [4.5]}$ and $J-W2$ as being more sensitive to metallicity than models have so far predicted, and caution that this must be considered when using these colours to estimate the relative properties of cool brown dwarfs. The expanded sample of late-T~dwarfs, and the new model grids from \citet{saumon2012} and \citet{morley2012} also now argue against the previous conclusion of \citet{sandy10} that the sample is dominated by young low-mass objects. This is supported by the kinematics of our sample. 

Our updated space density for late-T~dwarfs is consistent with the density reported by \citet{kirkpatrick2012} and confirms our previous conclusion that there are far fewer late-T~dwarfs than we would expect given the favoured forms for the IMF in young clusters. Whilst it is possible that this discrepancy arises from problems with either or both of the young and old evolutionary models for brown dwarfs, we also speculate that it could arise as a result of the dominant environment for low-mass star and brown dwarf formation being lower density regions than are currently probed, as suggested for low-mass stars by \citet{bressert2010}.

\appendix
\section{SQL database queries used for initial candidate selection}

The following code listings give the SQL code we used to query the WSA for our candidates. These include a crossmatch against SDSS~DR7.  To access SDSS DR8 sky not covered by SDSS DR7 we independently crossmatched candidate lists that had passed our near-infrared colour cuts against SDSS~DR8. Our final $J-K < 0.1$ colour cut was applied to the resulting candidate lists.

\lstset{ %
language=sql,               % choose the language of the code
basicstyle=\scriptsize,       % the size of the fonts that are used for the code
numbers=none,                   % where to put the line-numbers
backgroundcolor=\color{white},  % choose the background color. You must add \usepackage{color}
showspaces=false,               % show spaces adding particular underscores
showstringspaces=false,         % underline spaces within strings
showtabs=false,                 % show tabs within strings adding particular underscores
frame=single,           % adds a frame around the code
tabsize=2,          % sets default tabsize to 2 spaces
captionpos=b,           % sets the caption-position to bottom
breaklines=true,        % sets automatic line breaking
breakatwhitespace=false,    % sets if automatic breaks should only happen at whitespace
escapeinside={\%*}{*)}          % if you want to add a comment within your code
}

\begin{Query}
\begin{lstlisting}
SELECT las.ra,las.dec,ymj_1Pnt,ymj_1PntErr,j_1mhPnt,j_1mhPntErr,yAperMag3,yAperMag3Err,j_1AperMag3,j_1AperMag3Err,hAperMag3, hAperMag3Err,kAperMag3, kAperMag3Err,psfMag_u,psfMagErr_u,psfMag_g,psfMagErr_g,psfMag_r,psfMagErr_r,psfMag_i,psfMagErr_i,psfMag_z,psfMagErr_z,(psfMag_i- psfMag_z),(psfMag_z - j_1AperMag3)
FROM   lasSource AS las, BestDR7..PhotoObj AS dr7, lasSourceXDR7PhotoObj AS x
WHERE
masterObjID=las.sourceID AND slaveObjID=dr7.ObjID AND distanceMins<0.033333 AND
 sdssPrimary=1 AND distanceMins IN (
SELECT MIN(distanceMins)
FROM lasSourceXDR7PhotoObj
WHERE masterObjID=x.masterObjID AND sdssPrimary=1) AND
/*    Good sdss detections	*/
      (psfMagErr_i < 0.5 or psfMagErr_z < 0.5) AND
/*    Colour cuts for mid-T & later: */
      j_1mhPnt < 0.1        AND
      ((psfMag_z - j_1AperMag3) > 2.5 or
      (psfMag_i - psfMag_z > 1.5)) AND
/*     Unduplicated or primary duplicates only: */
       (priOrSec = 0 OR priOrSec = frameSetID) AND
/*     Generally good quality: */
       yppErrBits   < 256 AND
       j_1ppErrBits < 256 AND
       hppErrBits   < 256 AND

/*     Point-like morphological classification: */
       mergedClass BETWEEN -3.0 and -0.5 and
       mergedClassStat BETWEEN -3.0 AND +3.0 AND
/*     Reasonably circular images: */
       yEll < 0.45 AND j_1Ell < 0.45 and
/*     YJ measured to 3 sigma and */
       yAperMag3Err   < 0.30 AND
       j_1AperMag3Err < 0.30 AND
       yAperMag3 > 14.5 AND
       j_1AperMag3 > 14.5 AND
       hAperMag3 > 14.5
ORDER BY las.ra
\end{lstlisting}
\caption{The SQL query used to select candidates via our $YJH$ channel in the case where they have been detected in SDSS.
\label{cod:yjhsdss}}
\end{Query}

\begin{Query}
\begin{lstlisting}
SELECTra,dec,ymj_1Pnt,ymj_1PntErr,j_1mhPnt,j_1mhPntErr, yAperMag3,yAperMag3Err,j_1AperMag3,j_1AperMag3Err,hAperMag3, hAperMag3Err,kAperMag3, kAperMag3Err
FROM   lasSource
WHERE
/*     Colour cuts for mid-T & later: */
       j_1mhPnt < 0.1        AND
/*     Unduplicated or primary duplicates only: */
       (priOrSec = 0 OR priOrSec = frameSetID) AND
/*     Generally good quality: */
       yppErrBits   < 256 AND
       j_1ppErrBits < 256 AND
       hppErrBits   < 256 AND
/*     Source not detected above 2sigma in SDSS-DR7 i' or z' within 2 arcsec: */
       sourceID NOT IN (
       SELECT masterObjID
       FROM   lasSourceXDR7PhotoObj AS x,
              BestDR7..PhotoObj     AS p
       WHERE  p.objID = x.slaveObjID     AND
              (psfMagErr_i < 0.5 OR psfMagErr_z < 0.5) AND
              x.distanceMins < 2.0/60.0
       ) AND
/*     Use only frame sets overlapping with SDSS-DR7: */
       frameSetID IN (
       SELECT DISTINCT(frameSetID)
       FROM   lasSource             AS s,
              lasSourceXDR7PhotoObj AS x
       WHERE  s.sourceID = x.masterObjID
       ) AND
/*     Point-like morphological classification: */
       mergedClass BETWEEN -3.0 and -0.5 and
       mergedClassStat BETWEEN -3.0 AND +3.0 AND
/*     Reasonably circular images: */
       yEll < 0.45 AND j_1Ell < 0.45 and
/*     YJ measured to 3 sigma and */
       yAperMag3Err   < 0.30 AND
       j_1AperMag3Err < 0.30 AND
       yAperMag3 > 14.5 AND
       j_1AperMag3 > 14.5 AND
       hAperMag3 > 14.5
ORDER BY ra
\end{lstlisting}
\caption{The SQL query used to select candidates via our $YJH$ channel in the case where they have not been detected in SDSS.
\label{cod:yjhnosdss}}
\end{Query}

\begin{Query}
\begin{lstlisting}
SELECT las.ra,las.dec,ymj_1Pnt,ymj_1PntErr,j_1mhPnt,j_1mhPntErr,yAperMag3,yAperMag3Err,j_1AperMag3,j_1AperMag3Err,hAperMag3, hAperMag3Err,kAperMag3, kAperMag3Err,psfMag_u,psfMagErr_u,psfMag_g,psfMagErr_g,psfMag_r,psfMagErr_r,psfMag_i,psfMagErr_i,psfMag_z,psfMagErr_z,(psfMag_i- psfMag_z),(psfMag_z - j_1AperMag3)
FROM   lasSource AS las, BestDR7..PhotoObj AS dr7, lasSourceXDR7PhotoObj AS x
WHERE
masterObjID=las.sourceID AND slaveObjID=dr7.ObjID AND distanceMins<0.033333 AND
 sdssPrimary=1 AND distanceMins IN (
SELECT MIN(distanceMins)
FROM lasSourceXDR7PhotoObj
WHERE masterObjID=x.masterObjID AND sdssPrimary=1) AND
/*    Good sdss detections	*/
      (psfMagErr_i < 0.5 or psfMagErr_z < 0.5) AND
/*    Colour cuts for mid-T & later: */
      ((psfMag_z - j_1AperMag3) > 2.5 or
      (psfMag_i - psfMag_z > 1.5)) AND
/*	H and K dropout: */
	hAPerMag3Err < 0 AND kAperMag3Err < 0 AND
	hAperMag3 < 0 AND kAperMag3 < 0 AND
/*     Unduplicated or primary duplicates only: */
       (priOrSec = 0 OR priOrSec = frameSetID) AND
/*     Generally good quality: */
       yppErrBits   < 256 AND
       j_1ppErrBits < 256 AND
/*     Point-like morphological classification: */
       mergedClass BETWEEN -3.0 and -0.5 and
       mergedClassStat BETWEEN -3.0 AND +3.0 AND
/*     Reasonably circular images: */
       yEll < 0.45 AND j_1Ell < 0.45 and
/*     YJ measured to 3 sigma and */
       yAperMag3Err   < 0.30 AND
       j_1AperMag3Err < 0.30 AND
       yAperMag3 > 14.5 AND
       j_1AperMag3 > 14.5
\end{lstlisting}
\caption{The SQL query used to select candidates via our $YJ$-only channel in the case where they have been detected in SDSS.
\label{cod:yj_sdss}}
\end{Query}

\begin{Query}
\begin{lstlisting}
SELECT 	s.ra,s.dec,s.ymj_1Pnt,s.ymj_1PntErr,s.j_1mhPnt,s.j_1mhPntErr,
       	s.yAperMag3,s.yAperMag3Err,s.j_1AperMag3,s.j_1AperMag3Err,my.mjdObs - mj.mjdObs, s.mergedClass
	/* Only frames with full coverage */
FROM   lasYJHKSource As s, lasYJHKmergeLog AS l, Multiframe AS my, Multiframe AS mj
WHERE
/*     	Colour cuts for mid-T & later or bright enough that its not M dwarf: */
       	(s.ymj_1Pnt > 0.5   OR s.j_1AperMag3 < 18.5) AND
/*     	Unduplicated or primary duplicates only: */
       	(s.priOrSec = 0 OR s.priOrSec = s.frameSetID) AND
/*     	Generally good quality: */
      	s.yppErrBits   < 256 AND
       	s.j_1ppErrBits < 256 AND
	s.frameSetID = l.frameSetID AND
/* 	Pick out the YJHK frames to get the mjds*/
   	l.ymfID = my.multiframeID AND
   	l.j_1mfID = mj.multiframeID AND
/*     Source not detected above 2sigma in SDSS-DR7 i' or z' within 2 arcsec: */
       sourceID NOT IN (
       SELECT masterObjID
       FROM   lasSourceXDR7PhotoObj AS x,
              BestDR7..PhotoObj     AS p
       WHERE  p.objID = x.slaveObjID     AND
              (psfMagErr_i < 0.5 OR psfMagErr_z < 0.5) AND
              x.distanceMins < 1.0/60.0
       ) AND
/*     Use only frame sets overlapping with SDSS-DR7: */
       s.frameSetID IN (
       SELECT DISTINCT(s.frameSetID)
       FROM   lasSource             AS n,
              lasSourceXDR7PhotoObj AS x
       WHERE  n.sourceID = x.masterObjID
       ) AND
/*     	Star-like morphological classification: */
       	s.mergedClass BETWEEN -2.0 and -1.0 AND
       	s.mergedClassStat BETWEEN -3.0 AND +3.0 AND
/*     	Reasonably circular images: */
       	s.yEll < 0.45 AND s.j_1Ell < 0.45 AND
/*     	IR pairs within 0.75 arcsec: */
	(((s.yXi  BETWEEN -0.75 AND +0.75) AND
	(s.yEta   BETWEEN -0.75 AND +0.75) AND 
       	(s.j_1Xi  BETWEEN -0.75 AND +0.75) AND
       	(s.j_1Eta BETWEEN -0.75 AND +0.75)) OR
/*	Or MJD of OBS separated by more than 1 days */
	((my.mjdObs - mj.mjdObs) > 1 or (mj.mjdObs - my.mjdObs) > 1)) AND 	
/*	YJ measured to 3: */
       	s.yAperMag3Err   < 0.30 AND s.yAperMag3Err   > 0 AND
       	s.j_1AperMag3Err < 0.30 AND s.j_1AperMag3Err > 0 AND
/*	H and K dropout: */
	s.hAPerMag3Err < 0 AND s.kAperMag3Err < 0 AND
	s.hAperMag3 < 0 AND s.kAperMag3 < 0 AND
/*	J brighter than 19.5 */
	s.j_1AperMag3 < 19.30
ORDER BY ra
\end{lstlisting}
\caption{The SQL query used to select candidates via our $YJ$-only channel in the case where they have not been detected in SDSS.
\label{cod:yj_nosdss}}
\end{Query}

\section{Summary of photometric observations}

\begin{table*}
\begin{tabular}{l c c c c c }
\hline
 Object & Filter & Instrument &  Date & Program ID & $T_{int}$ \\
\hline 
 ULASJ0128+0633  & $z'$ & ACAM & 2011-01-08 & W/10B/P16 & 1800s\\
  ULASJ0130+0804 & MKO Y & WFCAM & 2010-11-22 & U/10B/8 &  280s \\
  		 & MKO J & WFCAM & 2010-11-22 & U/10B/8 &  120s \\
		 & MKO H & WFCAM & 2010-11-22 & U/10B/8 &  1000s \\
		 & MKO K & WFCAM & 2010-11-22 & U/10B/8 &  1000s \\
  ULASJ0226+0702 & MKO Y & WFCAM & 2010-11-22 & U/10B/8 &  280s \\
  		 & MKO J & WFCAM & 2010-11-22 & U/10B/8 &  120s \\ 
		 & MKO H & WFCAM & 2010-11-22 & U/10B/8 &  1000s \\ 
		 & MKO K & WFCAM & 2010-11-22 & U/10B/8 &  1000s \\ 
  ULASJ0245+0653 & MKO J & WFCAM & 2009-12-16 &	U/09B/7 & 120s \\
  ULASJ0255+0616 & MKO J & WFCAM & 2009-12-16 &	U/09B/7 & 120s \\
  		 & MKO H & WFCAM & 2009-12-16 &	U/09B/7 & 1000s \\
  ULASJ0329+0430 & MKO Y & WFCAM & 2010-11-24 & U/10B/8 & 280s \\	
  		 & MKO J & WFCAM & 2010-11-24 & U/10B/8 & 120s \\
		 & MKO H & WFCAM & 2010-11-24 & U/10B/8 & 1000s \\
		 & MKO K & WFCAM & 2010-11-24 & U/10B/8 & 1000s \\
  ULASJ0746+2355 & $z'$  & DOLORES & 2010-12-27 & A22TAC\_96 & 1200s \\
  				& $H$ & LIRIS & 2011-01-09 & W/10B/P16 & 3000s \\
				& CH$_4$l & LIRIS & 2011-01-09 & W/10B/P16 & 3000s\\	
  ULASJ0747+2455 & MKO Y & WFCAM & 2010-04-19 & U/10A/6 & 280s \\
  		 & MKO J & WFCAM & 2010-04-19 & U/10A/6 & 120s \\
		 & MKO H & WFCAM & 2010-04-19 & U/10A/6 & 1000s \\
		 & MKO K & WFCAM & 2010-04-19 & U/10A/6 & 1000s \\
  ULASJ0758+2225 & MKO J & WFCAM & 2009-12-16 & U/09B/7 & 120s \\
  		 & MKO H & WFCAM & 2009-12-16 & U/09B/7 & 1000s \\
  ULASJ0759+1855 & $z'$  & DOLORES & 2010-12-18 & A22TAC\_96 & 1200s \\
  ULASJ0800+1908 & $z'$  & DOLORES & 2010-12-28 & A22TAC\_96 & 1200s \\
  ULASJ0809+2126 & MKO J & WFCAM & 2010-01-08 & U/09B/7 & 120s \\
  		 & MKO H & WFCAM & 2010-01-08 & U/09B/7 & 1000s \\
  ULASJ0814+2452 & $z'$ & DOLORES & 2010-12-28 & A22TAC\_96 & 1200s \\ 
  				& $H$ & LIRIS & 2011-01-09	& W/2010B/P16 & 740s\\
				& CH$_4$l & LIRIS & 2011-01-09	& W/2010B/P16 & 740s\\
  ULASJ0815+2711 & MKO J & WFCAM & 2010-01-10 & U/09B/7	& 120s \\
  		 & MKO H & WFCAM & 2010-01-10 & U/09B/7	& 1000s \\
  ULASJ0821+2509 & MKO J & WFCAM & 2010-01-11 & U/09B/7	& 120s \\
  		 & MKO H & WFCAM & 2010-01-11 & U/09B/7	& 1000s \\
  ULASJ0847+0350 & $z'$ & ACAM & 2011-01-12 &  W/10B/P16 &1200s \\ 
  		& MKO Y & WFCAM & 2010-11-22 & U/10B/8	& 280s \\ 
  		 & MKO J & WFCAM & 2010-11-22 & U/10B/8 & 120s \\
		 & MKO H & WFCAM & 2010-11-22 & U/10B/8 & 1000s \\
		 & MKO K & WFCAM & 2010-11-22 & U/10B/8 & 1000s \\
 ULASJ0927+3413 & $z'$ & DOLORES & 2011-05-11 & A23TAC\_28 & 1200s \\
  ULASJ0929+0409 & MKO Y & WFCAM & 2010-11-22 & U/10B/8	& 280s \\
  		 & MKO J & WFCAM & 2010-11-22 & U/10B/8 & 120s \\ 
		 & MKO H & WFCAM & 2010-11-22 & U/10B/8 & 1000s \\ 
		 & MKO K & WFCAM & 2010-11-22 & U/10B/8 & 1000s \\ 		  
  ULASJ0950+0117 & MKO Y & WFCAM & 2010-01-08 & U/09B/7 & 280s \\
  		 & MKO J & WFCAM & 2009-12-16 & U/09B/7 & 120s \\
		 & MKO H & WFCAM & 2009-12-16 & U/09B/7 & 1000s \\
		 & MKO K & WFCAM & 2010-01-08 & U/09B/7 & 1000s \\ 
  ULASJ0954+0623 & MKO Y & WFCAM & 2010-11-23 & U/10B/8 & 280s \\
  		 & MKO J & WFCAM & 2010-11-23 & U/10B/8 & 120s \\
		 & MKO H & WFCAM & 2010-11-23 & U/10B/8 & 1000s \\
		 & MKO K & WFCAM & 2010-11-23 & U/10B/8 & 1000s \\
  ULASJ1021+0544 & MKO Y & WFCAM & 2010-11-25 & U/10B/8 & 280s \\
  		 & MKO J & WFCAM & 2010-11-25 & U/10B/8 & 120s \\
		 & MKO H & WFCAM & 2010-11-25 & U/10B/8 & 1000s \\
		 & MKO K & WFCAM & 2010-11-25 & U/10B/8 & 1000s \\
  ULASJ1023+0447 & $z'$ & DOLORES & 2010-12-28 & A22TAC\_96 & 900s \\ 
  		 & MKO Y & WFCAM & 2010-11-26 & U/10B/8 & 280s \\
  		 & MKO J & WFCAM & 2010-11-26 & U/10B/8 & 120s \\
  		& $H$ & LIRIS & 2011-01-09	& W/2010B/P16 & 900s\\
		& CH$_4$l & LIRIS & 2011-01-09	& W/2010B/P16 & 900s\\
		 & MKO H & WFCAM & 2010-11-26 & U/10B/8 & 1000s \\
		 & MKO K & WFCAM & 2010-11-26 & U/10B/8 & 1000s \\
  ULASJ1029+0935 & MKO Y & WFCAM & 2010-11-23 & U/10B/8 & 280s \\
  		 & MKO J & WFCAM & 2010-11-23 & U/10B/8 & 120s \\
		 & MKO H & WFCAM & 2010-11-23 & U/10B/8 & 1000s \\
		 & MKO K & WFCAM & 2010-11-23 & U/10B/8 & 1000s \\
  ULASJ1043+1048 & MKO Y & WFCAM & 2010-12-06 & U/10B/8 & 280s \\
  		 & MKO J & WFCAM & 2010-12-06 & U/10B/8 & 120s \\
		 & MKO H & WFCAM & 2010-12-06 & U/10B/8 & 1000s \\
		 & MKO K & WFCAM & 2010-12-06 & U/10B/8 & 1000s \\
\hline
\end{tabular}
\caption{Summary of broad band photometric observations.
\label{tab:photobs}}
\end{table*}

\begin{table*}
\addtocounter{table}{-1}
\begin{tabular}{l c c c c c }
\hline
 Object & Filter & Instrument &  UT Date & Program ID & $T_{int}$ \\
\hline 

  ULASJ1051-0154 & MKO Y & WFCAM & 2010-12-06 & U/10B/8 & 280s \\
  		 & MKO J & WFCAM & 2010-12-06 & U/10B/8 & 120s \\ 
		 & MKO H & WFCAM & 2010-12-06 & U/10B/8 & 1000s \\ 
		 & MKO K & WFCAM & 2010-12-06 & U/10B/8 & 1000s \\ 
 ULASJ1137+1126 & $z'$ & DOLORES & 2011-05-08 & A23TAC\_28 & 900s \\
  ULASJ1204-0150 & MKO J & UFTI  & 2009-01-25 & U/08B/15 & 300s \\
  		 & MKO H & UFTI  & 2009-01-25 & U/08B/15 & 1800s \\
  ULASJ1206+1018 & MKO J & WFCAM & 2009-07-14 & U/09A/1	& 120s \\
  		 & MKO H & WFCAM & 2009-07-14 & U/09A/1	& 1000s \\
  ULASJ1212+1010 & MKO J & WFCAM & 2009-07-13 & U/09A/1	& 120s \\
  		 & MKO H & WFCAM & 2009-07-13 & U/09A/1	& 1000s \\
  ULASJ1258+0307 & MKO J & WFCAM & 2010-04-20 & U/10A/6	& 120s \\
  		 & MKO H & WFCAM & 2010-04-20 & U/10A/6	& 1000s \\
  ULASJ1302+1434 & $z'$ & DOLORES & 2011-07-07 & A23TAC\_28 & 1200s \\
  		 & MKO J & WFCAM & 2009-07-12 & U/09A/1 & 120s \\
  		 & MKO H & WFCAM & 2009-07-12 & U/09A/1 & 1000s \\
  ULASJ1335+1506 & MKO Y & UFTI & 2009-01-11 & U/08B/15 & 540s \\
  		 & MKO J & UFTI & 2009-01-11 & U/08B/15 & 300s \\
		 & MKO H & UFTI & 2009-01-11 & U/08B/15 & 1800s \\
  ULASJ1338-0142 & MKO Y & WFCAM & 2010-05-13 & U/10A/6 & 280s \\
  		 & MKO J & WFCAM & 2010-05-13 & U/10A/6 & 120s \\
		 & MKO H & WFCAM & 2010-05-13 & U/10A/6 & 1000s \\ 
		 & MKO K & WFCAM & 2010-05-13 & U/10A/6 & 1000s \\
  ULASJ1339-0056 & MKO Y & WFCAM & 2010-05-12 & U/10A/6 & 280s \\
  		 & MKO J & WFCAM & 2010-05-12 & U/10A/6 & 120s \\
		 & MKO H & WFCAM & 2010-05-12 & U/10A/6 & 1000s \\ 
		 & MKO K & WFCAM & 2010-05-12 & U/10A/6 & 1000s \\ 
  ULASJ1421+0136 & MKO J & WFCAM & 2009-07-12 & U/09A/1 & 120s \\
  		 & MKO H & WFCAM & 2009-07-12 & U/09A/1 & 1000s \\
ULASJ1425+0451 & $z'$ & DOLORES & 2011-05-12 & A23TAC\_28 & 1200s \\
  ULASJ1516+0110 & MKO J & WFCAM & 2010-04-20 & U/10A/6 & 120s \\
  		 & MKO H & WFCAM & 2010-04-20 & U/10A/6 & 1000s \\
ULASJ1534+0556 & $z'$ & DOLORES & 2011-05-12 & A23TAC\_28 & 1800s \\
  ULASJ1549+2621 & MKO J & WFCAM & 2010-05-09 & U/10A/6 & 120s \\
  		 & MKO H & WFCAM & 2010-05-09 & U/10A/6 & 1000s \\
  ULASJ1601+2646 & $z'$ & DOLORES & 2011-05-14 & A23TAC\_28 & 1650s \\
  		 & MKO Y & WFCAM & 2010-05-15 & U/10A/6 & 280s \\
  		 & MKO J & WFCAM & 2010-05-15 & U/10A/6 & 120s \\ 	
		 & MKO H & WFCAM & 2010-05-15 & U/10A/6 & 1000s \\ 
		 & MKO K & WFCAM & 2010-05-15 & U/10A/6 & 1000s \\ 
ULASJ1614+2442 & $z'$ & DOLORES & 2011-05-09 & A23TAC\_28 & 1200s \\
  ULASJ1619+2358 & MKO J & WFCAM & 2010-05-11 & U/10A/6 & 120s \\  
  		 & MKO H & WFCAM & 2010-05-11 & U/10A/6 & 1000s \\ 
  ULASJ1619+3007 & MKO J & WFCAM & 2010-05-12 & U/10A/6 & 120s \\
  		 & MKO H & WFCAM & 2010-05-12 & U/10A/6 & 1000s \\
  ULASJ1626+2524 & MKO J & WFCAM & 2010-06-13 & U/10A/6 & 120s \\
  		 & MKO H & WFCAM & 2010-06-13 & U/10A/6 & 1000s \\
ULASJ2116-0101 & $z'$ & DOLORES & 2011-05-11 & A23TAC\_28 & 900s \\
  ULASJ2237+0642 & MKO Y & WFCAM & 2010-07-11 & U/10B/8 & 280s \\
  		 & MKO J & WFCAM & 2010-11-22 & U/10B/8 & 260s \\
		 & MKO H & WFCAM & 2010-11-22 & U/10B/8 & 2000s \\
		 & MKO K & WFCAM & 2010-11-03 & U/10B/8 & 2000s \\
  ULASJ2300+0703 & $z'$ & DOLORES & 2011-07-08 & A23TAC\_28 & 900s \\
  	        & MKO Y & WFCAM & 2009-07-14 & U/09A/1 & 120s \\
  		 & MKO J & WFCAM & 2009-07-14 & U/09A/1 & 120s \\
		 & MKO H & WFCAM & 2009-07-14 & U/09A/1 & 400s \\
		 & MKO K & WFCAM & 2009-07-14 & U/09A/1 & 400s \\
  ULASJ2318+0433 & $z'$ & DOLORES & 2011-11-16 & A24TAC\_49 & 900s\\
  ULASJ2326+0509 & $z'$ & DOLORES & 2012-01-13 & A24TAC\_49 & 900s\\
  ULASJ2331+0426 & $z'$ & DOLORES & 2011-11-16 & A24TAC\_49 & 900s\\
  ULASJ2342+0856 & MKO Y & WFCAM & 2010-11-25 & U/10B/8 & 280s \\
  		 & MKO J & WFCAM & 2010-11-25 & U/10B/8 & 120s \\ 	
		 & MKO H & WFCAM & 2010-11-25 & U/10B/8 & 1000s \\ 
		 & MKO K & WFCAM & 2010-11-25 & U/10B/8 & 1000s \\ 
  ULASJ2357+0132 & $z'$ & DOLORES & 2011-07-08 & A23TAC\_28 & 900s \\
  		 & MKO Y & WFCAM & 2010-11-22 & U/10B/8 & 280s \\
  		 & MKO J & WFCAM & 2010-11-22 & U/10B/8 & 120s \\
		 & MKO H & WFCAM & 2010-11-22 & U/10B/8 & 1000s \\
		 & MKO K & WFCAM & 2010-11-22 & U/10B/8 & 1000s \\
\hline
\end{tabular}
\caption{Continued.
\label{tab:photobs}}
\end{table*}

\begin{table*}
\begin{tabular}{l c c c c c}
\hline
Name & TNG program & UT Date & $T_{int}$ (s) & $N_{coadds}$ & $N_{dither}$ \\
\hline
  ULASJ0007+0112 & A24TAC\_49 & 2011-10-27 & 30.0 & 1 & 30\\
  ULASJ0127+1539 & A24TAC\_49 & 2011-11-19 & 20.0 & 2 & 30\\
  ULASJ0128+0633 & A22TAC\_96 & 2010-11-06 & 20.0 & 3 & 30\\
  ULASJ0130+0804 & A22TAC\_96 & 2010-12-25 & 30.0 & 1 & 30\\
  ULASJ0133+0231 & A24TAC\_49 & 2011-11-18 & 20.0 & 2 & 30\\
  ULASJ0139+1503 & A24TAC\_49 & 2011-10-27 & 45.0 & 1 & 30\\
  ULASJ0200+0908 & A24TAC\_49 & 2011-10-28 & 30.0 & 1 & 30\\
  ULASJ0745+2332 & A24TAC\_49 & 2011-10-28 & 60.0 & 1 & 30\\
  ULASJ0759+1855 & A24TAC\_49 & 2011-10-27 & 60.0 & 1 & 30\\
  ULASJ0811+2529 & A22TAC\_96 & 2010-12-27 & 30.0 & 1 & 30\\
  ULASJ0847+0350 & A23TAC\_28 & 2011-05-07 & 30.0 & 1 & 30\\
  ULASJ0926+0402 & A25TAC\_32 & 2012-04-29 & 30.0 & 4 & 10\\
  ULASJ0927+3413 & A23TAC\_28 & 2011-05-12 & 30.0 & 2 & 30\\
  ULASJ0929+0409 & A23TAC\_28 & 2011-05-07 & 30.0 & 1 & 30\\
  ULASJ0954+0623 & A23TAC\_28 & 2011-05-09 & 30.0 & 1 & 30\\
  ULASJ1021+0544 & A24TAC\_49 & 2012-01-16 & 30.0 & 4 & 10\\
  ULASJ1029+0935 & A23TAC\_28 & 2011-05-09 & 30.0 & 1 & 30\\
  ULASJ1042+1212 & A24TAC\_49 & 2012-01-16 & 30.0 & 4 & 10\\
  ULASJ1043+1048 & A23TAC\_28 & 2011-05-09 & 20.0 & 2 & 30\\
  ULASJ1051-0154 & A23TAC\_28 & 2011-05-10 & 30.0 & 1 & 30\\
  ULASJ1053+0157 & A24TAC\_49 & 2012-01-16 & 30.0 & 4 & 10\\
  ULASJ1137+1126 & A23TAC\_28 & 2011-05-10 & 20.0 & 2 & 30\\
  ULASJ1152+0359 & A22TAC\_96 & 2010-12-26 & 30.0 & 1 & 30\\
  ULASJ1152+1134 & A24TAC\_49 & 2012-01-17 & 30.0 & 4 & 10\\
  ULASJ1223-0131 & A25TAC\_32 & 2012-04-29 & 30.0 & 4 & 10\\
  ULASJ1228+0407 & A23TAC\_28 & 2011-05-11 & 30.0 & 1 & 30\\
  ULASJ1254+1222 & A24TAC\_49 & 2012-01-14 & 30.0 & 4 & 10\\
  ULASJ1259+2933 & A24TAC\_49 & 2012-02-01 & 30.0 & 4 & 10\\
  ULASJ1302+1434 & A24TAC\_49 & 2012-01-16 & 30.0 & 4 & 10\\
  ULASJ1335+1506 & A23TAC\_28 & 2011-05-10 & 30.0 & 1 & 30\\
  ULASJ1417+1330 & A23TAC\_28 & 2011-05-07 & 30.0 & 1 & 30\\
  ULASJ1425+0451 & A23TAC\_28 & 2011-05-13 & 30.0 & 2 & 30\\
  ULASJ1449+1147 & A23TAC\_28 & 2011-05-07 & 30.0 & 1 & 30\\
  ULASJ1516+0110 & A23TAC\_28 & 2011-07-09 & 30.0 & 1 & 30\\
  ULASJ1534+0556 & A23TAC\_28 & 2011-05-13 & 30.0 & 2 & 30\\
  ULASJ1549+2621 & A23TAC\_28 & 2011-07-10 & 30.0 & 1 & 30\\
  ULASJ1614+2442 & A23TAC\_28 & 2011-05-10 & 30.0 & 1 & 30\\
  ULASJ1617+2350 & A23TAC\_28 & 2011-05-07 & 30.0 & 1 & 30\\
  ULASJ1619+2358 & A23TAC\_28 & 2011-07-11 & 30.0 & 2 & 30\\
  ULASJ1619+3007 & A23TAC\_28 & 2011-05-08 & 30.0 & 2 & 30\\
  ULASJ2116-0101 & A24TAC\_49 & 2011-10-26 & 45.0 & 1 & 30\\
  ULASJ2300+0703 & A23TAC\_28 & 2011-07-09 & 30.0 & 1 & 30\\
  ULASJ2315+0344 & A24TAC\_49 & 2011-10-27 & 60.0 & 1 & 30\\
  ULASJ2318+0433 & A24TAC\_49 & 2012-01-16 & 30.0 & 4 & 10\\
  ULASJ2326+0201 & A24TAC\_49 & 2011-10-27 & 45.0 & 1 & 30\\
  ULASJ2326+0509 & A24TAC\_49 & 2012-01-16 & 30.0 & 4 & 10\\
  ULASJ2352+1244 & A24TAC\_49 & 2011-10-28 & 30.0 & 1 & 30\\
\hline
\end{tabular}
\caption{Summary of methane photometric observations using the TNG. $T_{int}$ gives the integration time for each co-add, $N_{coadd}$ is the number of co-added images at each dither point, $N_{dither}$ is the number of dither points in the mosaic.
\label{tab:ch4obs}}
\end{table*}

\section{Summary of spectroscopic observations}

\begin{table*}
\begin{tabular}{c c c c }
\hline
  \multicolumn{1}{ c }{Target} &
  \multicolumn{1}{c }{Instrument} &
  \multicolumn{1}{c }{UT Date} &
  \multicolumn{1}{c }{Programme ID} \\
\hline
  ULAS~J0007+0112 & GNIRS & 2011-11-14 & GN-2011B-Q-5\\
  ULAS~J0127+1539 & GNIRS & 2011-12-31 & GN-2011B-Q-43\\
  ULAS~J0128+0633 & GNIRS & 2010-12-15 & GN-2010B-Q-41\\
  ULAS~J0130+0804 & GNIRS & 2011-10-15 & GN-2011B-Q-5\\
  ULAS~J0133+0231 & GNIRS & 2011-12-17 & GN-2011B-Q-43\\
  ULAS~J0139+1503 & GNIRS & 2011-11-24 & GN-2011B-Q-43\\
  ULAS~J0200+0908 & GNIRS & 2011-11-26 & GN-2011B-Q-43\\
  ULAS~J0226+0702 & GNIRS & 2011-10-15 & GN-2011B-Q-5\\
  ULAS~J0245+0653 & IRCS & 2009-12-30 & o09164\\
  ULAS~J0255+0616 & X-shooter &  2010-12-01 & 086.C-0450(A) \\
  ULAS~J0329+0430 & NIRI & 2009-11-03 & GN-2009B-Q-62\\
  ULAS~J0745+2332 & GNIRS & 2011-11-22 & GN-2011B-Q-43\\
  ULAS~J0746+2355 & GNIRS & 2011-04-23 & GN-2011A-Q-73\\
  ULAS~J0747+2455 & IRCS & 2009-12-30 & o09164\\
  ULAS~J0758+2225 & IRCS & 2009-12-31 & o09164\\
  ULAS~J0759+1855 & GNIRS & 2011-11-25 & GN-2011B-Q-5\\
  ULAS~J0809+2126 & IRCS & 2009-12-30 & o09164\\
  ULAS~J0811+2529 & GNIRS & 2010-12-30 & GN-2010B-Q-41\\
  ULAS~J0814+2452 & GNIRS & 2011-04-19 & GN-2011A-Q-73\\
  ULAS~J0815+2711 & IRCS & 2009-12-30 & o09164\\
  ULAS~J0819+2103 & NIRI & 2009-11-01 & GN-2009B-Q-62\\
  ULAS~J0821+2509 & NIRI & 2009-12-31 & GN-2009B-Q-62\\
  ULAS~J0926+0402 & FIRE & 2012-05-09 & \\
  ULAS~J0927+2524 & GNIRS & 2011-06-16 & GN-2011A-Q-73\\
  ULAS~J0929+0409 & IRCS & 2011-01-24 & o10148\\
  ULAS~J0932+3102 & NIRI & 2009-12-31 & GN-2009B-Q-62\\
  ULAS~J0950+0117 & IRCS & 2009-05-07 & o09118\\
  ULAS~J0950+0117 & NIRI(H) & 2009-12-08 & GN-2009B-Q-62\\
  ULAS~J0950+0117 & NIRI(K) & 2009-12-30 & GN-2009B-Q-62\\
  ULAS~J0954+2452 & FIRE & 2012-05-09 & \\
  ULAS~J1021+0544 & IRCS & 2011-01-23 & o10148\\
  ULAS~J1023+0447 & GNIRS & 2011-06-12 & GN-2011A-Q-73\\
  ULAS~J1029+0935 & IRCS & 2011-01-23 & o10148\\
  ULAS~J1042+1212 & GNIRS & 2012-03-05 & GN-2012A-Q-84\\
  ULAS~J1043+1048 & GNIRS & 2011-06-17 & GN-2011A-Q-73\\
  ULAS~J1051+0154 & IRCS & 2011-01-24 & o10148\\
  ULAS~J1053+0157 & FIRE & 2012-05-09 & \\
  ULAS~J1111+0518 & FIRE & 2012-05-09 & \\
  ULAS~J1152+0359 & IRCS & 2011-01-24 & o10148\\
  ULAS~J1152+1134 & GNIRS & 2012-06-06 & GN-2012A-Q-84\\
  ULAS~J1155+0445 & NIRI & 2009-12-31 & GN-2009B-Q-62\\
  ULAS~J1204+0150 & NIRI & 2009-04-16 & GN-2009A-Q-16\\
  ULAS~J1206+1018 & NIRI & 2010-02-06 & GN-2010A-Q-44\\
  ULAS~J1212+1010 & NIRI & 2010-01-28 & GN-2009B-Q-62\\
  ULAS~J1223-0131 & FIRE & 2012-05-09 & \\
  ULAS~J1228+0407 & FIRE & 2012-05-09 & \\
  ULAS~J1258+0307 & IRCS & 2010-04-05 & o10121\\
  ULAS~J1259+2933 & GNIRS & 2012-04-20 & GN-2012A-Q-84\\
  ULAS~J1302+1434 & FIRE & 2012-05-09 & \\
  ULAS~J1335+1506 & IRCS & 2009-05-06 & o09118\\
  ULAS~J1338-0142 & NIRI & 2010-05-01 & GN-2010A-Q-44\\
  ULAS~J1339-0056 & IRCS & 2010-04-05 & o10121\\
  ULAS~J1339+0104 & IRCS & 2010-04-05 & o10121\\
  ULAS~J1417+1330 & GNIRS & 2011-05-16 & GN-2011A-Q-73\\
  ULAS~J1421+0136 & NIRI & 2010-01-28 & GN-2009B-Q-62\\
  ULAS~J1425+0451 & GNIRS & 2011-07-09 & GN-2011A-Q-73\\
  ULAS~J1449+1147 & GNIRS & 2011-05-15 & GN-2011A-Q-73\\
  ULAS~J1516+0110 & IRCS & 2010-04-06 & o10121\\
  ULAS~J1517+0529 & GNIRS & 2011-08-13 & GN-2011B-Q-5\\
  ULAS~J1534+0556 & GNIRS & 2011-07-09 & GN-2011A-Q-73\\
  ULAS~J1536+0155 & IRCS & 2010-04-05 & o10121\\
  ULAS~J1549+2621 & IRCS & 2010-04-06 & o10121\\
\hline\end{tabular}
\caption{Dates, instruments and programme numbers for spectra obtained for this work. 
\label{tab:specobs}}
\end{table*}

\begin{table*}
\addtocounter{table}{-1}
\begin{tabular}{c c c c }
\hline
  \multicolumn{1}{ c }{Target} &
  \multicolumn{1}{c }{Instrument} &
  \multicolumn{1}{c }{UT Date} &
  \multicolumn{1}{c }{Programme ID} \\
\hline
  ULAS~J1601+2646 & GNIRS & 2011-03-17 & GN-2011A-Q-73\\
  ULAS~J1614+2442 & GNIRS & 2011-07-10 & GN-2011A-Q-73\\
  ULAS~J1617+2350 & GNIRS & 2011-05-15 & GN-2011A-Q-73\\
  ULAS~J1619+2358 & NIRI & 2010-04-27 & GN-2010A-Q-44\\
  ULAS~J1619+3007 & GNIRS & 2012-05-31 & GN-2012A-Q-84\\
  ULAS~J1626+2524 & IRCS & 2010-04-05 & o10121\\
  ULAS~J1639+3232 & NIRI & 2010-04-30 & GN-2010A-Q-44\\
  ULAS~J2116-0101 & GNIRS & 2011-11-14 & GN-2011B-Q-5\\
  ULAS~J2237+0642 & GNIRS & 2010-12-07 & GN-2010B-Q-41\\
  ULAS~J2300+0703 & NIRI & 2009-06-15 & GN-2009A-Q-16\\
  ULAS~J2315+0344 & GNIRS & 2011-12-24 & GN-2011B-Q-43\\
  ULAS~J2318+0433 & GNIRS & 2012-06-02 & GN-2012A-Q-84\\
  ULAS~J2326+0201 & GNIRS & 2011-11-28 & GN-2011B-Q-43\\
  ULAS~J2331+0426 & GNIRS & 2012-06-06 & GN-2012A-Q-84\\
  ULAS~J2342+0856 & NIRI & 2009-10-31 & GN-2009B-Q-62\\
  ULAS~J2352+1244 & GNIRS & 2011-11-16 & GN-2011B-Q-43\\
  ULAS~J2357+0132 & GNIRS & 2010-12-08 & GN-2010B-Q-41\\
\hline\end{tabular}
\caption{Dates, instruments and programme numbers for spectra obtained for this work. 
\label{tab:specobs}}
\end{table*}

\section*{Acknowledgements}
We thank our referee, J. Davy Kirkpatrick, for a helpful review which substantially improved the quality of this manuscript.
Based on observations made under project A22TAC$\_$96 on the Italian Telescopio Nazionale Galileo (TNG) operated on the island of La Palma by the Fundaci—n Galileo Galilei of the INAF (Istituto Nazionale di Astrofisica) at the Spanish Observatorio del Roque de los Muchachos of the Instituto de Astrofisica de Canarias. Based on observations obtained at the Gemini Observatory,
which is operated by the
Association of Universities for Research in Astronomy, Inc., under a cooperative agreement
with the NSF on behalf of the Gemini partnership: the National Science Foundation (United
States), the Science and Technology Facilities Council (United Kingdom), the
National Research Council (Canada), CONICYT (Chile), the Australian Research Council (Australia),
Minist\'{e}rio da Ci\^{e}ncia e Tecnologia (Brazil)
and Ministerio de Ciencia, Tecnolog\'{i}a e Innovaci\'{o}n Productiva (Argentina).
We would like to acknowledge the support of
the Marie Curie 7th European Community Framework Programme grant n.247593
Interpretation and Parameterization of Extremely Red COOL dwarfs (IPERCOOL)
International Research Staff Exchange Scheme.
ADJ is supported by a Fondecyt postdoctorado fellowship, under project number 3100098, and is also partially supported by the proyecto Basal PB06 (CATA) and the Joint Committee ESO-Government Chile.
AHA thanks CNPq grant PQ306775/2009-3 and SHAO/CAS Visiting Professorship grant.
CGT is supported by ARC grant DP0774000.
SKL's research is supported by the Gemini Observatory.
JG is supported by RoPACS, a Marie Curie Initial
Training Network funded by the European CommissionÕs Seventh Framework
Programme. 
NL acknowledges funding from the Spanish Ministry of Science and Innovation through the Ram\'on y Cajal fellowship number 08-303-01-02 and the project number AYA2010-19136.
This research has made use of the NASA/ IPAC Infrared Science Archive, which is operated by the Jet Propulsion Laboratory, California Institute of Technology, under contract with the National Aeronautics and Space Administration.
This research has made use of the SIMBAD database,
operated at CDS, Strasbourg, France, and has benefited from the SpeX
Prism Spectral Libraries, maintained by Adam Burgasser at
http://www.browndwarfs.org/spexprism.
The authors wish to recognise and acknowledge the very significant cultural role and reverence that the summit of Mauna Kea has always had within the indigenous Hawaiian community.  We are most fortunate to have the opportunity to conduct observations from this mountain.
\bibliographystyle{mn2e}
\bibliography{refs}

\end{document}